\documentclass[aps, prx, reprint, amsmath, amssymb, superscriptaddress, longbibliography]{revtex4-2}
\usepackage[T1]{fontenc}
\usepackage{lmodern}
\usepackage{graphicx}
\usepackage{bm}
\usepackage{hyperref}
\usepackage{booktabs}
\newcommand{\paperfigure}[2][]{%
  \IfFileExists{#2}{\includegraphics[#1]{#2}}{%
    \fbox{\parbox[c][35mm][c]{0.82\linewidth}{\centering
    Original figure required: \texttt{\detokenize{#2}}}}}}

\newcommand{\dd}{\mathrm{d}}

\begin{document}
\title{Emergence of Memory-Dressed Collective Field Dynamics in Neural Systems}
\author{Byung Gyu Chae}
\affiliation{Electronics and Telecommunications Research Institute, 218 Gajeong-ro, 
Yuseong-gu, Daejeon 34129, Republic of Korea}
\email{bgchae@etri.re.kr}

\begin{abstract}
We establish a microscopic framework for the emergence of
memory-dressed collective field dynamics in interacting nonlinear
neural systems. The dynamical Hubbard--Stratonovich formulation
and Mori--Zwanzig projection provide a common microscopic basis
for collective ordering and memory feedback, with the static
memory self-energy entering the collective instability condition.
For a neural model with a collective quartic interaction, we derive
an exact macroscopic field equation incorporating local cubic
stabilization, retarded memory feedback, external driving, and
effective fluctuations. We also obtain an exact relation between
the response-weighted relaxation spectrum and the irreducible
memory spectrum. In the flat gapless continuum, an integrable
memory kernel, $K_\Sigma(t)\propto[t\ln^2(\Lambda t)]^{-1}$,
produces a nonintegrable $1/t$ reference response through repeated
memory feedback. The equilibrium fluctuation analysis reveals
that collective ordering and temporal persistence depend on
distinct inverse-rate spectral moments. For infrared spectral
exponents $0<\beta\leq1$, the integrated correlation time diverges
even when the static collective susceptibility remains finite.
Controlled numerical spectral constructions demonstrate the
infrared control of collective response and ordering thresholds,
while Transformer Jacobian spectra reveal near-marginal infrared
scaling and transient spectral enhancement during learning,
consistent with the spectral picture developed here.
The present framework identifies the infrared organization of
internal relaxation modes as a microscopic mechanism shaping
collective ordering, memory dynamics, and temporal persistence.
\end{abstract}

\maketitle

\section{Introduction}
\label{sec:intro}

Collective neural dynamics has long been investigated using
neuronal network models and statistical-mechanical approaches,
with experimental studies revealing low-dimensional population
dynamics and signatures of criticality in cortical activity
~\cite{gerstner2014,mante2013,chialvo2010,beggs2003}.
Experimental studies have further revealed long-range temporal
correlations, scale-free fluctuations, and their relationships
with behavioral dynamics, as well as the modulation of temporal
organization through neurofeedback training
~\cite{linkenkaer2001,palva2013,ros2016}.
Together with the hierarchy of intrinsic timescales identified
through electrophysiological measurements and large-scale circuit
models~\cite{murray2014,chaudhuri2015}, these observations suggest
that the organization of collective neural dynamics across broad
temporal scales is a fundamental aspect of cognition.

Wilson--Cowan dynamics and subsequent neural field formulations
describe interacting neural populations, with field-theoretic
extensions incorporating fluctuations and correlations beyond
deterministic population dynamics
~\cite{wilson1972,amari1977,buice2007,cook2022}.
Hopfield networks established how microscopic interactions can
generate collective attractors and associative memory, while
subsequent statistical-mechanical and generating-functional
approaches extended these descriptions to nonequilibrium
collective dynamics
~\cite{hopfield1982,hopfield1984,amit1985,coolen1993,mimura2009}.
Studies of random recurrent neural networks have characterized
chaotic dynamics, temporal correlations, and collective
fluctuations through dynamical mean-field and field-theoretic
approaches~\cite{sompolinsky1988,crisanti2018,dick2024}.
Extensions incorporating structured connectivity have further
elucidated the formation and stability of low-dimensional
collective states and their role in neural computation
~\cite{mastrogiuseppe2018,schuessler2020}.

A distinct question concerns how the emergence of macroscopic
collective order is related to the temporal persistence of
neural activity.
A stable collective state does not uniquely determine its
relaxation spectrum, while slow relaxation alone does not
establish an ordered phase.

Spectral and stability analyses of recurrent networks have
revealed the importance of long-time dependencies, gating,
and slow dynamical modes
~\cite{bengio1994,pascanu2013,wainrib2016,can2020,chen2018}.
Chen and Bialek further investigated the emergence of long
timescales in random neural networks and the constraints
required for a broad spectrum of slow modes~\cite{chen2024}.
Related developments in structured state-space models and
Jacobian analyses of learned neural architectures motivate
examining infrared relaxation
~\cite{gu2022,chae2026a,chae2026b}.
However, the eigenvalue density alone does not determine
the collective response; mode overlaps with the collective
interaction channel must also be taken into account.
This distinction becomes particularly important for
nonnormal dynamical operators, for which eigenvalues
alone do not determine transient dynamics or driven
responses~\cite{trefethen1997}.

These considerations raise a more fundamental question:
how are the formation and subsequent dynamics of a
macroscopic collective field related to memory feedback
generated by its own microscopic degrees of freedom?
Elimination of unresolved microscopic degrees of freedom
can generate retarded feedback and effective fluctuations,
so that the resulting collective dynamics need not be
Markovian even when the underlying microscopic dynamics
is Markovian. Such mechanisms have been extensively
investigated within projection-operator methods,
generalized Langevin equations, and statistical
coarse-graining~\cite{zwanzig1961,mori1965,mori1965cf,
chorin2002}.
A microscopic theory of collective neural dynamics must
therefore account not only for the emergence of collective
order but also for the temporal organization of its
effective dynamics.

Here we develop a microscopic framework for the emergence
of memory-dressed collective field dynamics in interacting
nonlinear neural systems. Starting from reciprocal stochastic
neural dynamics with a collective interaction channel,
we employ the MSRJD response formalism and a dynamical
Hubbard--Stratonovich transformation to investigate collective
instability, field formation, and nonlinear collective
vertices~\cite{martin1973,janssen1976,dedominicis1976,
hubbard1959,stratonovich1957}.

\begin{figure*}[t]
\centering
\paperfigure[scale=0.58, trim=0.3cm 1.9cm 0cm 0cm]{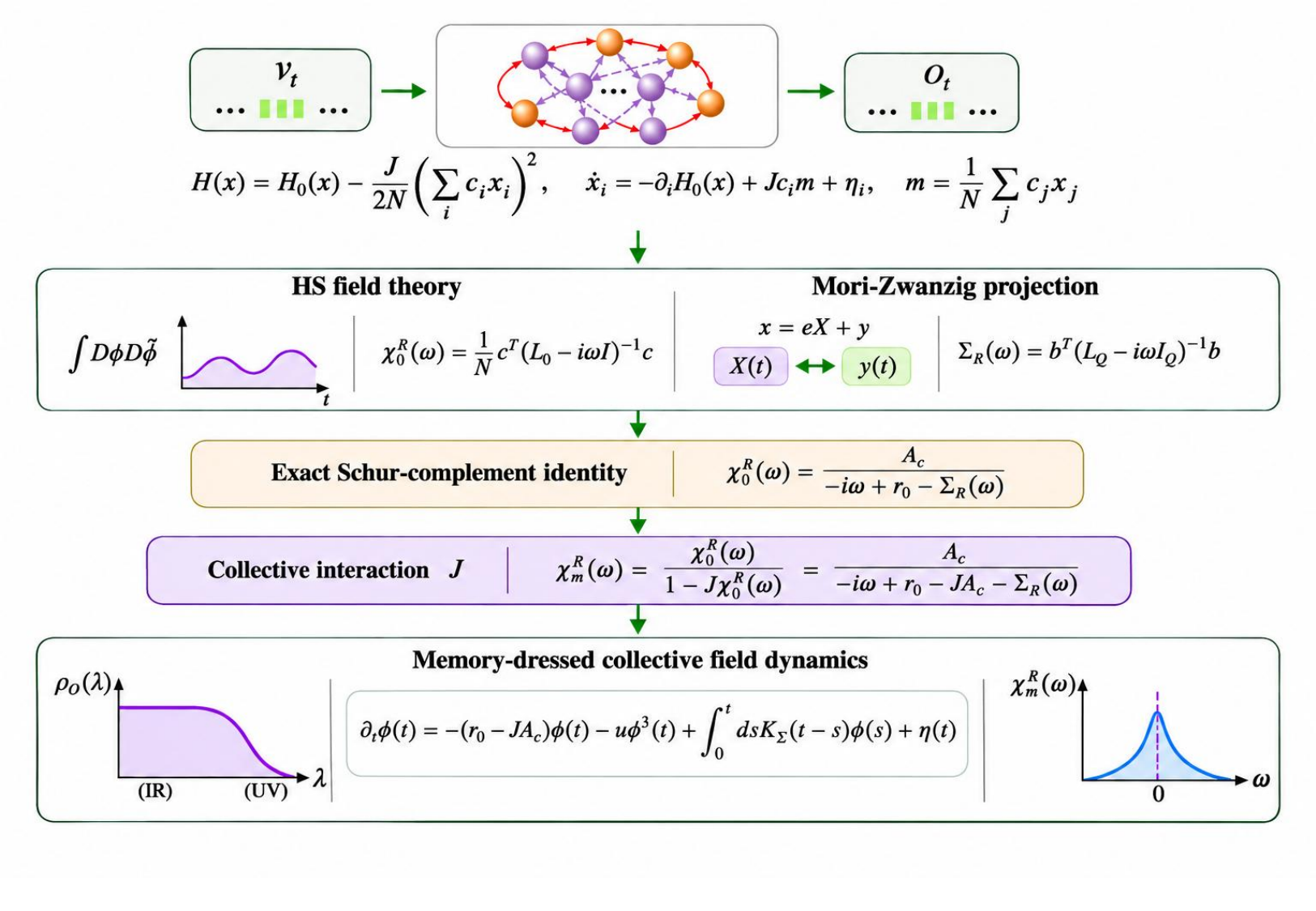}
\caption{
Microscopic framework for the emergence of memory-dressed
collective field dynamics in neural systems.
High-dimensional neural dynamics with rank-one collective
coupling is described through two complementary formulations:
the Hubbard--Stratonovich field theory, which yields the
reference collective response $\chi_0^R(\omega)$, and the
Mori--Zwanzig projection, which reveals the memory self-energy
$\Sigma_R(\omega)$ generated by internal relaxation modes.
The two formulations are connected through an exact
Schur-complement identity in the linearized reference dynamics.
Resummation of the collective interaction $J$ yields the
full collective susceptibility $\chi_m^R(\omega)$.
The resulting macroscopic dynamics incorporates nonlinear
stabilization, retarded memory feedback, and effective
fluctuations, as illustrated schematically in the bottom panel.
}
\label{fig:microscopic_framework}
\end{figure*}

We employ Mori--Zwanzig projection to resolve the microscopic
dynamics into the collective direction and its complementary
relaxation degrees of freedom~\cite{zwanzig1961,mori1965}.
This formulation provides a microscopic description of
retarded memory feedback and its associated self-energy.
Within the symmetric linearized reference dynamics,
we connect the reference collective response to the memory
self-energy through an exact Schur-complement identity,
consistent with established memory-function representations
of dynamical correlations and resolvent identities in
spectral theory~\cite{mori1965cf,gesztesy1997}.
As illustrated in Fig.~\ref{fig:microscopic_framework},
the two formulations provide complementary descriptions
of collective field formation and memory-dressed dynamics
within the same microscopic system.

Within this framework, we investigate how the infrared
organization of microscopic relaxation modes governs
collective instability and temporal persistence.
Particular attention is given to the distinction between
the response-weighted relaxation spectrum and the
irreducible memory spectrum, their spectral relation,
and the resulting long-time behavior of the corresponding
kernels. We further examine how nonlinear collective
interactions combine with microscopic memory feedback,
including an exactly reducible model with a collective
quartic interaction and quadratic complementary dynamics.

Using the equilibrium fluctuation--dissipation
relation~\cite{kubo1966}, we investigate the distinct roles
of inverse-rate spectral moments in collective ordering
and temporal correlations, and examine the conditions
under which long-time persistence can arise independently
of proximity to a collective critical point, in contrast
to conventional critical slowing down~\cite{hohenberg1977}.
The analysis also considers temporal persistence within
a locally stable ordered phase.

Finally, we examine these spectral mechanisms through
controlled numerical constructions of microscopic
relaxation operators and analyses of learning-induced
infrared organization in Transformer Jacobian spectra.
The latter characterize computational propagation rather
than directly determining the physical collective-channel
response, and are considered in this context as motivation
for investigating related mechanisms in learned neural
architectures.

The remainder of this paper is organized as follows.
Section~\ref{sec:micro} develops the microscopic field theory
of collective neural dynamics, including the effective
collective action, Gaussian instability, and spectral
representation of the reference response.
Section~\ref{sec:projection} establishes the microscopic
origin of memory feedback through Mori--Zwanzig projection.
Section~\ref{sec:infrared} investigates the infrared
organization of the response and memory spectra.
Section~\ref{sec:nonlinear} develops the nonlinear
memory-dressed macroscopic field dynamics, while
Sec.~\ref{sec:correlations} examines equilibrium collective
fluctuations and temporal persistence.
Section~\ref{sec:numerics} presents numerical tests
of infrared spectral control and examines learning-induced
infrared organization in Transformers.
Section~\ref{sec:discussion} discusses the physical
implications, limitations, and possible extensions
of the framework.
Finally, Sec.~\ref{sec:conclusion} summarizes the main results.
The Supplemental Material provides detailed spectral derivations,
numerical constructions and diagnostics, and a phenomenological
description of the learning trajectory.

\medskip
\section{Microscopic field theory of collective neural dynamics}
\label{sec:micro}

\subsection{Collective action and Gaussian instability}

Consider $N$ real neural activities $x_i(t)$ with reciprocal
interactions $W_{ij}=W_{ji}$. Following continuous-state
energy-based neural models~\cite{hopfield1984}, monotone invertible
activation functions admit local potentials satisfying
$U_i'(x_i)=\sigma_i^{-1}(x_i)$.
We separate a collective interaction channel,
$W_{ij}=W_{ij}^{(0)}+Jc_ic_j/N$, where $J>0$ and
$W^{(0)}$ retains the remaining microscopic interactions.
Defining
$H_0(\bm x)=\sum_iU_i(x_i)
-\frac12\sum_{ij}W_{ij}^{(0)}x_ix_j$
and $M=\sum_i c_ix_i$, the microscopic energy becomes
\begin{equation}
H(\bm x)=H_0(\bm x)-\frac{J}{2N}M^2,
\qquad m=M/N.
\label{eq:energy}
\end{equation}
The selected rank-one channel does not restrict the
full connectivity to rank one.

We consider overdamped gradient dynamics with additive noise,
\begin{equation}
\dot x_i=-\partial_iH_0(\bm x)+Jc_i m+\eta_i,
\label{eq:langevin}
\end{equation}
where $\langle\eta_i(t)\eta_j(s)\rangle
=2D\delta_{ij}\delta(t-s)$.
The reciprocal model has the stationary distribution
$P_{\rm st}\propto e^{-H/D}$ when normalizable.
This equilibrium relation does not extend automatically
to general neural activation-rate dynamics.

In the real-response It\^o MSRJD convention
~\cite{martin1973,janssen1976,dedominicis1976},
the microscopic action reads
\begin{equation}
\begin{split}
S[\bm x,\tilde{\bm x}]
={}&\int dt\sum_i
\left[
\tilde x_i(\dot x_i+\partial_iH_0)
-D\tilde x_i^2
\right]\\
&-\frac{J}{N}\int dt\,\widetilde M(t)M(t),
\end{split}
\label{eq:msrjd}
\end{equation}
where $\widetilde M=\sum_i c_i\tilde x_i$ is the
collective response variable. The last term couples
collective activity to its response counterpart.

A dynamical Hubbard--Stratonovich transformation
~\cite{hubbard1959,stratonovich1957} introduces a collective drive $\phi$
and its conjugate response field $\tilde\phi$.
Integrating out the full reference network yields
\begin{equation}
S_{\rm eff}[\phi,\tilde\phi]
=\frac{N}{J}\int dt\,\tilde\phi\phi
-\ln\left\langle e^{\mathcal A}\right\rangle_0,
\label{eq:Seff}
\end{equation}
where $\mathcal A=\int dt\,
(\tilde\phi M+\phi\widetilde M)$ and
$\langle\cdots\rangle_0$ denotes the normalized
MSRJD measure of the interacting reference network
$H_0$, which need not factorize into independent neurons.
The auxiliary-field contours implement the causal HS
identity, with $\phi=J\langle m\rangle$ at the
physical saddle point.

Equation~(\ref{eq:Seff}) is formally exact at the
path-integral level. Expanding the logarithm in
connected cumulants,
$\ln\langle e^{\mathcal A}\rangle_0
=\sum_{n=1}^{\infty}
\langle\mathcal A^n\rangle_{0,c}/n!$,
generates a hierarchy of collective vertices.
Under reflection symmetry of the reference MSRJD
measure, odd total orders vanish.
The quadratic sector determines the collective
susceptibility and the retarded response kernel,
while fourth-order connected responses generate
nonlinear retarded and fluctuation vertices.
Saddle-point and Gaussian approximations require
appropriate large-$N$ scaling and stability conditions.

For a stationary, reflection-symmetric reference state
with $\langle m\rangle_0=0$, the retarded response
of the network governed by $H_0$ to a collective
drive $\phi\bm c$ is
\begin{equation}
\chi_0^R(t-s)
=
\left.
\frac{\delta\langle m(t)\rangle_\phi}
{\delta\phi(s)}
\right|_{\phi=0}
=
\frac1N
\langle M(t)\widetilde M(s)\rangle_{0,c}.
\label{eq:R0}
\end{equation}
Causality requires $\chi_0^R(t-s)=0$ for $t<s$.
Expanding Eq.~(\ref{eq:Seff}) to quadratic order
gives the retarded collective vertex
$\Gamma_2^R(\omega)=J^{-1}-\chi_0^R(\omega)$.
The corresponding Gaussian physical-channel
susceptibility is
\begin{equation}
\chi_m^R(\omega)
=
\frac{\chi_0^R(\omega)}
{1-J\chi_0^R(\omega)}.
\label{eq:physicalchi}
\end{equation}
The auxiliary-field propagator and physical
susceptibility differ but share the Gaussian
instability denominator.
The quadratic action also contains a collective
noise vertex proportional to the connected
correlation $\langle M(t)M(s)\rangle_{0,c}$;
the retarded kernel alone does not determine
fluctuations.

The symmetric Gaussian saddle becomes marginal
at $J_c\chi_0^R(0)=1$, equivalently
$r_L\equiv J^{-1}-\chi_0^R(0)=0$.
This condition identifies a mean-field instability,
not necessarily a continuous phase transition.
The reference response $\chi_0^R$ is defined
for the full nonlinear reference dynamics,
whereas Eq.~(\ref{eq:physicalchi}) is exact
for Gaussian collective dynamics and otherwise
represents the Gaussian/RPA approximation.

\medskip
\subsection{Spectral representation of the reference response}
\label{sec:response}
Linearizing the deterministic reference dynamics
about a stable fixed point $\bm x_*$ with a
positive-definite Hessian gives
$\delta\dot{\bm x}=-L_0\delta\bm x
+\bm c\,\delta\phi$, where
$L_0=\nabla\nabla H_0(\bm x_*)$.
For reciprocal gradient dynamics,
$L_0=L_0^{\mathsf T}>0$, with eigenpairs
$(\lambda_\alpha,\bm v_\alpha)$.
The collective response then has the spectral form
\begin{equation}
\chi_0^R(\omega)
=
\frac1N\bm c^{\mathsf T}
(L_0-i\omega I)^{-1}\bm c
=
\sum_\alpha
\frac{A_\alpha}{\lambda_\alpha-i\omega},
\label{eq:resolvent}
\end{equation}
where $A_\alpha=
|\bm c^{\mathsf T}\bm v_\alpha|^2/N\ge0$.
The eigenmodes describe relaxation of the full
$N$-neuron state space rather than individual neurons.

The unweighted relaxation-rate density, hereafter referred to as
the time-scale density of states (TDOS), is defined as
\begin{equation}
\rho(\lambda)
=
\frac{1}{N}
\sum_\alpha \delta(\lambda-\lambda_\alpha).
\label{eq:raw_tdos}
\end{equation}
In contrast, the collective observable probes the
response-weighted relaxation spectrum,
\begin{equation}
\rho_O(\lambda)
=
\sum_\alpha A_\alpha
\delta(\lambda-\lambda_\alpha),
\label{eq:weighted_tdos}
\end{equation}
where the weights are defined in Eq.~\eqref{eq:resolvent}.
Because the collective channel
need not overlap with every mode, these distributions can have
different infrared behavior. Slow modes contribute to the collective
instability only through their nonzero overlaps with the
interaction channel.

The weighted spectrum determines both the frequency response
and the causal reference response kernel,
\begin{equation}
\begin{aligned}
\chi_0^R(\omega)&=\int_0^\infty d\lambda\,
\frac{\rho_O(\lambda)}{\lambda-i\omega},\\
K_O(t)&=\Theta(t)\int_0^\infty d\lambda\,
\rho_O(\lambda)e^{-\lambda t}.
\end{aligned}
\label{eq:spectralIR}
\end{equation}

At the Gaussian level, the resulting collective
field obeys the linearized saddle equation
\begin{equation}
\frac1J\phi(t)
-\int_{-\infty}^{t}ds\,
K_O(t-s)\phi(s)=0,
\label{eq:memory_saddle}
\end{equation}
in the absence of an external collective source.
Thus the nonlocal reference response kernel $K_O(t)=\chi_0^R(t)$
emerges from integrating out microscopic neural
degrees of freedom rather than being imposed
phenomenologically.
This kernel is the reference response already
contained in $J^{-1}-\chi_0^R$ and must be
distinguished from the irreducible memory
self-energy generated by projection onto
the collective direction.

For nonlinear reference dynamics at finite noise,
the exact stochastic response need not equal the
deterministic Jacobian resolvent.
Nonreciprocal networks require left/right
eigenvector overlaps and may exhibit complex
relaxation modes.

\section{Microscopic origin of memory feedback}
\label{sec:projection}
\label{sec:supp_mori_zwanzig}

The linearized reference response derived in
Sec.~\ref{sec:response}
admits an equivalent representation in terms of
an irreducible memory self-energy. Here we derive
this representation by an exact Mori--Zwanzig
projection of the symmetric linearized microscopic
dynamics~\cite{zwanzig1961,mori1965}.
The response kernel $K_O(t)$ defined in
Eq.~\eqref{eq:spectralIR} and the projected
memory kernel $K_\Sigma(t)$ are distinguished
throughout. All results below concern the linear
reference dynamics unless otherwise stated.

\begin{figure*}[t]
\centering
\paperfigure[scale=0.54, trim=-0.5cm 3cm 0cm 0cm]{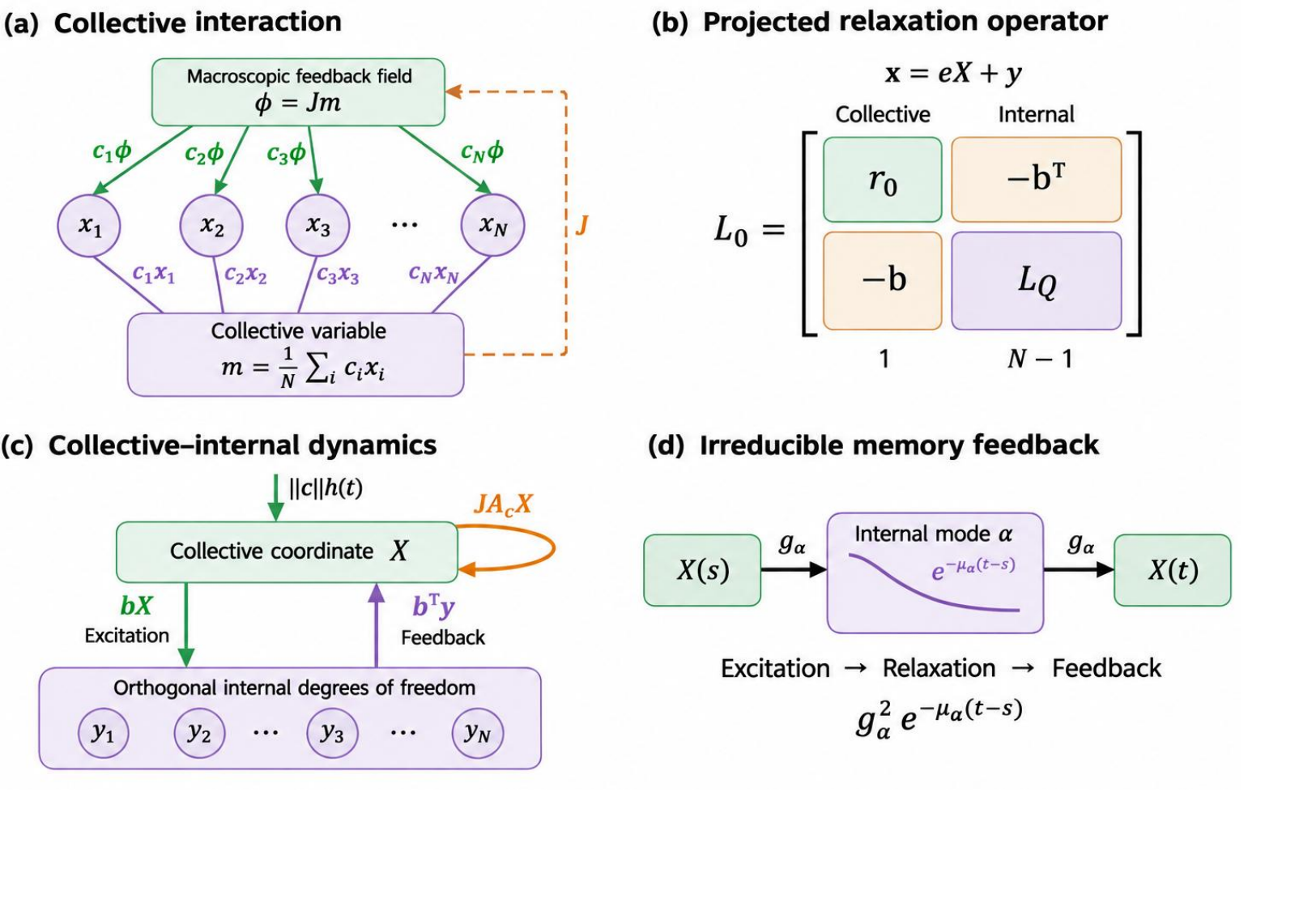}
\caption{
Microscopic origin of collective interaction
and irreducible memory feedback.
(a) Reciprocal collective interaction in which
microscopic activities $x_i$ contribute to
the collective variable
$m=N^{-1}\sum_i c_i x_i$,
while the instantaneous feedback field
$\phi=Jm$ acts back on the microscopic degrees
of freedom through the same collective channel.
(b) Block representation of the microscopic
reference relaxation operator $L_0$ in the
collective and orthogonal internal subspaces.
(c) Coupled dynamics of the collective
coordinate $X$ and the orthogonal internal
degrees of freedom $\mathbf{y}$, distinguishing
the instantaneous collective feedback $JA_cX$
from the dynamical feedback $\mathbf{b}^T\mathbf{y}$.
(d) Elementary excitation--relaxation--feedback
process underlying the irreducible memory kernel.
A collective perturbation excites an internal
mode $\alpha$ through the coupling $g_\alpha$.
The mode relaxes at rate $\mu_\alpha$ and
feeds back into the collective coordinate
through the same coupling.
Summation over the orthogonal internal modes
generates the irreducible memory kernel
$K_\Sigma(t)=\Theta(t)\sum_\alpha
g_\alpha^2 e^{-\mu_\alpha t}$.
}
\label{fig:microscopic_memory}
\end{figure*}

\subsection{Projection onto the collective direction}

Let $L_0=L_0^{\mathsf T}>0$ denote the reference
relaxation operator.
The microscopic equation in the presence of a
collective source $h(t)$ is
\begin{equation}
\dot{\bm x}=-L_0\bm x
+\frac{J}{N}\bm c\bm c^{\mathsf T}\bm x
+\bm c h(t)+\bm\eta(t).
\end{equation}
Here $\bm x$ denotes deviations from the reference
state and $\bm\eta$ is the microscopic noise.
Figure~\ref{fig:microscopic_memory}(a) illustrates
the reciprocal structure of the collective
interaction. Microscopic activities contribute
to the collective variable $m$, while the
resulting feedback field $\phi=Jm$ acts back
on the microscopic degrees of freedom through
the same collective channel $\mathbf{c}$.

The external source is chosen to couple directly to the
collective channel. A general input direction can generate
an additional memory-mediated input-transfer kernel
without modifying the intrinsic memory self-energy
at fixed linear reference dynamics, as derived in
Sec.~I F of the Supplemental Material.

Define
\begin{equation}
\bm e=\frac{\bm c}{\|\bm c\|},\quad
A_c=\frac{\|\bm c\|^2}{N},\quad
\mathcal P=\bm e\bm e^{\mathsf T},\quad
\mathcal Q=I-\mathcal P.
\end{equation}
The collective coordinate and its orthogonal
complement are
\begin{equation}
X=\bm e^{\mathsf T}\bm x,\qquad
\bm y=\mathcal Q\bm x,\qquad
\bm x=\bm eX+\bm y.
\end{equation}
The physical collective variable is
$m=\|\bm c\|X/N$. Introduce
\begin{equation}
r_0=\bm e^{\mathsf T}L_0\bm e,\quad
L_Q=\mathcal QL_0\mathcal Q
\big|_{\operatorname{ran}\mathcal Q},\quad
\bm b=-\mathcal QL_0\bm e.
\end{equation}
As illustrated in
Fig.~\ref{fig:microscopic_memory}(b),
the projection resolves the microscopic
relaxation operator into a direct collective
relaxation channel, an orthogonal internal
relaxation sector, and their mutual coupling.
The latter provides the microscopic origin
of retarded memory feedback.

Using $\mathcal Q\bm e=0$ and the symmetry of $L_0$,
the microscopic dynamics becomes
\begin{align}
\dot X&=-(r_0-JA_c)X+\bm b^{\mathsf T}\bm y
+\|\bm c\|h+\eta_X,\\
\dot{\bm y}&=-L_Q\bm y+\bm bX+\bm\eta_Q,
\label{eq:supp_mz_internal}
\end{align}
where $\eta_X=\bm e^{\mathsf T}\bm\eta$ and
$\bm\eta_Q=\mathcal Q\bm\eta$. The rank-one
interaction modifies only the collective diagonal
block, leaving the orthogonal dynamics and its
coupling unchanged at fixed $L_0$.
Figure~\ref{fig:microscopic_memory}(c)
illustrates the two distinct feedback
mechanisms contained in these equations.
The collective interaction produces an
instantaneous feedback $JA_cX$, whereas
the coupling to the orthogonal degrees of
freedom generates dynamical feedback through
their internal relaxation.

Diagonalizing $L_Q\bm u_\alpha=\mu_\alpha\bm u_\alpha$
with orthonormal eigenvectors and expanding
$\bm y=\sum_\alpha y_\alpha\bm u_\alpha$
gives the microscopic coupling constants
\begin{equation}
g_\alpha=\bm u_\alpha^{\mathsf T}\bm b
=-\bm u_\alpha^{\mathsf T}L_0\bm e.
\end{equation}
Thus $\dot y_\alpha=-\mu_\alpha y_\alpha+g_\alpha X
+\eta_\alpha$, where
$\eta_\alpha=\bm u_\alpha^{\mathsf T}\bm\eta_Q$,
while the feedback into $X$ is
$\sum_\alpha g_\alpha y_\alpha$. Neither an independent
memory reservoir nor phenomenological coupling
constants are required in this linear model.

\subsection{Exact elimination and memory self-energy}

The solution of Eq.~\eqref{eq:supp_mz_internal} is
\begin{equation}
\begin{split}
\bm y(t)={}&e^{-L_Qt}\bm y(0)\\
&+\int_0^t ds\,e^{-L_Q(t-s)}
[\bm bX(s)+\bm\eta_Q(s)].
\end{split}
\end{equation}
Substitution into the collective equation gives
\begin{equation}
\begin{split}
\dot X(t)={}&-(r_0-JA_c)X(t)
+\int_0^t ds\,K_\Sigma(t-s)X(s)\\
&+\|\bm c\|h(t)+\eta_{\rm eff}(t),
\end{split}
\label{eq:supp_mz_gle}
\end{equation}
where
\begin{align}
K_\Sigma(t)&=\Theta(t)\bm b^{\mathsf T}
e^{-L_Qt}\bm b =\Theta(t)\sum_\alpha g_\alpha^2e^{-\mu_\alpha t},
\label{eq:memory_kernel}
\\
\eta_{\rm eff}(t)&=\eta_X(t)
+\bm b^{\mathsf T}e^{-L_Qt}\bm y(0)\nonumber\\
&\quad+\int_0^t ds\,\bm b^{\mathsf T}
e^{-L_Q(t-s)}\bm\eta_Q(s).
\end{align}
The effective fluctuating force includes the
unresolved initial state and microscopic noise;
it is not generally white. For the retarded
response, initial perturbations are set to zero.
The elementary physical process underlying
Eq.~(\ref{eq:memory_kernel}) is illustrated in
Fig.~\ref{fig:microscopic_memory}(d).
A collective perturbation at time $s$ excites
an orthogonal internal mode $\alpha$ through
the coupling $g_\alpha$.
The induced amplitude relaxes at rate
$\mu_\alpha$ and subsequently feeds back
into the collective coordinate through the
same coupling.
This excitation--relaxation--feedback process
generates the elementary memory contribution
$g_\alpha^2 e^{-\mu_\alpha(t-s)}$.
Summation over the orthogonal internal modes
yields the full irreducible memory kernel.

For $\operatorname{Re}z>0$, the Laplace transform of
the memory kernel is
\begin{equation}
\Sigma(z)=\bm b^{\mathsf T}(zI_Q+L_Q)^{-1}\bm b
=\int_0^\infty d\lambda\,
\frac{\rho_\Sigma(\lambda)}{z+\lambda},
\label{eq:supp_mz_sigma}
\end{equation}
where $I_Q$ is the identity on the orthogonal
subspace and
\begin{equation}
\rho_\Sigma(\lambda)=\sum_\alpha g_\alpha^2
\delta(\lambda-\mu_\alpha).
\end{equation}
The retarded self-energy is
$\Sigma_R(\omega)=\Sigma(-i\omega+0^+)$. Its
spectral measure is nonnegative for the symmetric
reference operator considered here.

\subsection{Equivalence to the HS collective response}

In the orthogonal decomposition, the inverse
reference resolvent has the block representation
\begin{equation}
zI+L_0=\begin{pmatrix}
z+r_0&-\bm b^{\mathsf T}\\
-\bm b&zI_Q+L_Q
\end{pmatrix}.
\end{equation}
The Schur complement therefore gives
\begin{equation}
G_X(z)=\bm e^{\mathsf T}(zI+L_0)^{-1}\bm e
=\frac{1}{z+r_0-\Sigma(z)}.
\end{equation}
Defining $\chi_0(z)=N^{-1}\bm c^{\mathsf T}
(zI+L_0)^{-1}\bm c$, one obtains the exact identity
\begin{equation}
\chi_0(z)=\frac{A_c}{z+r_0-\Sigma(z)}.
\label{eq:supp_mz_pi}
\end{equation}
Equivalently,
\begin{equation}
\Sigma(z)=z+r_0-\frac{A_c}{\chi_0(z)}.
\label{eq:supp_mz_inverse}
\end{equation}
A detailed spectral reconstruction of the irreducible
memory self-energy from the reference collective response
is presented in Sec.~I of the Supplemental Material.
The HS quadratic vertex is therefore $\Gamma_2(z)=J^{-1}-\chi_0(z)$,
and the physical collective susceptibility becomes
\begin{equation}
\chi_m(z)=\frac{\chi_0(z)}{1-J\chi_0(z)}
=\frac{A_c}{z+r_0-JA_c-\Sigma(z)}.
\label{eq:supp_mz_chi}
\end{equation}
In particular, the inverse HS quadratic vertex satisfies
$\Gamma_2^{-1}=J+J^2\chi_m$ and is therefore distinct
from the physical susceptibility itself.
The collective Gaussian instability occurs at
\begin{equation}
J_c=\frac{1}{\chi_0(0)}
=\frac{r_0-\Sigma(0)}{A_c}.
\label{eq:collective_threshold}
\end{equation}
For the symmetric linear model, this condition coincides
with the loss of positive definiteness of the full
relaxation operator $L_0-JA_c\mathcal P$.

For a positive definite finite reference operator,
$r_0-\Sigma(0)>0$ by the Schur complement.
This quantity is a static inverse-response scale,
not necessarily the smallest dynamical decay rate.
The self-energy must not be added independently
to $J^{-1}-\chi_0$, since its effect is already
contained in the reference response.

\section{Infrared organization of response and memory spectra}
\label{sec:infrared}

\subsection{Infrared scaling and collective threshold}

Assume a positive weighted spectrum
$\rho_O(\lambda)=A\lambda^\beta$ over
$\lambda_{\rm IR}\le\lambda\le\Lambda$,
with $A>0$ and $\beta>-1$, and negligible
contributions outside this window.
For $\Lambda^{-1}\ll t\ll\lambda_{\rm IR}^{-1}$,
the reference response kernel obeys
\begin{equation}
K_O(t)\simeq
A\Gamma(1+\beta)t^{-(1+\beta)}.
\label{eq:Kscale}
\end{equation}
For the marginal exponent $\beta=0$, the
finite-window expression is
$K_O(t)=A(e^{-\lambda_{\rm IR}t}
-e^{-\Lambda t})/t$, yielding $K_O(t)\sim A/t$
in the intermediate regime.

The same weighted spectrum determines the
Gaussian ordering threshold through
$J_c^{-1}=\chi_0^R(0)
=A\int_{\lambda_{\rm IR}}^\Lambda
d\lambda\,\lambda^{\beta-1}$, giving
\begin{equation}
J_c^{-1}=
\begin{cases}
\dfrac{A}{\beta}
(\Lambda^\beta-\lambda_{\rm IR}^\beta),
&\beta\ne0,\\[4pt]
A\ln(\Lambda/\lambda_{\rm IR}),
&\beta=0.
\end{cases}
\label{eq:Jc}
\end{equation}
At fixed $A$ and $\Lambda$, decreasing
$\lambda_{\rm IR}$ yields
$J_c\propto\lambda_{\rm IR}^{-\beta}$
for $-1<\beta<0$,
$J_c\propto[\ln(\Lambda/\lambda_{\rm IR})]^{-1}$
for $\beta=0$, and a finite limiting
threshold for $\beta>0$.

The marginal case $\beta=0$ separates
power-law infrared divergence from a finite
static response, with logarithmic enhancement
at the boundary.
Thus the $1/t$ reference-response regime and logarithmic
infrared enhancement of the static response
arise from the same weighted relaxation spectrum.
Finite cutoffs regularize these asymptotics,
while changes in spectral normalization or
mode weights can modify their cutoff dependence.

\subsection{Spectral relations and memory dressing}

The unweighted and response-weighted relaxation spectra
introduced in Sec.~II B differ from the irreducible
memory spectrum $\rho_\Sigma$.
The first two involve eigenvalues of $L_0$, whereas the
last involves eigenvalues of $L_Q$ and the coupling
weights $g_\alpha^2$. In particular,
\begin{equation}
\int d\lambda\,\rho_O=A_c,\qquad
\int d\lambda\,\rho_\Sigma=\|\bm b\|^2.
\end{equation}
For an absolutely continuous response spectrum, the Laplace
representation of Eq.~\eqref{eq:spectralIR} reads
$\chi_0(z)=\int_0^\infty d\lambda\,\rho_O(\lambda)/(z+\lambda)$.
On the upper edge of the negative real axis,
\begin{equation}
\chi_0(-\lambda+i0^+)=P_O(\lambda)
-i\pi\rho_O(\lambda),
\end{equation}
where
\begin{equation}
P_O(\lambda)=\operatorname{PV}\int_0^\infty
\frac{\rho_O(s)}{s-\lambda}\,ds.
\end{equation}
Taking the imaginary part of
Eq.~\eqref{eq:supp_mz_inverse} gives
\begin{equation}
\rho_\Sigma(\lambda)=
\frac{A_c\rho_O(\lambda)}
{P_O(\lambda)^2+\pi^2\rho_O(\lambda)^2}.
\label{eq:supp_mz_density}
\end{equation}
The inverse-resolvent structure underlying this spectral
relation is consistent with established continued-fraction
representations of dynamical correlations and resolvent
identities in spectral theory
~\cite{mori1965cf,gesztesy1997}.
Further details of the continuous spectral transformation,
discrete memory-mode reconstruction, and spectral moment
sum rules are given in Sec.~I of the Supplemental Material.

This identity applies to the absolutely continuous
part wherever the boundary value is regular and
nonzero. Discrete spectral contributions must be
handled through the resolvent; zeros of $\chi_0(z)$
in spectral gaps can generate additional poles of
$\Sigma(z)$. For a simple zero at $z=-\mu$, its
residue is $-A_c/\chi_0'(-\mu)>0$.

Consider an idealized flat response spectrum
$\rho_O(\lambda)=A$ on $[\lambda_{\rm IR},\Lambda]$
and zero elsewhere, with
$A_c=A(\Lambda-\lambda_{\rm IR})$. Then
\begin{equation}
\chi_0(z)=A\ln\frac{z+\Lambda}{z+\lambda_{\rm IR}},
\qquad r_0=\frac{\Lambda+\lambda_{\rm IR}}{2}.
\end{equation}
The second relation follows from the first moment
of the normalized response spectrum. Within the
spectral band,
\begin{equation}
P_O(\lambda)=A\ln\frac{\Lambda-\lambda}
{\lambda-\lambda_{\rm IR}},
\end{equation}
so that
\begin{equation}
\rho_\Sigma(\lambda)=\frac{A_c}{A}
\left[\ln^2\frac{\Lambda-\lambda}
{\lambda-\lambda_{\rm IR}}+\pi^2\right]^{-1}.
\end{equation}
There are no zeros of $\chi_0$ outside this single
continuous band. In the gapless continuum limit,
$\lambda_{\rm IR}\to0$, the infrared asymptotics is
\begin{equation}
\rho_\Sigma(\lambda)\sim
\frac{A_c}{A\ln^2(\Lambda/\lambda)}.
\end{equation}
Changing variables to $u=\lambda t$ in the Laplace
integral for $K_\Sigma$ yields, at large $\Lambda t$,
\begin{equation}
K_\Sigma(t)\sim\frac{A_c}{At}
\int_0^{\sqrt{\Lambda t}} du\,
\frac{e^{-u}}{\ln^2(\Lambda t/u)}
\sim\frac{A_c}{At\ln^2(\Lambda t)}.
\end{equation}
The upper limit isolates the infrared region;
the omitted contribution is exponentially small.
The leading result follows from slow variation of
the logarithmic factor and $\int_0^\infty e^{-u}du=1$.
Thus a flat response spectrum does not imply a
flat irreducible memory spectrum.

Having established the logarithmically corrected
irreducible memory kernel, we now examine the
recovery of the full reference response through
memory dressing. For $J=0$, the bare projected
propagator is $G_b(z)=(z+r_0)^{-1}$.
Equation~\eqref{eq:supp_mz_gle}
gives the Dyson identity
\begin{equation}
G_X=G_b+G_b\Sigma G_X.
\end{equation}
Where the geometric series converges, this reads
\begin{equation}
G_X=G_b+G_b\Sigma G_b
+G_b\Sigma G_b\Sigma G_b+\cdots.
\end{equation}
Its resolvent form is valid throughout the common analytic domain.
Substituting the flat-band reference response into the exact
self-energy relation in Eq.~\eqref{eq:supp_mz_inverse}, we recover
\begin{equation}
\begin{split}
G_X(z)&=\frac{1}{z+r_0-\Sigma(z)}\\
&=\frac{A}{A_c}\ln\frac{z+\Lambda}
{z+\lambda_{\rm IR}}
=\frac{A}{A_c}\int_{\lambda_{\rm IR}}^\Lambda
\frac{d\lambda}{z+\lambda}.
\end{split}
\end{equation}
Inverse Laplace transformation yields
\begin{equation}
G_X(t)=\Theta(t)\frac{A}{A_c}
\frac{e^{-\lambda_{\rm IR}t}-e^{-\Lambda t}}{t}.
\label{eq:supp_mz_recovery}
\end{equation}
Hence $A_cG_X(t)=K_O(t)$ exactly, and
$G_X(t)\simeq A/(A_ct)$ for
$\Lambda^{-1}\ll t\ll\lambda_{\rm IR}^{-1}$.
In the gapless limit this becomes a long-time
$1/t$ response, even though the irreducible memory
kernel decays as $1/[t\ln^2(\Lambda t)]$.

For finite $\lambda_{\rm IR}$,
\begin{equation}
r_0-\Sigma(0)=\frac{A_c}
{A\ln(\Lambda/\lambda_{\rm IR})}>0.
\end{equation}
In the gapless limit $\Sigma(0)=r_0$ and
$\int_0^\infty K_\Sigma(t)dt=r_0$ remains finite,
whereas $\int_0^\infty G_X(t)dt$ diverges.
The cancellation of the static restoring term
accounts for the enhanced persistence of the
memory-dressed response. This is a property of
the marginal reference continuum; it does not
imply stability at fixed positive $J$, since
$J_c\to0$ in this limit. At finite $J$ the physical
response is instead Eq.~\eqref{eq:supp_mz_chi}.

\subsection{Collective ordering, memory feedback,
and learning observables}

If $\bm e$ is an eigenvector of $L_0$, then
$\bm b=0$ and $\Sigma(z)=0$. This includes the
uniform linearized Glauber mean-field dynamics
of the pure Curie--Weiss model and its Mattis
counterpart with identical spin-flip time scales.
For the latter, the transformation
$\sigma_i=c_is_i$, $c_i=\pm1$, maps the pure
rank-one Hamiltonian to the Curie--Weiss form.
Collective ordering can therefore occur without
irreducible memory feedback in this simple limit.
General interacting or heterogeneous spin dynamics
need not have this property.

For example, two independent reference modes with
rates $\lambda_1,\lambda_2>0$ and collective direction
$\bm e=(1,1)^{\mathsf T}/\sqrt{2}$ give
$r_0=(\lambda_1+\lambda_2)/2$ and
$g=(\lambda_2-\lambda_1)/2$. Their projected
self-energy is $g^2/(z+r_0)$, and
\begin{equation}
G_X(z)=\frac{1}{z+r_0-g^2/(z+r_0)}
=\frac12\left(\frac1{z+\lambda_1}
+\frac1{z+\lambda_2}\right).
\end{equation}
This illustrates explicitly how unresolved
microscopic time scales generate memory without
an additional physical reservoir.

The inverse-rate spectral estimator previously introduced
in Ref.~\cite{chae2026a} as a memory self-energy estimator is more
naturally interpreted within the present framework as a
proxy for the reference static response, under approximately
uniform response-overlap weights and an appropriate
identification of the relaxation generator. It should
therefore be distinguished from the irreducible memory
self-energy in Eq.~\eqref{eq:supp_mz_sigma}.

The actual projected self-energy requires a specified
collective channel and its coupling to the unresolved
dynamics. For a nonnormal operator, left and right couplings
need not coincide, and the positive spectral-measure
relations above do not generally apply
~\cite{trefethen1997}. Moreover,
interlayer Transformer propagation is not in itself
physical-time evolution. These distinctions delimit
the connection between the exact linear projection and
the learning-induced spectral organization reported in
Sec.~\ref{sec:supp_learning_critical}.

\section{Nonlinear collective order and macroscopic field dynamics}
\label{sec:nonlinear}

\subsection{Nonlinear collective vertices and ordering}

The dynamical Hubbard--Stratonovich effective action generates
nonlinear collective vertices through connected MSRJD response
functions. For a reflection-symmetric reference state, the
fourth-order connected cumulant contains a term linear in the
collective response field and cubic in the collective drive,
\begin{equation}
W_0^{(4)}
\supset
\frac{1}{3!}
\int dt\,d^3\bm s\,
\tilde\phi(t)
R_0^{(3)}(t;\bm s)
\prod_{a=1}^3\phi(s_a),
\end{equation}
where the retarded third-order response vertex is
\begin{equation}
R_0^{(3)}(t;\bm s)
=
\left\langle
M(t)\widetilde M(s_1)
\widetilde M(s_2)
\widetilde M(s_3)
\right\rangle_{0,c}.
\end{equation}
The factor $1/3!$ follows from the combinatorics of the
fourth connected cumulant.

At the physical saddle $\tilde\phi=0$, variation with respect
to $\tilde\phi$ gives the collective equation through cubic
order,
\begin{equation}
\begin{split}
0={}&J^{-1}\phi(t)
-\int ds\,\chi_0^R(t-s)\phi(s)\\
&-\frac{1}{3!N}\int d^3\bm s\,
R_0^{(3)}(t;\bm s)
\prod_{a=1}^3\phi(s_a)
+O(\phi^5).
\end{split}
\label{eq:nonlocal}
\end{equation}
At this saddle, $\phi=J\langle m\rangle$. Other fourth-order
vertices containing two or more collective response fields
govern nonlinear noise and fluctuations. Equation~\eqref{eq:nonlocal}
is therefore a saddle-point equation, not an exact stochastic
equation for the collective field.

If the cubic response is regular and short-lived relative to the
slow collective dynamics, its convolution can be approximated by a
local cubic term $u\phi^3(t)$, with
$u=-(6N)^{-1}\int d^3\bm s\,R_0^{(3)}(t;\bm s)$.
Separating the static contribution of the linear response
kernel then gives
\begin{equation}
\begin{split}
0={}&r_L\phi(t)
+\int_0^\infty d\tau\,
K_O(\tau)[\phi(t)-\phi(t-\tau)]\\
&+u\phi^3(t),
\end{split}
\label{eq:memorylandau}
\end{equation}
where $r_L=J^{-1}-\chi_0^R(0)$ and
$K_O(t)=\chi_0^R(t)$.

A local first-derivative approximation to the linear
convolution additionally requires a finite temporal moment
\begin{equation}
Z_1=\int_0^\infty d\tau\,\tau K_O(\tau)
=\int d\lambda\,
\frac{\rho_O(\lambda)}{\lambda^2}.
\end{equation}
For $\rho_O(\lambda)\propto\lambda^\beta$, this moment
diverges as $\lambda_{\rm IR}\to0$ for $\beta\le1$,
preventing a uniform local derivative expansion.
If the cubic response is itself long-lived, the full
nonlocal equation~\eqref{eq:nonlocal} must be retained.

For the reciprocal equilibrium reference network, the
static HS free-energy density follows directly from the
microscopic partition function,
\begin{equation}
f_{\rm HS}(\phi)
=\frac{\phi^2}{2J}
-\frac{D}{N}\ln z_0(\phi),
\label{eq:freeenergy}
\end{equation}
where
\begin{equation}
z_0(\phi)=\int d^Nx\,
\exp[-(H_0(\bm x)-\phi M)/D].
\end{equation}
Assuming $M\mapsto-M$ symmetry, its connected-cumulant
expansion yields
\begin{equation}
f_{\rm HS}(\phi)
=f_0+\frac{r_L}{2}\phi^2
+\frac{u_{\rm stat}}{4}\phi^4+O(\phi^6),
\label{eq:Landau}
\end{equation}
with
\begin{equation}
r_L=J^{-1}-\frac{\kappa_2(M)}{ND},
\qquad
u_{\rm stat}=-\frac{\kappa_4(M)}{6ND^3},
\end{equation}
where $\kappa_n(M)=\langle M^n\rangle_{0,c}$.
Equilibrium static fluctuation--dissipation gives
$\chi_0^R(0)=\kappa_2(M)/(ND)$, ensuring consistency
between the static and dynamical quadratic coefficients.

The corresponding relation between the static and
dynamical quartic vertices follows from the integrated
third-order response. For stationary equilibrium dynamics,
assuming that the static nonlinear response is regular
and that the relevant time integrals converge, one obtains
\begin{equation}
\mathcal R_3
=\int d^3\bm s\,
R_0^{(3)}(t;\bm s)=\left.
\frac{\partial^3\langle M\rangle_\phi}
{\partial\phi^3}
\right|_{\phi=0}=\frac{\kappa_4(M)}{D^3}.
\end{equation}
The last equality follows from the equilibrium generating
function $z_0(\phi)$ introduced above. Consequently, the
local dynamical cubic coefficient satisfies
\begin{equation}
u_{\rm dyn}
=-\frac{\mathcal R_3}{6N}
=-\frac{\kappa_4(M)}{6ND^3}
=u_{\rm stat}.
\end{equation}
Thus the local dynamical cubic coefficient agrees with
the quartic coefficient of the static HS free energy.
We henceforth denote their common value by $u$.
The nonlinear collective vertex is therefore determined
by microscopic connected correlations rather than
introduced phenomenologically.

This equality concerns the zero-frequency nonlinear
response and does not imply that the full third-order
response is local in time. The local nonlinear reduction
requires the temporal support of the cubic response to
be short relative to the collective time scale.

Nonlinear stabilization is not automatic. A continuous
symmetric transition requires $u>0$, equivalently
$\kappa_4(M)<0$, together with regular higher-order terms.
For a fixed reference network and $J>J_c$, sufficiently
close to the transition, the locally stable branch
satisfies
\begin{equation}
\phi_*^2
=\frac{J-J_c}{uJJ_c}
+O((J-J_c)^2),
\label{eq:critical}
\end{equation}
giving the mean-field order-parameter exponent
$\beta_{\rm L}=1/2$ and static susceptibility
$\chi_\phi^{\rm stat}\propto|J-J_c|^{-1}$,
with $\gamma_{\rm L}=1$.
The Landau exponent $\beta_{\rm L}$ is distinct
from the infrared spectral exponent $\beta$.

The ordered-state curvature is
$f_{\rm HS}''(\phi_*)=-2r_L+O(r_L^2)>0$,
but this static restoring curvature does not by itself
determine the dynamical forgetting rate, which depends
on the ordered-state retarded response.

\subsection{Memory-dressed macroscopic field dynamics}

The HS saddle and the projected coordinate describe the same
collective channel, but the auxiliary drive is not an independent
microscopic coordinate. At the physical saddle,
$\phi=J\langle m\rangle$.

To make the nonlinear reduction explicit, write the causal cubic
functional in Eq.~\eqref{eq:nonlocal} as $\mathcal C_3[\phi]$,
including its factor $1/(3!N)$ but not the minus sign. With a
collective external drive $h$, the saddle equation becomes
\begin{equation}
J^{-1}\phi
=
\chi_0*(\phi+h)
+\mathcal C_3[\phi+h]
+O((\phi+h)^5),
\end{equation}
where $*$ denotes causal convolution and the order symbol assumes
that both drives are small. Applying the inverse linear response
and using Eq.~\eqref{eq:supp_mz_pi} gives, for $J>0$,
\begin{equation}
\begin{split}
\dot\phi={}&-(r_0-JA_c)\phi
+K_\Sigma*\phi+JA_c h\\
&+JA_c\chi_0^{-1}*\mathcal C_3[\phi+h]
+JA_c\chi_0^{-1}*O((\phi+h)^5).
\end{split}
\label{eq:nonlinear_projected_saddle}
\end{equation}
The inverse response acts on the entire cubic functional;
therefore, locality of the HS cubic vertex alone does not
establish locality of the cubic term in this differential
equation. At zero source, if
$\mathcal C_3[\phi]\simeq-u\phi^3$ and the inverse response
can be replaced by its static value on the nonlinear frequency
range, the effective cubic coefficient in the differential
equation is
\begin{equation}
u_{\rm eff}
=
\frac{JA_c u}{\chi_0(0)},
\end{equation}
rather than simply $u$. This additional approximation is
controlled only when the relevant frequency dependence is
small, and need not be uniform as the infrared cutoff
vanishes. The saddle equation does not determine the
nonlinear fluctuating force.

Although a local cubic reduction of the general nonlinear
saddle requires additional approximations, an exactly
reducible microscopic model demonstrates that local cubic
stabilization and nonlocal linear memory can coexist without
postulating a separate reservoir. To establish this explicitly,
consider a symmetric positive reference matrix $L_0$ and
the energy
\begin{equation}
H
=
\frac12\bm x^{\mathsf T}L_0\bm x
-\frac{J}{2N}M^2
+\frac{vN}{4}m^4,
\qquad v>0.
\label{eq:exact_nonlinear_energy}
\end{equation}
This is a special choice of nonlinear reference interaction,
not a claim that arbitrary local neuronal potentials have
this form. The quartic force is $-v\bm c m^3$, so its
orthogonal projection vanishes exactly.

For this microscopic model, define the instantaneous
collective feedback field as
\begin{equation}
\phi(t)
\equiv
Jm(t)
=
\frac{J}{N}\bm c^{\mathsf T}\bm x(t)
=
\frac{J\|\bm c\|}{N}X(t).
\end{equation}
Its ensemble average corresponds to the physical HS saddle.
This definition expresses the projected stochastic dynamics
in the same collective-field normalization as the saddle
equation, without identifying the fluctuating microscopic
field with an independent HS integration variable.

Multiplying the projected $X$ equation by
$J\|\bm c\|/N$ and eliminating the orthogonal coordinates
gives the exact nonlinear memory-dressed field equation,
\begin{equation}
\begin{aligned}
\dot\phi(t)={}&-(r_0-JA_c)\phi(t)
-\frac{vA_c}{J^2}\phi^3(t)\\
&+\int_0^t ds\,K_\Sigma(t-s)\phi(s)
+JA_c h(t)+\xi_\phi(t).
\end{aligned}
\label{eq:exact_nonlinear_memory}
\end{equation}
Here
$\xi_\phi=J\|\bm c\|\eta_{\rm eff}/N$,
with the initial-state and noise contributions specified
in Sec.~\ref{sec:projection}. The orthogonal dynamics
remains linear, and its elimination is exact even at
finite collective-field amplitude. The quartic interaction
has been explicitly chosen along the collective channel;
generic nonlinear reference interactions instead produce
nonlinear memory and, in general, state-dependent noise.

Integrating the Gaussian orthogonal coordinates in the
equilibrium measure and expressing the resulting constrained
free-energy density in terms of $\phi=Jm$ gives, up to an
additive constant,
\begin{equation}
f_{\rm con}(\phi)
=
\frac{1}{2J^2}
\left[
\frac{r_0-\Sigma(0)}{A_c}-J
\right]\phi^2
+\frac{v}{4J^4}\phi^4.
\end{equation}
This is the constrained free-energy density of the
microscopic collective coordinate expressed in the
feedback-field normalization, rather than the auxiliary
HS free-energy density of Eq.~\eqref{eq:freeenergy}.

The same stationary condition follows from the long-time deterministic
limit of Eq.~\eqref{eq:exact_nonlinear_memory}, using
$\Sigma(0)=\int_0^\infty dt\,K_\Sigma(t)$. At zero external drive,
the large-$N$ saddle becomes unstable at the threshold in
Eq.~\eqref{eq:collective_threshold}, and the nonzero stationary branches
for $J>J_c$ obey
\begin{equation}
\phi_*^2
=
\frac{J^2}{v}(J-J_c).
\end{equation}
At finite $N$, the equilibrium distribution remains
reflection symmetric in the absence of a symmetry-breaking
source.

The nonlinear example therefore provides an exact
microscopic realization of a memory-dressed scalar
collective field with nonlinear stabilization. It does
not constitute a derivation of the full geometric or
nonreciprocal dynamics of a general neural system.

\begin{figure}[t]
\centering
\paperfigure[scale=0.43, trim=0.3cm 2.3cm 0cm 0cm]{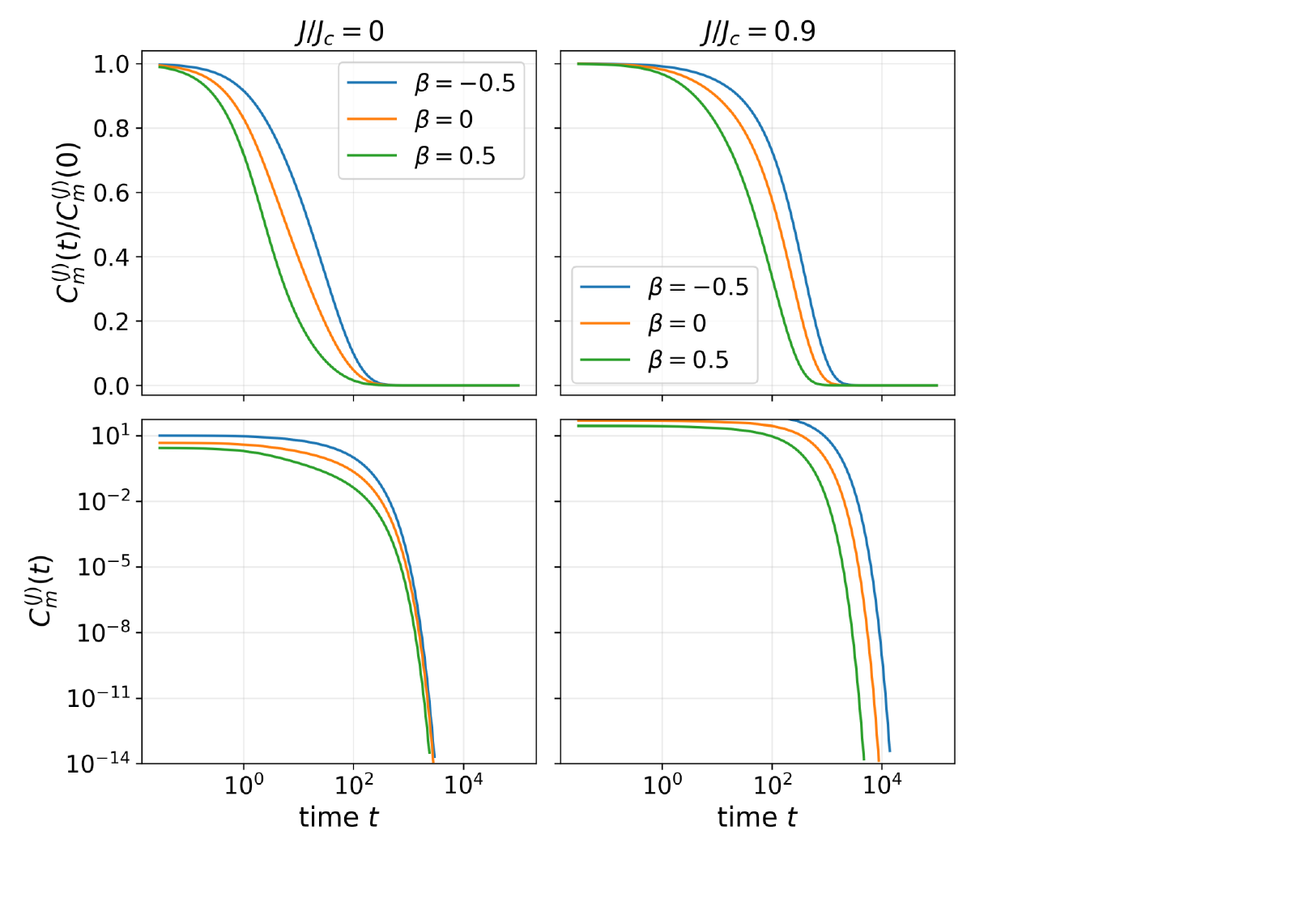}
\caption{Collective temporal correlations for three weighted
infrared spectral exponents $\beta=-0.5,0,0.5$, evaluated at
relative couplings $J/J_c=0$ and $0.9$. Upper panels show
$C_m^{(J)}(t)/C_m^{(J)}(0)$ and lower panels show
$C_m^{(J)}(t)$. Greater infrared spectral weight produces
more persistent correlations, while collective coupling
enhances fluctuations and slows their decay.}
\label{fig:supp_correlation_beta}
\end{figure}

\begin{figure}[t]
\centering
\paperfigure[scale=0.53, trim=-0.3cm 1cm 0cm 0cm]{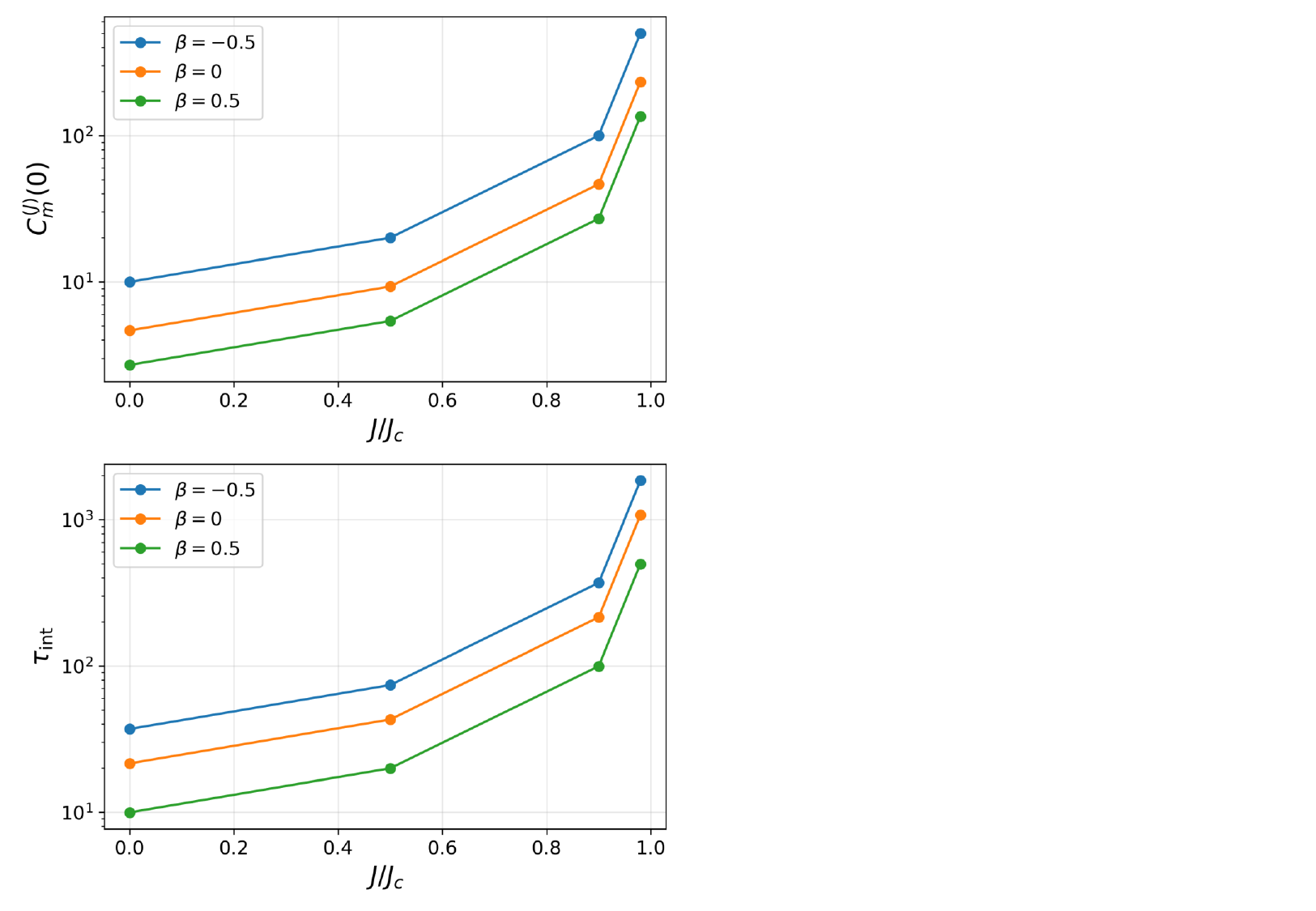}
\caption{Equal-time collective fluctuations $C_m^{(J)}(0)$
(top) and integrated correlation times $\tau_{\mathrm{int}}$
(bottom) versus $J/J_c$ for the three infrared spectral
exponents. The Gaussian relations
$C_m^{(J)}(0)=DM_1/(1-J/J_c)$ and
$\tau_{\mathrm{int}}=(M_2/M_1)/(1-J/J_c)$
account for the critical enhancement and for the different
spectral prefactors.}
\label{fig:supp_critical_beta}
\end{figure}

\section{Equilibrium collective fluctuations and temporal persistence}
\label{sec:correlations}
\label{sec:supp_temporal_correlations}

The preceding sections established that the collective ordering
threshold is governed by the relaxation spectrum projected onto
the collective interaction channel, and that the same spectrum
determines the memory-dressed collective response. We now examine
the corresponding temporal fluctuations, first on the disordered
side of the transition and subsequently within the ordered phase.

We restrict this analysis to the equilibrium Gaussian limit of
overdamped dynamics, where the fluctuation--dissipation theorem
(FDT) relates the collective response to its connected correlation
function~\cite{kubo1966}.

For the collective variable
$m(t)=N^{-1}\mathbf c^{\mathsf T}\mathbf x(t)$,
we define the $N$-rescaled connected correlation function
\begin{equation}
C_m^{(J)}(t)
=
N\left[
\langle m(t)m(0)\rangle
-\langle m\rangle^2
\right].
\label{eq:supp_Cm_definition}
\end{equation}
For $J>0$, the corresponding $N$-rescaled correlation function
of the collective feedback field $\phi=Jm$ is
$C_\phi^{(J)}(t)=J^2C_m^{(J)}(t)$, with the same normalized
temporal decay and integrated correlation time. We retain
$m$ as the collective observable to include the reference
limit $J=0$.

On the disordered side, we use the reference response
in Eq.~(\ref{eq:resolvent}) and the
interaction-dressed susceptibility derived in Sec.~III.C.
With the equilibrium white-noise convention
$\langle\eta_i(t)\eta_j(t')\rangle
=2D\delta_{ij}\delta(t-t')$, the classical FDT gives
\begin{equation}
\begin{split}
S_m^{(J)}(\omega)
&=
\frac{2D}{\omega}
\operatorname{Im}\chi_m^R(\omega),\\
C_m^{(J)}(t)
&=
\int_{-\infty}^{\infty}
\frac{d\omega}{2\pi}\,
e^{-i\omega t}S_m^{(J)}(\omega).
\end{split}
\label{eq:supp_fdt_correlation}
\end{equation}
The normalization in Eq.~\eqref{eq:supp_Cm_definition}
is consistent with the collective response to a field
coupled as $-Nh(t)m(t)$.

Equation~\eqref{eq:supp_fdt_correlation} assumes an
equilibrium gradient Langevin model with unit mobility.
For general nonequilibrium neural dynamics, the fluctuation
spectrum additionally depends on the effective noise kernel
and cannot be determined from the retarded response alone.

To distinguish static and temporal spectral sensitivity, introduce the
inverse-rate moments of the weighted spectrum in
Eq.~\eqref{eq:weighted_tdos},
\begin{equation}
M_n=\int d\lambda\,\frac{\rho_O(\lambda)}{\lambda^n}
=\sum_\alpha\frac{A_\alpha}{\lambda_\alpha^n}.
\label{eq:supp_spectral_moments}
\end{equation}
The ordering threshold is
$J_c^{-1}=\chi_0^R(0)=M_1$.
Defining $\epsilon=1-JM_1=1-J/J_c>0$,
the equal-time correlation follows from the static FDT:
\begin{equation}
C_m^{(J)}(0)
=
D\chi_m^R(0)
=
\frac{DM_1}{\epsilon}.
\label{eq:supp_equal_time}
\end{equation}

For a finite infrared cutoff, the low-frequency expansion
$\chi_0^R(\omega)=M_1+i\omega M_2+O(\omega^2)$
gives
\begin{equation}
\chi_m^R(\omega)
=
\frac{M_1}{\epsilon}
+i\omega\frac{M_2}{\epsilon^2}
+O(\omega^2).
\end{equation}
Consequently,
\begin{equation}
\lim_{\omega\to0}
\frac{
\operatorname{Im}\chi_m^R(\omega)
}{\omega}
=
\frac{M_2}{\epsilon^2}.
\label{eq:supp_low_freq_slope}
\end{equation}
Since
$S_m^{(J)}(0)=2\int_0^\infty dt\,C_m^{(J)}(t)$,
Eqs.~\eqref{eq:supp_fdt_correlation}
and \eqref{eq:supp_low_freq_slope} give
\begin{equation}
\int_0^\infty dt\,C_m^{(J)}(t)
=
\frac{DM_2}{\epsilon^2}.
\label{eq:supp_integrated_C}
\end{equation}
The integrated correlation time therefore obeys the
exact Gaussian relation
\begin{equation}
\tau_{\mathrm{int}}(J)
\equiv
\int_0^\infty dt\,
\frac{C_m^{(J)}(t)}{C_m^{(J)}(0)}
=
\frac{M_2}{M_1}
\frac{1}{1-J/J_c}.
\label{eq:supp_tau_int}
\end{equation}
Thus $M_1$ determines the collective ordering threshold,
whereas the additional inverse-rate moment $M_2$ controls
temporal persistence through the ratio $M_2/M_1$.
In particular, equal relative distances from the instability
need not imply equal correlation times for different
infrared spectra.

This distinction is directly related to the infrared
organization discussed in Sec.~\ref{sec:infrared}.
For a weighted spectral density $\rho_O(\lambda)=A\lambda^\beta$
over the interval $[\lambda_{\rm IR},\Lambda]$, the inverse-rate
moments become $M_n=A\int_{\lambda_{\rm IR}}^\Lambda
 d\lambda\,\lambda^{\beta-n}$.
At fixed spectral amplitude and ultraviolet cutoff,
$M_1$ diverges as $\lambda_{\rm IR}\to0$ for
$\beta\leq0$, whereas $M_2$ diverges for $\beta\leq1$,
with logarithmic divergences at $\beta=0$ and $\beta=1$,
respectively.

In particular, for $0<\beta\leq1$, the collective
ordering threshold remains finite, while the integrated
correlation time diverges even at a fixed relative
coupling $J/J_c<1$. Unlike conventional critical slowing
down associated with approaching a collective instability
~\cite{hohenberg1977}, this mechanism originates from
the infrared organization of the relaxation spectrum
and can persist at a finite relative distance from
the collective instability.

Moreover, the moment $M_2$ coincides with the temporal
moment $Z_1$ introduced in Sec.~\ref{sec:nonlinear}.
The same infrared spectral contribution that prevents
a uniform local derivative expansion thus also controls
the persistence of equilibrium collective fluctuations.

We illustrate this distinction by numerically evaluating the
Gaussian spectral expressions derived above. We consider
response-weighted spectra $\rho_O(\lambda)\propto\lambda^\beta$
for $\beta=-1/2,0,1/2$ over a common rate interval
$0.01\leq\lambda\leq1$. Each spectrum is discretized into
$N=2048$ relaxation modes using deterministic quantiles
of the prescribed distribution, with uniform mode weights
$A_\alpha=1/N$ and noise strength $D=1$. For each spectrum,
$J_c$ is computed from its discrete $M_1$.

The temporal correlations are evaluated using the spectral
decomposition of the interaction-dressed relaxation operator.
Writing the corresponding susceptibility as
$\chi_m^R(\omega)=\sum_\nu b_\nu/(\mu_\nu-i\omega)$,
where $\mu_\nu$ and $b_\nu$ denote the dressed relaxation
rates and their collective-channel spectral weights,
respectively, the FDT yields
\begin{equation}
C_m^{(J)}(t)
=
D\sum_\nu
\frac{b_\nu}{\mu_\nu}
e^{-\mu_\nu|t|}.
\label{eq:collective_spectral_correlation}
\end{equation}

Figure~\ref{fig:supp_correlation_beta} compares the temporal
correlations at $J/J_c=0$ and $0.9$, showing that
infrared-enhanced spectra exhibit slower normalized temporal
decay at fixed relative coupling. Increasing $J/J_c$
further enhances the correlation amplitude and produces
Gaussian critical slowing down.

For the finite-dimensional stable spectra considered
here, the ultimate long-time decay is exponential,
governed by the smallest dressed relaxation rate with
nonzero collective-channel spectral weight.
This behavior is consistent with the distinction between
intermediate-time spectral scaling and asymptotic
relaxation discussed in Sec.~VII.
Further details of finite-size infrared resolution
and temporal crossover scales are provided in
Sec.~II F of the Supplemental Material.

Figure~\ref{fig:supp_critical_beta} summarizes the
corresponding equal-time fluctuations and integrated
correlation times at $J/J_c=0,0.5,0.9,0.98$.
Both grow as $(1-J/J_c)^{-1}$ within the Gaussian
approximation, but their spectral prefactors differ:
$C_m^{(J)}(0)$ is proportional to $M_1$, whereas
$\tau_{\mathrm{int}}$ is proportional to $M_2/M_1$.

At $J/J_c=0.9$, the representative spectra give
$\tau_{\mathrm{int}}\simeq370,215,100$ for
$\beta=-0.5,0,0.5$, respectively. These results
visualize the temporal consequences of the same
projected relaxation spectrum that governs the
collective instability.

The preceding spectral distinction also extends to
fluctuations within the ordered phase. Let $\mathbf{x}_*$
denote a locally stable ordered configuration with
$m_*\neq0$. Linearization about this configuration gives
the relaxation operator
\begin{equation}
\mathcal{L}_*
=
\left.\nabla^2 H_0\right|_{\mathbf{x}_*}
-
\frac{J}{N}\mathbf{c}\mathbf{c}^{\mathsf T}.
\label{eq:ordered_relaxation_operator}
\end{equation}
The collective susceptibility retains the structure of
Eq.~\eqref{eq:physicalchi}, with the reference
polarization now evaluated from the spectrum of the
reference Hessian $\left.\nabla^2 H_0\right|_{\mathbf{x}_*}$.
Denoting its collective-channel weighted spectral density
by $\rho_{O,*}(\lambda)$, we define the corresponding
inverse-rate moments
$M_n^*=\int d\lambda\,\rho_{O,*}(\lambda)/\lambda^n$.

Within the Gaussian approximation about a single stable
ordered configuration, the preceding FDT analysis directly
gives
\begin{equation}
\tau_{\mathrm{int}}^{\mathrm{ord}}
=
\frac{M_2^*}{M_1^*}
\frac{1}{1-JM_1^*}.
\label{eq:ordered_temporal_persistence}
\end{equation}
Here $\epsilon_*=1-JM_1^*>0$ characterizes local
collective stability in the ordered background, rather
than the distance from the original ordering threshold.
The static collective susceptibility is correspondingly
given by $M_1^*/\epsilon_*$.

If the ordered background retains a broad infrared
relaxation spectrum, temporal persistence can remain
strongly enhanced after collective order has been
established. In particular, for
$\rho_{O,*}(\lambda)\sim A_*\lambda^{\beta_*}$ with
$0<\beta_*\leq1$, the integrated correlation time can
diverge in the gapless spectral limit while the static
collective susceptibility remains finite, provided that
$\epsilon_*$ remains finite and positive. For nearly
flat weighted spectra with a finite infrared cutoff,
the strong infrared sensitivity of $M_2^*/M_1^*$
similarly allows substantial temporal persistence
within a locally stable ordered phase.

The present analysis therefore distinguishes collective
ordering from temporal persistence on both sides of
the transition at the level of equilibrium Gaussian
fluctuations. Although both are governed by the projected
relaxation spectrum, temporal persistence is additionally
sensitive to the second inverse-rate moment. A broad
infrared relaxation spectrum can thus support long-time
collective dynamics even after the formation of a stable
collective order parameter, without requiring proximity
to the original ordering threshold. The persistence of
such infrared spectral organization within the ordered
phase depends on the underlying nonlinear dynamics.
Extensions beyond Gaussian fluctuations and to generic
nonequilibrium neural dynamics require additional analysis.

\section{Numerical investigations of infrared spectral organization}
\label{sec:numerics}

\subsection{Numerical tests of spectral control}

\begin{figure*}[t]
\centering
\paperfigure[scale=0.75, trim=-1.3cm 5.2cm 0cm 0cm]{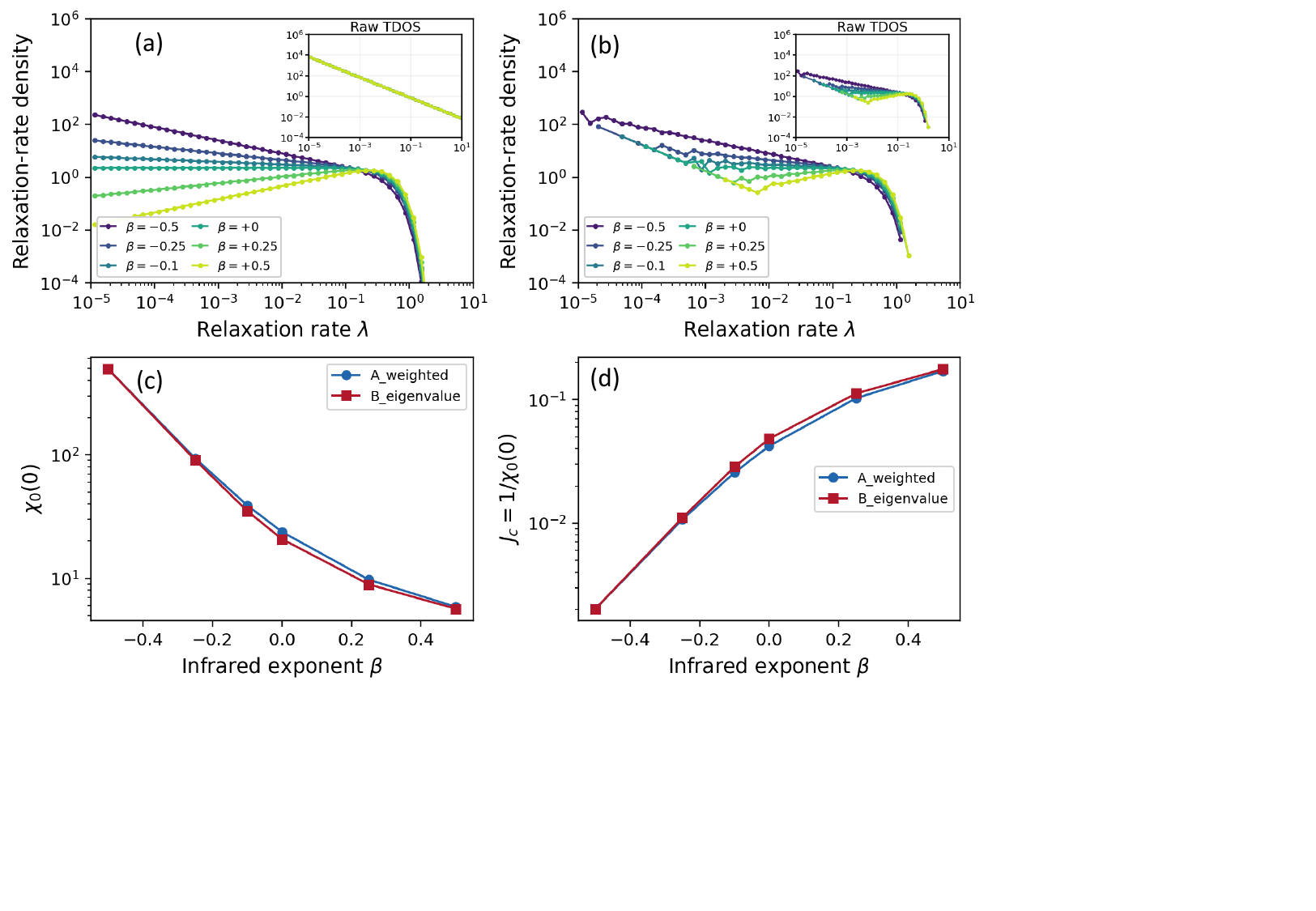}
\caption{
Spectral organization and collective response
in microscopic relaxation matrices.
Two constructions of symmetric, non-diagonal
operators with $N=2048$ are compared.
(a,b) Response-weighted relaxation spectra
$\rho_O(\lambda)$ for six infrared exponents
$\beta=-0.5,-0.25,-0.1,0,0.25,0.5$.
Construction A varies collective-mode overlaps
at fixed eigenvalues, whereas construction B
varies the eigenvalue density with uniform overlaps.
Insets show the corresponding unweighted
relaxation-rate densities.
(c) Static reference response $\chi_0^R(0)$.
(d) Gaussian ordering threshold
$J_c=1/\chi_0^R(0)$.
}
\label{fig:microscopic_response}
\end{figure*}

\begin{figure}[t]
\centering
\paperfigure[scale=0.49, trim=0.5cm 5.2cm 0cm 0cm]{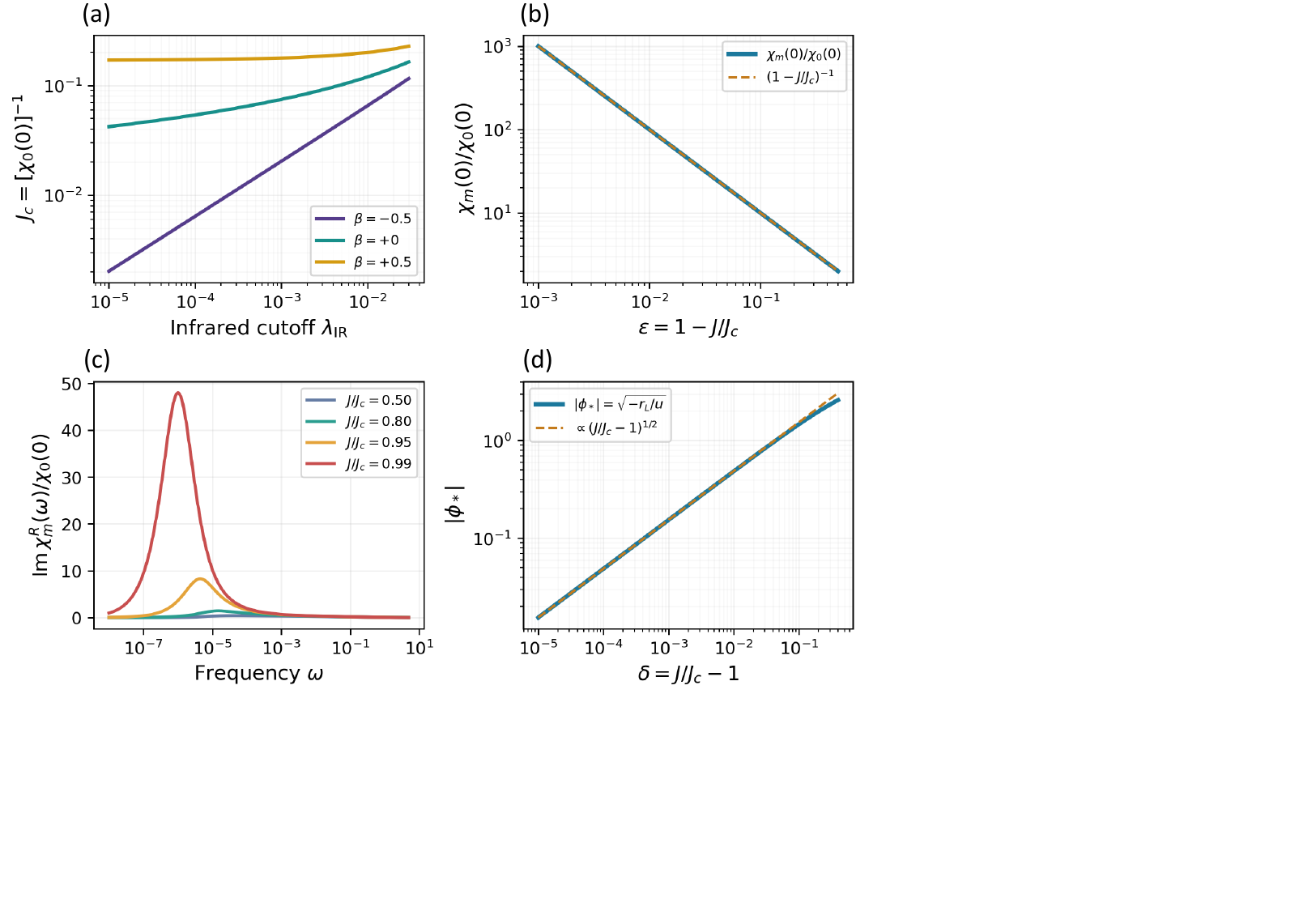}
\caption{
Infrared spectral control of collective criticality.
(a) Dependence of the Gaussian ordering threshold
$J_c=[\chi_0^R(0)]^{-1}$ on the infrared cutoff
$\lambda_{\rm IR}$ for representative infrared exponents.
(b) Normalized static collective susceptibility
$\chi_m(0)/\chi_0^R(0)$ approaching the Gaussian
instability at $J=J_c$.
(c) Low-frequency imaginary part of the collective
response for $\beta=0$, illustrating dynamical
softening as $J/J_c$ approaches unity.
(d) Illustrative Landau order-parameter scaling,
$|\phi_*|=\sqrt{-r_L/u}\propto(J/J_c-1)^{1/2}$,
assuming $u>0$; the quartic coefficient is not
independently evaluated from microscopic
fourth-order cumulants.
}
\label{fig:infrared_scaling}
\end{figure}

To test the spectral formulation of collective criticality,
we construct symmetric, non-diagonal relaxation operators
of dimension $N=2048$ as
$L_0=U\,\mathrm{diag}(\lambda_\alpha)U^{\mathsf T}$,
where $U$ is obtained by QR decomposition of a Gaussian
random matrix. For a collective channel $\bm c$, the weighted spectrum
is constructed using the mode overlaps in Sec.~II.B.
We prescribe a normalized target measure
$\rho_O(\lambda)\propto
\lambda^\beta e^{-(\lambda/\lambda_{\rm UV})^2}$,
with $\lambda_{\rm UV}=0.5$,
$\lambda_{\rm IR}=10^{-5}$, $\lambda_{\max}=10$,
and six exponents $\beta=-0.5,-0.25,-0.1,0,0.25,0.5$.
Details of the spectral constructions, eigenmode
diagnostics, collective response, and finite-size
temporal response kernels are provided in
Sec.~II of the Supplemental Material.

Figure~\ref{fig:microscopic_response} compares two
complementary constructions. In construction A,
logarithmically spaced relaxation rates are fixed while
the collective-mode overlaps are varied. The unweighted
density is thus independent of $\beta$, whereas the
weighted density follows
$\rho_O(\lambda)\propto\lambda^\beta$ in the infrared
[Fig.~\ref{fig:microscopic_response}(a)].
In construction B, relaxation rates follow the quantiles
of the target distribution with uniform weights
$A_\alpha=1/N$, so both densities exhibit the prescribed
infrared scaling [Fig.~\ref{fig:microscopic_response}(b)].

The two constructions approximate the same weighted
continuum measure through distinct finite-size discretizations.
In particular, the quantile-based B construction may have
a smallest sampled relaxation rate well above the nominal
infrared cutoff, whereas the logarithmically spaced A
construction retains modes close to that cutoff.
Consequently, their temporal kernels can show different
crossover times at finite $N$, even when their continuum
limits agree. The observed A--B discrepancy is consistent
with finite-size infrared discretization effects rather
than requiring different underlying continuum memory laws.
Quantitative confirmation of convergence would require
increasing $N$ and comparing both constructions with the
continuum spectral integral.

The static reference response and Gaussian threshold are
reconstructed from the microscopic eigenmodes as
$\chi_0^R(0)=\sum_\alpha A_\alpha/\lambda_\alpha$
and $J_c=[\chi_0^R(0)]^{-1}$.
Figures~\ref{fig:microscopic_response}(c,d) show that
enhanced slow-mode weight increases the static response
and lowers the collective threshold.
For $\beta=-0.5$, constructions A and B yield
$\chi_0^R(0)\simeq493.12$ and $492.72$, respectively,
corresponding to $J_c\simeq2.03\times10^{-3}$.
The corresponding static responses decrease to $23.77$
and $20.78$ at $\beta=0$, and to $5.89$ and $5.68$
at $\beta=0.5$.
Apart from finite-size discretization effects, these
results demonstrate that the collective response is
governed by the weighted spectrum rather than the
unweighted relaxation-rate density alone.

Figure~\ref{fig:infrared_scaling}(a) shows the infrared
cutoff dependence of $J_c$ for $\beta=-0.5,0,$ and $0.5$,
exhibiting power-law enhancement of the static response,
logarithmic sensitivity, and infrared convergence,
respectively. Unlike the fixed-amplitude asymptotics
discussed in Sec.~\ref{sec:infrared}, these calculations
retain the ultraviolet regularization and cutoff-dependent
normalization of the prescribed measure.

The collective susceptibility is calculated from
Eq.~\eqref{eq:physicalchi}.
Defining $\epsilon=1-J/J_c$, the normalized static
susceptibility obeys
$\chi_m^R(0)/\chi_0^R(0)=\epsilon^{-1}$
[Fig.~\ref{fig:infrared_scaling}(b)], corresponding to
the Gaussian exponent $\gamma_{\rm L}=1$ rather than
an independently measured fluctuation-corrected exponent.

Figure~\ref{fig:infrared_scaling}(c) shows the dynamical
response $\operatorname{Im}\chi_m^R(\omega)/\chi_0^R(0)$
for $\beta=0$.
As $J/J_c$ increases from $0.50$ to $0.99$, the peak
frequency decreases from approximately $4.3\times10^{-5}$
to $1.0\times10^{-6}$, while its normalized amplitude
increases from $0.39$ to $48$, consistent with collective
critical slowing down.
The peak frequency characterizes the response scale
but need not correspond to an isolated relaxation pole.
The equilibrium fluctuation analysis in
Sec.~\ref{sec:correlations} further distinguishes
the collective ordering threshold, governed by the first
inverse-rate spectral moment, from temporal persistence,
which also depends on the second inverse-rate moment.

Finally, Fig.~\ref{fig:infrared_scaling}(d) illustrates
the nonlinear ordering predicted by the static HS expansion
in Sec.~\ref{sec:nonlinear}.
For an assumed positive quartic coefficient, the ordered
solution obeys
$|\phi_*|=\sqrt{-r_L/u}\propto(J/J_c-1)^{1/2}$
near the threshold, illustrating the mean-field exponent
$\beta_{\rm L}=1/2$ without independently calculating
the microscopic fourth cumulant determining $u$.

Together, Figs.~\ref{fig:microscopic_response}
and~\ref{fig:infrared_scaling} establish a controlled
connection between microscopic relaxation spectra,
collective-mode overlaps, and infrared-enhanced
static and dynamic collective responses.
The spectral distributions are prescribed rather than
generated through learning, while the finite infrared
cutoff and discrete mode sampling restrict the observable
power-law time window.

\subsection{Microscopic reconstruction of irreducible memory}
\label{sec:numerical_memory}

\begin{figure*}[t]
\centering
\paperfigure[scale=0.75, trim=-1.5cm 5.3cm 0cm 0cm]{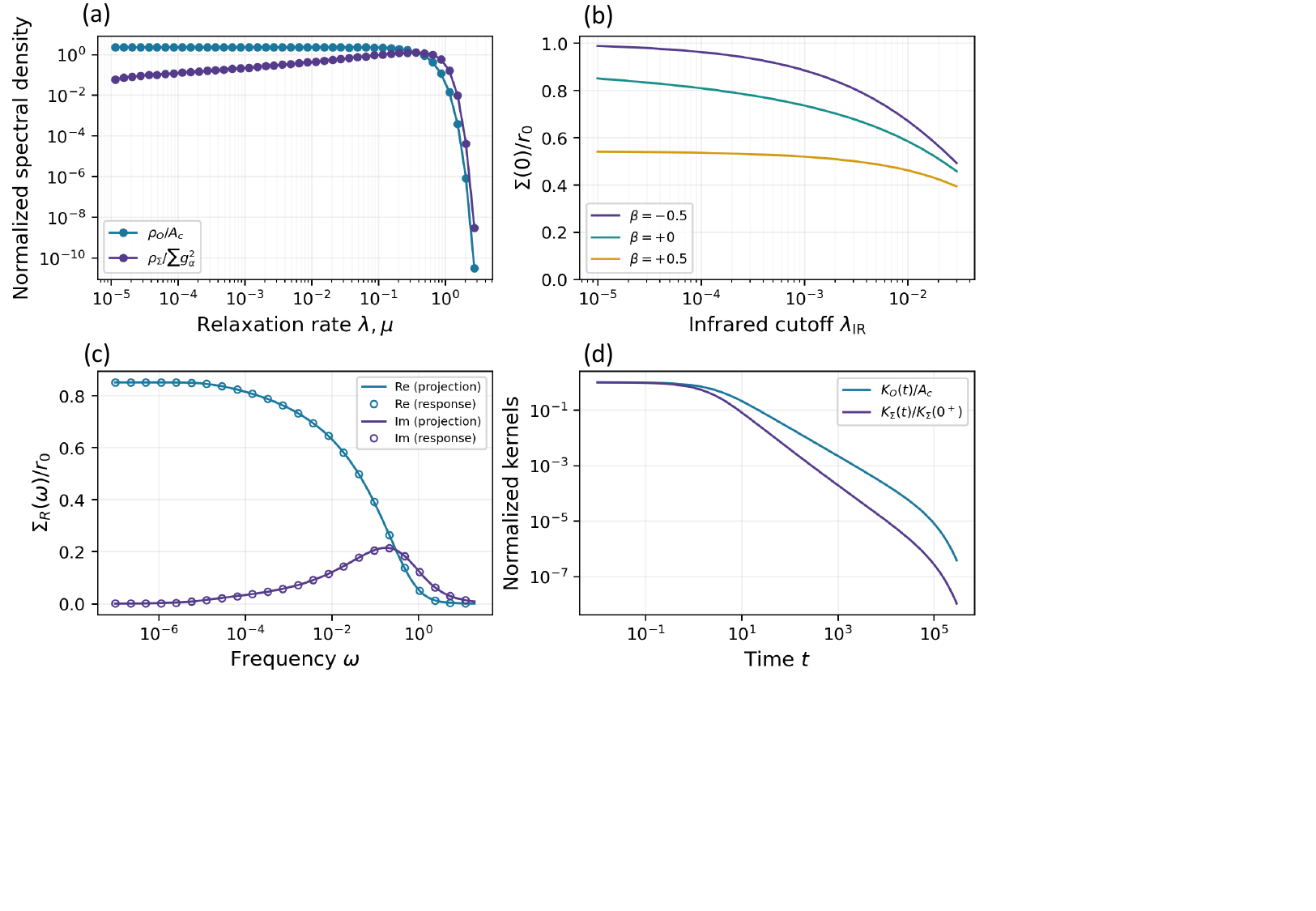}
\caption{
Microscopic reconstruction of irreducible memory
feedback from the relaxation spectrum.
(a) Normalized response-weighted spectrum
$\rho_O(\lambda)$ and irreducible memory spectrum
$\rho_\Sigma(\mu)$ for the A-weighted construction
with $N=2048$ and $\beta=0$.
(b) Relative static memory dressing
$\Sigma_R(0)/r_0$ as a function of the infrared
cutoff $\lambda_{\rm IR}$ for representative
infrared exponents.
(c) Real and imaginary parts of the retarded
memory self-energy $\Sigma_R(\omega)$.
Solid curves represent the direct Mori--Zwanzig
projection, while symbols denote reconstruction
from the microscopic reference response.
(d) Normalized reference-response kernel
$K_O(t)/K_O(0^+)$ and irreducible memory kernel
$K_\Sigma(t)/K_\Sigma(0^+)$ obtained from their
respective microscopic relaxation modes.
}
\label{fig:irreducible_memory}
\end{figure*}

Using the same spectral construction as in the preceding
subsection, we reconstruct the irreducible memory
self-energy through two independent numerical procedures:
direct Mori--Zwanzig projection of the microscopic
relaxation operator and inversion of the corresponding
reference response.

For a specified collective direction
$\bm e=\bm c/\|\bm c\|$, the orthogonal relaxation
operator $L_Q$ and the coupling vector $\bm b$
are constructed in the microscopic eigenmode basis.
Diagonalization of $L_Q$ yields the internal relaxation
rates $\mu_\alpha$ and their corresponding coupling
weights $g_\alpha^2$, from which the irreducible memory
spectrum is obtained as
\(
\rho_\Sigma(\mu)
=
\sum_{\alpha=1}^{N-1}
g_\alpha^2\delta(\mu-\mu_\alpha).
\)
Unlike the response-weighted spectrum $\rho_O(\lambda)$,
which describes relaxation in the full reference system,
$\rho_\Sigma(\mu)$ resolves the elementary memory-feedback
processes mediated by the orthogonal internal degrees
of freedom.

Figure~\ref{fig:irreducible_memory}(a) compares the
two spectra for the A-weighted construction with
$\beta=0$.
Although the response-weighted spectrum is nearly flat
in the infrared, the irreducible memory spectrum exhibits
a progressive suppression toward small relaxation rates.
This behavior is consistent with the logarithmically
suppressed infrared spectrum
$\rho_\Sigma(\mu)\propto
[\ln(\Lambda/\mu)]^{-2}$
derived in Sec.~\ref{sec:infrared}.
The two spectra therefore exhibit distinct infrared
structures even though they originate from the same
microscopic relaxation operator.
Their normalization in the figure facilitates comparison
of their spectral shapes rather than their absolute
spectral weights.

The static memory self-energy can be reconstructed
directly from the response-weighted spectral moments as
\begin{equation}
\Sigma_R(0)
=
r_0-\frac{A_c}{\chi_0^R(0)},
\qquad
r_0=
\frac{1}{A_c}
\sum_\alpha A_\alpha\lambda_\alpha,
\end{equation}
where $A_c=\sum_\alpha A_\alpha$.
Figure~\ref{fig:irreducible_memory}(b) shows the
infrared cutoff dependence of the relative memory
dressing $\Sigma_R(0)/r_0$ for $\beta=-0.5,0,$
and $0.5$.

For $\beta<0$, the static memory self-energy rapidly
approaches the bare projected relaxation rate as the
infrared cutoff is reduced, whereas the marginal case
$\beta=0$ exhibits a slower logarithmic approach.
For $\beta>0$, the relative memory dressing remains
below unity in the gapless continuum limit.
These results illustrate that the infrared enhancement
of the static response does not require a divergent
irreducible memory self-energy.
Instead, it originates from the progressive cancellation
of the bare projected relaxation by internal memory
feedback.

For the A-weighted construction with $\beta=0$,
$N=2048$, and $\lambda_{\rm IR}=10^{-5}$,
we obtain $r_0\simeq0.28210$ and
$\Sigma_R(0)\simeq0.24003$, yielding
$r_0-\Sigma_R(0)\simeq0.04208$.
Thus, approximately $85\%$ of the bare projected
relaxation is compensated by the static memory
self-energy before the explicit collective interaction
$J$ is introduced.
For the present normalization $A_c=1$, the remaining
static restoring scale directly determines the
collective ordering threshold through
$J_c=[r_0-\Sigma_R(0)]/A_c$.

The dynamical self-energy is independently evaluated from the projected
internal relaxation modes using Eq.~\eqref{eq:supp_mz_sigma}, and from
the microscopic reference response using Eq.~\eqref{eq:supp_mz_inverse},
with $z=-i\omega+0^+$.
As shown in Fig.~\ref{fig:irreducible_memory}(c),
the two numerical evaluations coincide for both
the real and imaginary parts of the self-energy.
The maximum absolute difference, normalized by
the maximum self-energy magnitude over the sampled
frequency range, is approximately $1.2\times10^{-13}$.
This agreement provides a numerical consistency check
of the exact Schur-complement relation connecting
the microscopic reference response and irreducible
memory feedback.

Finally, Fig.~\ref{fig:irreducible_memory}(d) compares the temporal
kernels reconstructed independently from the two sets of relaxation
modes using Eqs.~\eqref{eq:spectralIR} and~\eqref{eq:memory_kernel}.
The normalized irreducible memory kernel exhibits
a faster decay than the reference-response kernel
over the resolved long-time regime.
This difference is consistent with the distinct
continuum asymptotics
$K_O(t)\propto t^{-1}$ and
$K_\Sigma(t)\propto[t\ln^2(\Lambda t)]^{-1}$
for a flat, gapless response-weighted spectrum.

The finite infrared cutoff ultimately produces
exponential relaxation in both kernels.
The present calculations therefore illustrate the
predicted distinction between the two temporal
structures, rather than establish their asymptotic
scaling in the gapless continuum limit.
A quantitative test of the logarithmic correction
would require systematic finite-size and infrared
cutoff analysis.

Together, these results resolve the microscopic
spectral origin of memory dressing and demonstrate
how infrared-enhanced collective response emerges
from the interplay between direct projected relaxation
and irreducible internal memory feedback.
The collective instability condition can consequently
be expressed in two equivalent forms,
\(
J_c
=
\frac{1}{\chi_0^R(0)}
=
\frac{r_0-\Sigma_R(0)}{A_c},
\)
thereby connecting the spectral control of collective
ordering to its underlying microscopic memory dynamics.

\subsection{Learning-induced infrared organization}
\label{sec:supp_learning_critical}

\begin{figure*}[t]
  \centering
  \paperfigure[scale=0.68, trim=-1.3cm 9.7cm 0cm 0cm]{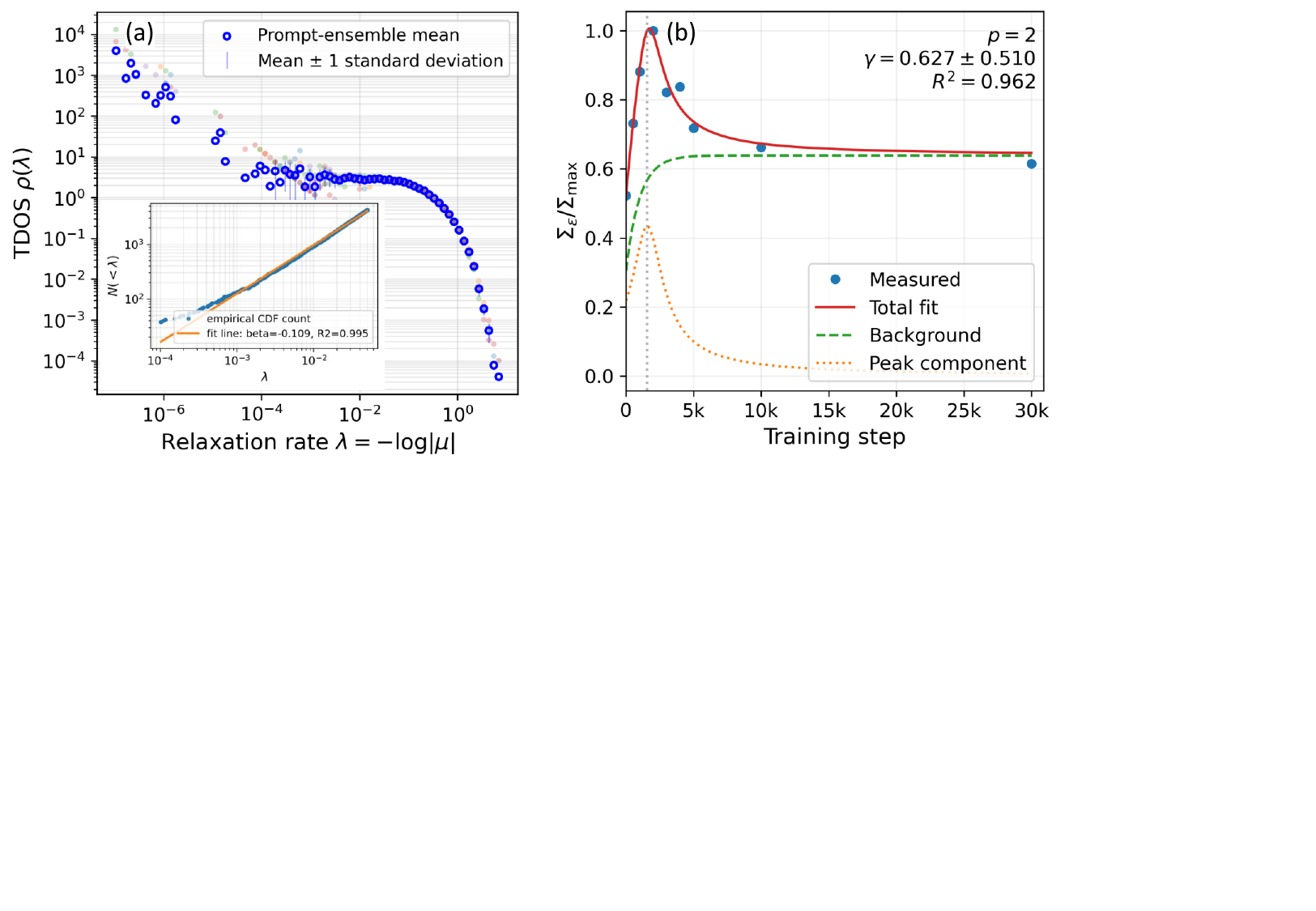}
  \caption{
  Learning-induced infrared spectral organization and transient
  critical-like response in the Pythia-410M Transformer.
  (a) TDOS averaged over five distinct input prompts.
  Error bars indicate one standard deviation across the prompt ensemble.
  The inset shows the cumulative mode count and its power-law fit,
  yielding an infrared spectral exponent $\beta=-0.109$.
  (b) Evolution of the normalized static response estimator during
  learning for layer $5\to6$.
  The solid curve represents the total phenomenological fit, while
  the dashed and dotted curves represent the background and transient
  contributions, respectively.
  The vertical dotted line indicates the center $s_c$ of the transient
  response. The fitted effective response exponent is
  $\gamma=0.627\pm0.510$.
  }
  \label{fig:ir_criticality}
\end{figure*}

Having established the spectral control of collective response, ordering,
and irreducible memory feedback in controlled microscopic models, we now
examine whether analogous infrared spectral organization emerges during
learning in actual neural architectures. Learned neural architectures can
be represented as high-dimensional nonlinear maps between hidden states,
$\mathbf{h}_{\ell+1}=\mathcal{F}_{\ell}(\mathbf{h}_{\ell})$, whose full
Jacobians capture the local linearized response without decomposing their
internal computational operations. Recent Transformer and Mamba studies
motivate spectral comparison with the present microscopic framework
\cite{chae2026a,chae2026b}.

Our prediction concerns the \emph{response-weighted} low-rate spectrum of
a specified collective observable. Transformer layer-to-layer Jacobians
describe computational propagation, requiring a validated effective
evolution law to identify physical relaxation rates. Non-normal or
nonreciprocal operators also require appropriate left/right mode overlaps.
A nearly flat unweighted TDOS alone establishes neither a physical-time
$1/t$ response kernel nor spontaneous collective ordering.

Recent measurements of Transformer Jacobian spectra reveal a robust
near-marginal infrared distribution, $\rho(\lambda)\sim\lambda^\beta$
with $\beta\simeq-0.1$, together with a pronounced transient maximum of
the infrared spectral estimator during early learning \cite{chae2026a}.
Despite substantial spectral redistribution during training, the infrared
scaling remains approximately invariant across learning stages and input
realizations. Figure~\ref{fig:ir_criticality}(a) shows the TDOS averaged
over five distinct input prompts for the Pythia-410M Transformer.
The spectra exhibit a common infrared structure despite prompt-dependent
fluctuations. The cumulative mode count follows
$N(\lambda)\sim\lambda^{1+\beta}$, yielding $\beta=-0.109$ with
$R^2=0.995$ [inset of Fig.~\ref{fig:ir_criticality}(a)].

Within the present microscopic framework, the measured infrared spectrum
is more naturally associated with the spectral representation of the
collective reference response than with that of the irreducible memory
self-energy. Under the approximation of uniform response weights and
neglecting the oscillatory components of the Jacobian modes, we introduce
the regularized static response estimator
\begin{equation}
  \chi_{{\rm est},\epsilon}(0;s)
  =\frac{A_c}{N}
  \sum_{\alpha:\lambda_\alpha(s)>0}
  \frac{1}{\lambda_\alpha(s)+\epsilon},
  \label{eq:learning_response_estimator}
\end{equation}
where $s$ denotes the training step, $\epsilon$ is a positive infrared
regularization parameter, and $N$ is the dimension of the Jacobian.
This estimator is proportional to the inverse-rate spectral quantity
previously introduced in Ref.~\cite{chae2026a} as a memory self-energy
estimator. The distinction established in the present microscopic theory
motivates its reinterpretation as a spectral proxy for the static reference
response. These observations suggest that learning-induced infrared
softening may enhance the collective reference response and reduce the
critical coupling for macroscopic ordering, provided that the slow modes
have appreciable overlap with the relevant collective channel.

Figure~\ref{fig:ir_criticality}(b) shows the evolution of the normalized
static response estimator
$\overline{\chi}_{\rm est}(s)
=\chi_{{\rm est},\epsilon}(0;s)/\chi_{\rm est,max}$.
We phenomenologically decompose the observed response estimator into an
evolving background and a transient critical-like contribution,
\begin{equation}
  \overline{\chi}_{\rm est}(s)
  =\overline{\chi}_f
  +(\overline{\chi}_i-\overline{\chi}_f)e^{-s/\tau}
  +\frac{B}{\left[1+((s-s_c)/w)^2\right]^\gamma}.
  \label{eq:learning_critical_fit}
\end{equation}
The exponential background describes the gradual reorganization of the
preexisting relaxation spectrum, whereas the second term provides a
phenomenological representation of a rounded critical-like enhancement.
Its functional form is motivated by the critical response scaling
$\chi_m(0)\sim\delta_J^{-\gamma}$ and a local quadratic dependence of
the effective distance $\delta_J$ from collective instability along
the learning trajectory. The fit yields an effective response exponent
$\gamma=0.627\pm0.510$ with $R^2=0.962$, consistent, within the fitting
uncertainty, with the mean-field value $\gamma_{\rm MF}=1$.
This numerical consistency does not independently establish a critical
exponent or a collective phase transition. The spectral decomposition,
theoretical motivation, and numerical fitting procedure are detailed
in Secs.~III.A--III.C of the Supplemental Material.

The possible implication of this infrared enhancement for collective
ordering follows from Eq.~\eqref{eq:collective_threshold}, which gives
the learning-dependent threshold $J_c(s)=1/\chi_0(0;s)$.
To estimate the learning-induced variation of the ordering threshold,
we assume a phenomenological relation
$\chi_0(0;s)=\chi_{\rm bg}+A\overline{\chi}_{\rm est}(s)$,
with a constant background $\chi_{\rm bg}$ and $A>0$.
The resulting critical coupling satisfies
$J_c(s)/J_c(s_{\rm peak})
=(b+1)/[b+\overline{\chi}_{\rm est}(s)]$,
with $b=\chi_{\rm bg}/A$ and
$\overline{\chi}_{\rm est}(s_{\rm peak})=1$.
Under this assumption, the observed maximum of the static response
estimator corresponds to a minimum of the estimated critical coupling,
indicating a reduced threshold for collective ordering.

The present framework also provides a route to reconstructing the
irreducible memory self-energy from the observed relaxation spectrum.
Unlike the static reference response, the irreducible memory self-energy
is determined through the inverse of the memory-dressed reference
propagator through Eq.~\eqref{eq:supp_mz_inverse}.
Within an effective symmetric relaxation model with uniform response
weights, the corresponding spectral estimate takes the form
\begin{equation}
  \Sigma_{\rm est}(z;s)
  =z+\langle\lambda\rangle_s
  -\left\langle\frac{1}{z+\lambda}\right\rangle_s^{-1},
  \label{eq:learning_memory_reconstruction}
\end{equation}
where $\langle\cdots\rangle_s$ denotes the normalized spectral average
over the retained stable modes at training step $s$. Its relation to
the full-Jacobian normalization in
Eq.~\eqref{eq:learning_response_estimator} is given in Sec.~III.A of
the Supplemental Material. In particular, its static limit is given by
\begin{equation}
  \Sigma_{\rm est}(0;s)
  =\langle\lambda\rangle_s
  -\frac{1}{\langle\lambda^{-1}\rangle_s},
  \label{eq:learning_static_memory}
\end{equation}
corresponding to the difference between the arithmetic and harmonic
means of the relaxation rates. The inverse-rate moment therefore controls
the static reference response within the symmetric surrogate, whereas
the irreducible memory self-energy depends on both the direct relaxation
scale and the inverse-rate spectral moment. Consequently, a transient
enhancement of the reference response does not necessarily imply a
corresponding maximum of the irreducible memory self-energy.

The irreducible memory self-energy may remain finite even when the
reference response becomes singular, with the infrared singularity
instead appearing through the suppression of the static inverse-response
scale $r_0-\Sigma(0)=A_c/\chi_0(0)$. For a flat gapless response-weighted
spectrum, this distinction is further reflected in the logarithmically
suppressed irreducible memory spectrum
$\rho_\Sigma(\lambda)\propto[\ln(\Lambda/\lambda)]^{-2}$,
and the corresponding long-time memory kernel
$K_\Sigma(t)\propto[t\ln^2(\Lambda t)]^{-1}$,
rather than the $1/t$ reference response.

The quantitative spectral reconstruction presented in Sec.~III.D of
the Supplemental Material shows that, within the symmetric surrogate,
the self-energy is evaluated at the same positive infrared regulator
as the reference-response estimator. The transient maximum of the
regularized reference-response estimator does not coincide with a maximum of the irreducible memory self-energy.
Instead, the reconstructed memory self-energy increases during later
learning, while the inverse-response scale is governed by the competition
between direct relaxation and memory feedback. The relative strength of
memory dressing also increases during this stage, despite the reduction
of the regularized reference response. Thus, the transient response
enhancement and the subsequent organization of memory feedback represent
distinct aspects of the learning-induced spectral evolution.

The reconstructed quantities characterize an effective symmetric
relaxation model with uniform response weights, rather than direct
measurements of the Transformer collective response or projected memory
self-energy. The distinction between infrared marginality and collective
instability, together with the sensitivity to individual near-zero
relaxation modes, is examined in Sec.~III.E of the Supplemental Material.
Direct perturbation-response measurements, explicit mode-overlap analysis,
controlled feedback variation, and comparisons between learned and
untrained networks are needed to test the collective ordering and
memory-feedback predictions of the present framework.

\section{Discussion}
\label{sec:discussion}

The present formulation provides complementary microscopic perspectives
on collective ordering and memory dynamics. The dynamical HS
transformation identifies the collective interaction channel and its
instability, whereas the Mori--Zwanzig projection reveals how the
remaining internal degrees of freedom generate memory feedback in the
projected dynamics. Their equivalence at the level of the linearized
collective response clarifies that collective interactions and memory
feedback need not be introduced as independent phenomenological
ingredients. Instead, they enter the same microscopic response through
distinct physical contributions: the collective interaction modifies
the restoring force, while the elimination of internal degrees of
freedom generates a frequency-dependent memory self-energy. The static
component of the memory self-energy therefore enters the collective
instability condition itself, linking the organization of internal
relaxation modes to the emergence of macroscopic collective order.

An important physical implication is the distinction between the
response-weighted relaxation spectrum and the spectral measure of the
irreducible memory self-energy. A broad distribution of microscopic
relaxation rates does not directly determine the irreducible memory
kernel, because the latter also depends on the projection onto the
collective coordinate. The marginal continuum provides an analytic
example of this distinction. Although the irreducible memory kernel
decays as $K_\Sigma(t)\sim[t\ln^2(\Lambda t)]^{-1}$, the memory-dressed
reference response recovers a nonintegrable $1/t$ tail in the gapless
limit. Thus, an integrable memory kernel can support a nonintegrable
reference response through repeated memory feedback. At a finite
infrared cutoff, the scale-free response is restricted to an
intermediate temporal window.

The equilibrium fluctuation analysis further clarifies the physical
distinction between collective ordering and temporal persistence.
Conventional critical slowing down enhances temporal correlations as
the collective interaction approaches the instability threshold
\cite{hohenberg1977}. In systems with broadly distributed relaxation
rates, however, temporal persistence is additionally controlled by
the infrared organization of the projected spectrum. This provides
a spectral perspective on the emergence of long time scales without
fine tuning to a collective instability \cite{chen2024}. The different
infrared sensitivities of $M_1$ and $M_2$ allow long-lived collective
fluctuations to develop without approaching the collective ordering
threshold. In particular, the regime $0<\beta\leq1$ illustrates that
a finite ordering threshold and a finite static collective
susceptibility can coexist with a divergent integrated correlation
time in the gapless continuum limit. The same distinction can persist
within a stable ordered phase, provided that the relaxation spectrum
about the ordered configuration retains the appropriate infrared
structure. Conversely, infrared accumulation alone does not establish
spontaneous collective order, which requires an appropriate collective
interaction and nonlinear stabilization.

The nonlinear analysis also distinguishes the emergence of collective
order from the locality of its effective dynamics. Although the
static Landau expansion identifies the conditions for continuous
ordering, the corresponding dynamical vertices generally retain
temporal nonlocality. A local nonlinear potential therefore does not,
by itself, justify a local-in-time collective field equation. The
exact nonlinear construction considered here demonstrates that local
cubic stabilization and nonlocal linear memory can coexist within a
single microscopic model. Its exact reducibility relies on the
alignment of the quartic interaction with the collective channel,
while more general nonlinear interactions may generate nonlinear
memory functionals and state-dependent fluctuating forces.

The controlled numerical constructions emphasize the importance of
distinguishing the density of microscopic relaxation rates from their
coupling to a specified collective observable. Similar collective
responses can arise from different microscopic eigenvalue distributions
when the corresponding response-weighted spectral measures are
equivalent. This observation suggests that the projected spectral
measure, rather than the unweighted density alone, provides a natural
macroscopic descriptor of collective relaxation dynamics within the
present framework.

The Transformer measurements provide a complementary perspective on
how infrared spectral organization may emerge through learning rather
than being prescribed at the microscopic level. The persistence of
near-marginal infrared scaling across different stages of learning,
together with the transient maximum of the infrared spectral
estimator, suggests that learning may organize a broad spectrum of
internal relaxation modes in a manner favorable to collective
response and temporal persistence. However, the measured Jacobian
spectra characterize computational propagation and do not
independently determine the physical collective-channel susceptibility
or its irreducible memory self-energy. Establishing the proposed
connection requires direct perturbation-response measurements,
identification of the relevant collective-mode overlaps, and
controlled variation of collective feedback.

While the present spectral analysis focuses on symmetric microscopic
relaxation operators, the dynamical HS formulation extends to
nonequilibrium reference dynamics. For diagonalizable non-normal
operators, the collective response involves biorthogonal left and
right eigenvectors and potentially complex relaxation spectra, with
spectral weights that need not be positive. Such extensions provide
a framework for investigating circulatory dynamics, oscillatory
collective instabilities, and non-normal transient amplification.
The eliminated internal degrees of freedom can also mediate the
transfer of external inputs to the collective coordinate, providing
a complementary route through which the internal relaxation
structure shapes the driven collective response. In addition, outside
equilibrium, the fluctuation spectrum cannot generally be inferred
from the retarded response alone and requires an independent
characterization of the effective noise kernel.

These considerations motivate further investigation of how learning,
adaptation, and nonreciprocal interactions organize the
collective-channel relaxation spectrum. Whether the spectral
mechanisms identified here govern collective ordering and temporal
persistence in biological neural systems remains an open empirical
question.

\section{Conclusion}
\label{sec:conclusion}

We have developed a microscopic field-theoretic framework for the
emergence of memory-dressed collective dynamics in stochastic neural
systems. The dynamical Hubbard--Stratonovich formulation and the
Mori--Zwanzig projection establish a common microscopic basis for
collective ordering and memory feedback. Within this framework,
the static memory self-energy enters the collective instability
condition, linking the organization of internal relaxation modes
to the emergence of macroscopic collective order.

The spectral analysis establishes an exact relation between the
response-weighted relaxation spectrum and the irreducible memory
spectrum. In the flat gapless continuum, an integrable memory kernel
with $K_\Sigma(t)\sim[t\ln^2(\Lambda t)]^{-1}$ generates, through
repeated memory feedback, a nonintegrable $1/t$ reference response.
The equilibrium fluctuation analysis further demonstrates that
collective ordering and temporal persistence are governed by
distinct inverse-rate spectral moments, allowing a divergent
integrated correlation time even when the static collective
susceptibility remains finite.

Beyond the Gaussian instability, an explicitly constructed nonlinear
microscopic model yields an exact collective field equation combining
local cubic stabilization, nonlocal memory feedback, external
driving, and effective fluctuating forces. The controlled numerical
constructions confirm the spectral dependence of the collective
response and ordering threshold, while the Transformer measurements
reveal learning-induced near-marginal infrared organization and
transient spectral enhancement. Although the latter do not directly
establish collective ordering, they motivate further investigation
of the proposed mechanisms in learned neural systems.

The present theory establishes a microscopic connection between
collective interactions, infrared spectral organization, memory
feedback, and temporal persistence. The organization of internal
relaxation modes thus emerges as a physical mechanism governing
both the formation and the dynamics of macroscopic collective
fields in neural systems and, more generally, in high-dimensional
dynamical systems with distributed relaxation modes.

\medskip
\emph{Acknowledgements}---This work was partially supported by the Institute of Information \& Communications Technology Planning \& Evaluation (IITP) grant 
funded by the Korea government (MSIT) (IITP-RS-2025-02214780).


\clearpage
\onecolumngrid

\setcounter{section}{0}
\setcounter{subsection}{0}
\setcounter{equation}{0}
\setcounter{figure}{0}
\setcounter{table}{0}

\renewcommand{\thesection}{S\arabic{section}}

\renewcommand{\thesubsection}{\thesection.\arabic{subsection}}

\numberwithin{equation}{section}
\renewcommand{\theequation}{S\arabic{section}.\arabic{equation}}

\renewcommand{\thefigure}{S\arabic{figure}}
\renewcommand{\thetable}{S\arabic{table}}

\setcounter{secnumdepth}{2}

\begin{center}
{\large\bfseries
Supplementary Materials
}
\end{center}

\section{Spectral reconstruction of the irreducible memory self-energy}
\label{app:spectral_reconstruction}

We provide the derivations connecting the collective response spectrum
of the reference system to the irreducible memory spectrum. Throughout
this section, $L_0$ is a real symmetric positive-definite relaxation
operator obtained by linearizing the unit-mobility gradient dynamics
about a stable reference configuration. The gapless results below are
understood as limits of such stable systems.

\subsection{Exact inverse-response representation}

Let $L_0\bm v_\alpha=\lambda_\alpha\bm v_\alpha$, with orthonormal
eigenvectors. For a nonzero collective direction $\bm c$, the response
weights and their normalization are
\begin{equation}
  A_\alpha=\frac{|\bm c^{\mathsf T}\bm v_\alpha|^2}{N},
  \qquad
  A_c=\sum_\alpha A_\alpha=\frac{\|\bm c\|^2}{N}.
\end{equation}
The weighted response spectrum and its Stieltjes transform are
\begin{align}
  \rho_O(\lambda)
  &=\sum_\alpha A_\alpha\delta(\lambda-\lambda_\alpha),\\
  \chi_0(z)
  &=\frac{1}{N}\bm c^{\mathsf T}(zI+L_0)^{-1}\bm c
  =\int_0^\infty\frac{\rho_O(\lambda)}{z+\lambda}\,\dd\lambda.
  \label{eq:sr_pi}
\end{align}
The retarded response is obtained at $z=0^+-i\omega$. The unweighted
eigenvalue density does not in general determine this projected response,
since the spectral weights depend on the selected collective direction.

Introduce $\bm e=\bm c/\|\bm c\|$, $P=\bm e\bm e^{\mathsf T}$, and
$Q=I-P$. Decomposing the microscopic perturbation as
$\delta\bm x=\bm eX+\bm y$, with $X=\bm e^{\mathsf T}\delta\bm x$
and $\bm y=Q\delta\bm x$, gives the block representation
\begin{equation}
  L_0=
  \begin{pmatrix}
    r_0&-\bm b^{\mathsf T}\\
    -\bm b&L_Q
  \end{pmatrix},
  \qquad
  r_0=\bm e^{\mathsf T}L_0\bm e,\quad
  \bm b=-QL_0\bm e,
\end{equation}
where $L_Q=QL_0Q|_{\operatorname{Ran}Q}$. For a probe $\delta\phi$
coupled along $\bm c$, the Laplace-transformed equations with vanishing
initial perturbations are
\begin{align}
  (z+r_0)X(z)
  &=\bm b^{\mathsf T}\bm y(z)+\|\bm c\|\delta\phi(z),\\
  (zI_Q+L_Q)\bm y(z)&=\bm bX(z).
\end{align}
Eliminating $\bm y$ yields
$[z+r_0-\Sigma(z)]X(z)=\|\bm c\|\delta\phi(z)$, where the
irreducible memory self-energy is
\begin{equation}
  \Sigma(z)=\bm b^{\mathsf T}(zI_Q+L_Q)^{-1}\bm b.
  \label{eq:sr_schur}
\end{equation}
For orthonormal eigenvectors satisfying
$L_Q\bm u_\alpha=\mu_\alpha\bm u_\alpha$, define
$g_\alpha=\bm u_\alpha^{\mathsf T}\bm b$. The corresponding
memory spectrum and self-energy are
\begin{align}
  \rho_\Sigma(\mu)
  &=\sum_\alpha g_\alpha^2\delta(\mu-\mu_\alpha),\\
  \Sigma(z)
  &=\int_0^\infty\frac{\rho_\Sigma(\mu)}{z+\mu}\,\dd\mu.
\end{align}
Since the collective observable satisfies $\delta m=\|\bm c\|X/N$,
the reference response and the irreducible memory self-energy obey
\begin{equation}
  \chi_0(z)=\frac{A_c}{z+r_0-\Sigma(z)},
  \qquad
  \Sigma(z)=z+r_0-\frac{A_c}{\chi_0(z)}.
  \label{eq:sr_inverse}
\end{equation}
This Schur-complement identity provides the common starting point for
both discrete and continuous spectral reconstruction. The two spectra
refer to different operators, $L_0$ and $L_Q$, and carry different
spectral weights.

\subsection{Transformation of continuous spectral densities}

For a Stieltjes transform
$F(z)=\int_0^\infty\rho(\lambda')/(z+\lambda')\,\dd\lambda'$,
the relaxation-rate spectrum is extracted on the negative real $z$ axis,
rather than on the real-frequency axis. Using the Sokhotski--Plemelj
identity
\begin{equation}
  \frac{1}{x+i0^+}
  =\mathcal P\frac{1}{x}-i\pi\delta(x),
\end{equation}
the boundary value of the reference response becomes
\begin{align}
  P_O(\lambda)
  &=\mathcal P\int_0^\infty
  \frac{\rho_O(\lambda')}{\lambda'-\lambda}\,\dd\lambda',\\
  \chi_0(-\lambda+i0^+)
  &=P_O(\lambda)-i\pi\rho_O(\lambda).
\end{align}
Inserting this boundary value into Eq.~\eqref{eq:sr_inverse}
and taking the imaginary part gives
\begin{equation}
  \operatorname{Im}\Sigma(-\lambda+i0^+)
  =-\frac{A_c\pi\rho_O(\lambda)}
  {P_O^2(\lambda)+\pi^2\rho_O^2(\lambda)}.
\end{equation}
Using Stieltjes inversion,
$\rho_\Sigma(\lambda)=-\pi^{-1}
\operatorname{Im}\Sigma(-\lambda+i0^+)$, the absolutely continuous
memory density is therefore
\begin{equation}
  \rho_\Sigma(\lambda)
  =\frac{A_c\rho_O(\lambda)}
  {P_O^2(\lambda)+\pi^2\rho_O^2(\lambda)}
  =\frac{A_c\rho_O(\lambda)}
  {|\chi_0(-\lambda+i0^+)|^2}.
  \label{eq:sr_density}
\end{equation}
This density is nonnegative in the reciprocal Gaussian system.
Equation~\eqref{eq:sr_density} applies wherever the boundary values
are regular and the denominator is nonzero; it is not a pointwise
formula for delta-function measures. Isolated zeros of $\chi_0$ in
spectral gaps can generate additional atomic contributions, while
more general singular spectral components require the full Stieltjes
inversion of Eq.~\eqref{eq:sr_inverse}.

\subsection{Reconstruction of discrete memory modes}

For a finite system, combine degenerate response poles and omit
zero-weight modes. Let $n$ distinct visible rates remain, with
$A_\beta>0$. According to Eq.~\eqref{eq:sr_inverse}, the poles of
the irreducible self-energy occur at the zeros of the response,
\begin{equation}
  \chi_0(-\mu_\alpha)
  =\sum_{\beta=1}^{n}
  \frac{A_\beta}{\lambda_\beta-\mu_\alpha}=0.
  \label{eq:sr_zeros}
\end{equation}
Near a simple zero,
$\chi_0(z)=\chi_0'(-\mu_\alpha)(z+\mu_\alpha)
+O((z+\mu_\alpha)^2)$. Comparing the singular part of
Eq.~\eqref{eq:sr_inverse} with $g_\alpha^2/(z+\mu_\alpha)$
determines the memory weight,
\begin{equation}
  g_\alpha^2
  =-\frac{A_c}{\chi_0'(-\mu_\alpha)}
  =\frac{A_c}
  {\displaystyle\sum_\beta
  A_\beta/(\lambda_\beta-\mu_\alpha)^2}.
  \label{eq:sr_residue}
\end{equation}
The second equality follows from
$\chi_0'(z)=-\sum_\beta A_\beta/(z+\lambda_\beta)^2$,
which also establishes the positivity of the reconstructed weights.

For ordered visible rates $0<\lambda_1<\cdots<\lambda_n$, the
function $f(\mu)=\chi_0(-\mu)$ is strictly increasing between
successive poles, since
$f'(\mu)=\sum_\beta A_\beta/(\lambda_\beta-\mu)^2>0$.
It tends to $-\infty$ at the left endpoint and $+\infty$ at the
right endpoint of each interval. There is therefore exactly one
zero per interval, yielding the interlacing relation
\begin{equation}
  \lambda_1<\mu_1<\lambda_2<\cdots<\mu_{n-1}<\lambda_n.
\end{equation}
This is consistent with Cauchy interlacing for the compressed
operator. When all $N$ eigenvalues are distinct and visible, $n=N$.
Otherwise, the reconstruction determines only the memory modes
coupled to the collective direction, not the entire orthogonal
subspace. Equations~\eqref{eq:sr_zeros} and \eqref{eq:sr_residue}
thus provide a finite-dimensional reconstruction without explicitly
diagonalizing $L_Q$.

As an elementary example, take
$L_0=\operatorname{diag}(\lambda_1,\lambda_2)$,
$\bm e=(1,1)^{\mathsf T}/\sqrt{2}$, and $A_c=1$.
With $r_0=(\lambda_1+\lambda_2)/2$, the reference response and
the reconstructed self-energy are
\begin{align}
  \chi_0(z)
  &=\frac12\left(\frac{1}{z+\lambda_1}
  +\frac{1}{z+\lambda_2}\right)
  =\frac{z+r_0}{(z+\lambda_1)(z+\lambda_2)},\\
  \Sigma(z)
  &=\frac{(\lambda_2-\lambda_1)^2}{4(z+r_0)}.
\end{align}
The corresponding spectra are
\begin{align}
  \rho_O(\lambda)
  &=\tfrac12[\delta(\lambda-\lambda_1)
  +\delta(\lambda-\lambda_2)],\\
  \rho_\Sigma(\lambda)
  &=\tfrac14(\lambda_2-\lambda_1)^2\delta(\lambda-r_0).
\end{align}
For $\lambda_1=0.01$ and $\lambda_2=1$, the memory rate is
$\mu=0.505$, its weight is $g^2=0.245025$, and
$\Sigma(0)\simeq0.485198$. The static restoring coefficient
$r_0-\Sigma(0)\simeq0.019802$ is neither the smallest reference
rate nor the memory rate, illustrating the distinction between
the static inverse-response scale and the spectral relaxation gap.

\subsection{Spectral moment sum rules}

Define the nonnegative-order spectral moments
$B_n=\int_0^\infty\lambda^n\rho_O(\lambda)\,\dd\lambda$,
distinct from the inverse-rate moments used for temporal persistence
in the main text. For a bounded spectrum, the large-$z$ expansion
of the reference response gives
\begin{align}
  \chi_0(z)
  &=\frac{B_0}{z}-\frac{B_1}{z^2}
  +\frac{B_2}{z^3}+O(z^{-4}),\\
  B_0&=A_c,\qquad
  B_1=\frac{\bm c^{\mathsf T}L_0\bm c}{N}=A_cr_0.
\end{align}
Expanding the inverse response and using Eq.~\eqref{eq:sr_inverse}
then yields
\begin{align}
  \frac{A_c}{\chi_0(z)}
  &=z+r_0+\frac{r_0^2-B_2/A_c}{z}+O(z^{-2}),\\
  \Sigma(z)
  &=\frac{B_2/A_c-r_0^2}{z}+O(z^{-2}).
\end{align}
Comparison with the large-$z$ expansion of the memory Stieltjes
transform gives the spectral sum rule
\begin{equation}
  \int_0^\infty\rho_\Sigma(\lambda)\,\dd\lambda
  =\frac{B_2}{A_c}
  -\left(\frac{B_1}{A_c}\right)^2.
  \label{eq:sr_sumrule}
\end{equation}
With $p_O=\rho_O/A_c$ and averages taken with respect to $p_O$,
the right-hand side is the weighted spectral variance
$\operatorname{Var}_O(\lambda)$.

The same result follows independently from the microscopic projection,
\begin{align}
  \|\bm b\|^2
  =\bm e^{\mathsf T}L_0QL_0\bm e=\bm e^{\mathsf T}L_0^2\bm e
  -(\bm e^{\mathsf T}L_0\bm e)^2
  =\langle\lambda^2\rangle_O-\langle\lambda\rangle_O^2.
\end{align}
Completeness in the orthogonal subspace gives
$\sum_\alpha g_\alpha^2=\|\bm b\|^2$, so that
\begin{equation}
  K_\Sigma(0^+)
  =\int_0^\infty\rho_\Sigma(\lambda)\,\dd\lambda
  =\|\bm b\|^2
  =\operatorname{Var}_O(\lambda).
\end{equation}
If $\bm e$ is an eigenvector of $L_0$, then $\bm b=0$ and the
response spectrum is concentrated at a single rate. The irreducible
memory therefore vanishes identically, as in the simplest Curie--Weiss
and Mattis-type linearizations. Collective ordering need not imply
memory dressing. More generally, the spectral variance determines
the initial memory weight, whereas long-time persistence depends on
the infrared distribution.

\subsection{Flat infrared response spectrum}

Consider the exactly flat weighted response density
$\rho_O(\lambda)=A$ on $\lambda_{\rm IR}<\lambda<\Lambda$,
and zero outside. Its normalization is
$A_c=A(\Lambda-\lambda_{\rm IR})$, and direct integration gives
\begin{equation}
  \chi_0(z)=A\ln\frac{z+\Lambda}{z+\lambda_{\rm IR}},
  \qquad
  \chi_0(0)=A\ln\frac{\Lambda}{\lambda_{\rm IR}}.
\end{equation}
The logarithm is continued analytically from positive real $z$.
Within the spectral interval, the principal-value integral is
\begin{equation}
  P_O(\lambda)
  =A\mathcal P\int_{\lambda_{\rm IR}}^\Lambda
  \frac{\dd\lambda'}{\lambda'-\lambda}
  =A\ln\frac{\Lambda-\lambda}{\lambda-\lambda_{\rm IR}}.
\end{equation}
Substitution into Eq.~\eqref{eq:sr_density} yields the exact
irreducible memory density
\begin{equation}
  \rho_\Sigma(\lambda)
  =\frac{A_c/A}
  {\ln^2[(\Lambda-\lambda)/(\lambda-\lambda_{\rm IR})]+\pi^2}.
  \label{eq:sr_flat}
\end{equation}
There are no additional poles outside this single interval, since
the reference response has a definite sign on either exterior real
interval. In the gapless continuum limit, with
$\lambda_{\rm IR}\to0^+$ and finite positive $A$ and $\Lambda$,
the deep infrared behavior becomes
\begin{equation}
  \rho_\Sigma(\lambda)
  \sim\frac{A_c}{A\ln^2(\Lambda/\lambda)},
  \qquad \lambda\ll\Lambda.
\end{equation}

The exact reference response kernel for $t>0$ is
\begin{equation}
  K_O(t)
  =A\int_{\lambda_{\rm IR}}^\Lambda
  e^{-\lambda t}\,\dd\lambda
  =\frac{A}{t}
  (e^{-\lambda_{\rm IR}t}-e^{-\Lambda t}),
\end{equation}
which gives $K_O(t)\sim A/t$ in the intermediate regime
$\Lambda^{-1}\ll t\ll\lambda_{\rm IR}^{-1}$. In contrast, the
irreducible memory kernel is determined by
$K_\Sigma(t)=\int_0^\infty\rho_\Sigma(\lambda)
e^{-\lambda t}\,\dd\lambda$. To obtain its gapless long-time
asymptotic behavior, restrict the infrared approximation to
$0<\lambda<\lambda_*\ll\Lambda$, since the remaining contribution
is exponentially suppressed. Setting $u=\lambda t$ gives
\begin{align}
  K_\Sigma(t)
  &\sim\frac{A_c}{At}
  \int_0^{\lambda_*t}
  \frac{e^{-u}\,\dd u}
  {[\ln(\Lambda t)-\ln u]^2}\nonumber\\
  &\sim\frac{A_c}{At\ln^2(\Lambda t)}
  \int_0^\infty e^{-u}\,\dd u
  =\frac{A_c}{At\ln^2(\Lambda t)}.
\end{align}
The last step follows from the slow variation of the logarithmic
factor, with the exponentially suppressed large-$u$ region making
no contribution to the leading asymptotic term. At fixed positive
infrared cutoff, the final decay is instead exponentially suppressed.
The gapless limit must therefore precede the long-time asymptotic limit.

The logarithmic correction changes the integrability of the kernel.
In the gapless limit, the long-time contributions satisfy
\begin{align}
  \int_{t_0}^T K_O(t)\,\dd t
  &\sim A\ln(T/t_0),\\
  \int_{t_0}^\infty K_\Sigma(t)\,\dd t
  &\sim\frac{A_c}{A}
  \int_{\ln(\Lambda t_0)}^\infty\frac{\dd s}{s^2}
  =\frac{A_c}{A\ln(\Lambda t_0)}.
\end{align}
The second equivalence holds asymptotically for large $t_0$.
Both first time moments diverge in the gapless limit, although
only the reference response kernel has a divergent zeroth moment.

The static self-energy follows independently from
Eq.~\eqref{eq:sr_inverse},
\begin{equation}
  \Sigma(0)
  =r_0-\frac{A_c}{A\ln(\Lambda/\lambda_{\rm IR})},
  \qquad
  r_0=\frac{\Lambda+\lambda_{\rm IR}}{2}.
\end{equation}
As the cutoff vanishes, $\Sigma(0)\to r_0$, while $\chi_0(0)$
diverges. Including the rank-one collective coupling $J$, the
corresponding instability threshold is
\begin{equation}
  J_c=\frac{r_0-\Sigma(0)}{A_c}
  =\frac{1}{A\ln(\Lambda/\lambda_{\rm IR})}.
\end{equation}
The divergence of the reference response thus results from the
cancellation of finite restoring terms, not from a divergent
irreducible self-energy. At fixed positive $J$, the disordered
state cannot remain stable all the way to this flat gapless limit.
The static restoring coefficient is not in general the spectral gap.

Finally, the flat spectrum provides an exact check on
Eq.~\eqref{eq:sr_flat}. Its normalized first two moments are
$(\Lambda+\lambda_{\rm IR})/2$ and
$(\Lambda^2+\Lambda\lambda_{\rm IR}+\lambda_{\rm IR}^2)/3$,
giving the variance $(\Lambda-\lambda_{\rm IR})^2/12$.
Equation~\eqref{eq:sr_sumrule} therefore requires
\begin{equation}
  \int_{\lambda_{\rm IR}}^\Lambda
  \rho_\Sigma(\lambda)\,\dd\lambda
  =\frac{(\Lambda-\lambda_{\rm IR})^2}{12}.
\end{equation}
With the substitution
$x=(\lambda-\lambda_{\rm IR})/(\Lambda-\lambda_{\rm IR})$,
this reduces to the dimensionless identity
\begin{equation}
  \int_0^1
  \frac{\dd x}{\ln^2[(1-x)/x]+\pi^2}
  =\frac{1}{12}.
\end{equation}
These results distinguish the collective reference response from
the irreducible memory spectrum even when the former is exactly flat.
The numerical constructions in Sec.~II address the reference response
spectrum rather than directly prescribing the irreducible memory
density. Extension to nonnormal or nonreciprocal dynamics requires
left and right eigenvector overlaps and, in general, complex
spectral weights.

\subsection{Memory-mediated transfer of external inputs}
\label{app:input_transfer}

We now extend the projected dynamics of Sec.~I.A to a general external
input, distinguishing the memory-mediated transfer of external
information from the intrinsic memory feedback generated by internal
relaxation modes. Consider the linearized microscopic dynamics
\begin{equation}
  \dot{\bm x}(t)
  =-L_0\bm x(t)+JA_cP\bm x(t)
  +B\bm I(t)+\bm\eta(t),
  \label{eq:supp_input_micro}
\end{equation}
where $\bm x$ denotes the microscopic perturbation about the reference
configuration, $\bm I(t)$ is a $d_I$-dimensional external input, and
$B\in\mathbb{R}^{N\times d_I}$ specifies its microscopic coupling.
Using the collective and orthogonal coordinates defined in Sec.~I.A,
the projected equations become
\begin{align}
  \dot X(t)
  &=-(r_0-JA_c)X(t)+\bm b^{\mathsf T}\bm y(t)
 +\bm e^{\mathsf T}B\bm I(t)+\eta_X(t),
  \label{eq:supp_input_collective}\\
  \dot{\bm y}(t)
  &=-L_Q\bm y(t)+\bm bX(t)
  +QB\bm I(t)+\bm\eta_Q(t),
  \label{eq:supp_input_internal}
\end{align}
where $\eta_X=\bm e^{\mathsf T}\bm\eta$ and
$\bm\eta_Q=Q\bm\eta$. The external input acts on the collective
coordinate through two distinct pathways: a direct coupling
$\bm e^{\mathsf T}B$ and an indirect coupling through the orthogonal
internal degrees of freedom.

The formal solution of Eq.~\eqref{eq:supp_input_internal} is
\begin{align}
  \bm y(t)
  =e^{-L_Qt}\bm y(0)
  +\int_0^t ds\,e^{-L_Q(t-s)}
  [\bm bX(s)+QB\bm I(s)+\bm\eta_Q(s)].
  \label{eq:supp_input_internal_solution}
\end{align}
Substituting this solution into Eq.~\eqref{eq:supp_input_collective}
yields the exact projected equation
\begin{align}
  \dot X(t)
  &=-(r_0-JA_c)X(t)
  +\int_0^t ds\,K_\Sigma(t-s)X(s)\nonumber\\
  &\quad+\bm e^{\mathsf T}B\bm I(t)
  +\int_0^t ds\,K_I(t-s)\bm I(s)
  +\eta_{\rm eff}(t),
  \label{eq:supp_input_projected}
\end{align}
where the intrinsic memory kernel and the memory-mediated
input-transfer kernel are respectively given by
\begin{align}
  K_\Sigma(t)
  &=\Theta(t)\bm b^{\mathsf T}e^{-L_Qt}\bm b,
  \label{eq:supp_input_intrinsic_kernel}\\
  K_I(t)
  &=\Theta(t)\bm b^{\mathsf T}e^{-L_Qt}QB,
  \label{eq:supp_input_transfer_kernel}
\end{align}
and the effective fluctuating force is
\begin{align}
  \eta_{\rm eff}(t)=\eta_X(t)+\bm b^{\mathsf T}e^{-L_Qt}\bm y(0)
  +\int_0^t ds\,
  \bm b^{\mathsf T}e^{-L_Q(t-s)}\bm\eta_Q(s).
\end{align}
The two kernels originate from the same orthogonal relaxation operator
$L_Q$ but describe physically distinct processes. The intrinsic memory
kernel $K_\Sigma$ describes the excitation of internal modes by the
collective coordinate and their subsequent feedback into that coordinate.
In contrast, $K_I$ describes the excitation of internal modes by an
external input and the subsequent transfer of their relaxation dynamics
into the collective coordinate.

For the retarded response, setting the initial perturbations to zero
and omitting the fluctuating force, Laplace transformation of
Eq.~\eqref{eq:supp_input_projected} gives
\begin{equation}
  [z+r_0-JA_c-\Sigma(z)]X(z)
  =T_I(z)\bm I(z),
  \label{eq:supp_input_laplace}
\end{equation}
where the intrinsic self-energy and the input-transfer function are
\begin{align}
  \Sigma(z)
  &=\bm b^{\mathsf T}(zI_Q+L_Q)^{-1}\bm b,
  \label{eq:supp_input_sigma}\\
  T_I(z)
  &=\bm e^{\mathsf T}B
  +\bm b^{\mathsf T}(zI_Q+L_Q)^{-1}QB.
  \label{eq:supp_input_transfer_function}
\end{align}
The collective response to a general external input is therefore
\begin{equation}
  X(z)=\frac{T_I(z)\bm I(z)}
  {z+r_0-JA_c-\Sigma(z)}.
  \label{eq:supp_input_dressed_response}
\end{equation}
The denominator describes the intrinsic memory-dressed collective
dynamics, whereas the numerator incorporates the direct and
memory-mediated transfer of the external input.

To elucidate the spectral structure of the input-transfer function,
use the orthonormal eigenvectors of $L_Q$ introduced in Sec.~I.A,
satisfying $L_Q\bm u_\alpha=\mu_\alpha\bm u_\alpha$. The
mode-dependent coupling coefficients are
\begin{equation}
  g_\alpha=\bm u_\alpha^{\mathsf T}\bm b,
  \qquad
  \bm q_\alpha=\bm u_\alpha^{\mathsf T}QB,
  \label{eq:supp_input_mode_couplings}
\end{equation}
where $\bm q_\alpha$ is a row vector in the external input space.
The amplitude $y_\alpha=\bm u_\alpha^{\mathsf T}\bm y$ of each
internal mode obeys
\begin{equation}
  \dot y_\alpha(t)
  =-\mu_\alpha y_\alpha(t)+g_\alpha X(t)
  +\bm q_\alpha\bm I(t)+\eta_\alpha(t),
  \label{eq:supp_input_mode_dynamics}
\end{equation}
with $\eta_\alpha=\bm u_\alpha^{\mathsf T}\bm\eta_Q$.
Consequently, the intrinsic memory self-energy and the input-transfer
function admit the spectral representations
\begin{align}
  \Sigma(z)
  &=\sum_\alpha\frac{g_\alpha^2}{z+\mu_\alpha},
  \label{eq:supp_input_sigma_spectral}\\
  T_I(z)
  &=\bm e^{\mathsf T}B
  +\sum_\alpha\frac{g_\alpha\bm q_\alpha}{z+\mu_\alpha}.
  \label{eq:supp_input_transfer_spectral}
\end{align}
In analogy with the irreducible memory spectral density
$\rho_\Sigma(\lambda)=\sum_\alpha
g_\alpha^2\delta(\lambda-\mu_\alpha)$, we introduce the
input-weighted spectral density
\begin{equation}
  \bm\rho_I(\lambda)
  =\sum_\alpha g_\alpha\bm q_\alpha
  \delta(\lambda-\mu_\alpha).
  \label{eq:supp_input_spectral_density}
\end{equation}
The memory-mediated input-transfer kernel and its Laplace
representation can then be expressed as
\begin{align}
  K_I(t)
  &=\Theta(t)\int_0^\infty
  \bm\rho_I(\lambda)e^{-\lambda t}\,\dd\lambda,
  \label{eq:supp_input_kernel_integral}\\
  T_I(z)
  &=\bm e^{\mathsf T}B
  +\int_0^\infty\frac{\bm\rho_I(\lambda)}
  {z+\lambda}\,\dd\lambda.
  \label{eq:supp_input_transfer_integral}
\end{align}
Unlike $\rho_\Sigma(\lambda)$, which is a nonnegative spectral
measure for the symmetric reference operator considered here,
$\bm\rho_I(\lambda)$ is generally vector-valued and need not be
componentwise nonnegative. Its spectral weights depend on both the
coupling of the internal modes to the collective coordinate and
their overlaps with the external input channels.

The collective source considered in the main text is recovered by
taking $B=\bm c$ and $\bm I(t)=h(t)$. Since $Q\bm c=0$ and
$\bm e^{\mathsf T}\bm c=\|\bm c\|$, one obtains $K_I(t)=0$
and $T_I(z)=\|\bm c\|$. Using $m=\|\bm c\|X/N$, the corresponding
physical collective susceptibility becomes
\begin{equation}
  \chi_m(z)=\frac{A_c}{z+r_0-JA_c-\Sigma(z)},
\end{equation}
in agreement with the result derived in the main text.

More generally, the intrinsic memory self-energy is determined by
the reference relaxation operator and the selected collective channel,
whereas the input-transfer function additionally depends on the
microscopic structure of the external coupling $B$. Thus, at fixed
linear reference dynamics, changing the external input channel
modifies the excitation and transfer of internal relaxation modes
without modifying the intrinsic memory self-energy. The resulting
collective response reflects both the memory-dressed collective
dynamics and the selective activation of internal relaxation modes
by external inputs.

\section{Numerical construction and spectral diagnostics}
\label{app:numerics}
\subsection{Numerical setup and spectral target}

We numerically construct finite-dimensional symmetric relaxation operators to
examine how microscopic relaxation eigenvalues and their overlaps with a
prescribed collective channel determine the collective spectral measure and
the associated static response. The calculation uses $N=2048$ degrees of
freedom and the normalized target density
\begin{equation}
  p_\beta(\lambda)=\frac{\lambda^\beta\exp[-(\lambda/\lambda_{\rm UV})^2]}
  {\displaystyle\int_{\lambda_{\rm IR}}^{\lambda_{\max}}ds\,s^\beta\exp[-(s/\lambda_{\rm UV})^2]},\qquad
  \lambda_{\rm IR}\leq\lambda\leq\lambda_{\max},
\label{eq:target}
\end{equation}
where $\lambda_{\rm IR}=10^{-5}$, $\lambda_{\max}=10$, $\lambda_{\rm UV}=0.5$,
and $\beta\in\{-0.5,-0.25,-0.1,0,0.25,0.5\}$. The cumulative distribution
$F_\beta(\lambda)=\int_{\lambda_{\rm IR}}^\lambda p_\beta(s)\,ds$ is evaluated
on a logarithmic grid of 30,001 points using trapezoidal integration. All
matrix calculations are performed in double precision. A pseudorandom Gaussian
$N\times N$ matrix is orthogonalized by QR factorization to obtain a real
orthogonal matrix $U$, with pseudorandom seed 20261110. The same $U$ is used
throughout the A/B comparison.

For each construction we define
\begin{equation}
  L_0=U\,\mathrm{diag}(\lambda_1,\ldots,\lambda_N)U^{\mathsf T},\qquad
  \bm c=\sqrt N\,U\begin{pmatrix}\sqrt{w_1}&\cdots&\sqrt{w_N}\end{pmatrix}^{\mathsf T},
\label{eq:construction}
\end{equation}
where $w_\alpha\geq0$ and $\sum_\alpha w_\alpha=1$. The collective response
and its spectral weights are defined by
\begin{equation}
  \chi_0^R(\omega)=\frac1N\bm c^{\mathsf T}(L_0-i\omega I)^{-1}\bm c
  =\sum_{\alpha=1}^N\frac{A_\alpha}{\lambda_\alpha-i\omega},\qquad
  A_\alpha=\frac1N\left|\bm c^{\mathsf T}\bm v_\alpha\right|^2,
\label{eq:polarization}
\end{equation}
with $L_0\bm v_\alpha=\lambda_\alpha\bm v_\alpha$. The unweighted and
collective-weighted spectral measures are, respectively,
\begin{equation}
  \rho_{\rm raw}(\lambda)=\frac1N\sum_\alpha\delta(\lambda-\lambda_\alpha),\qquad
  \rho_O(\lambda)=\sum_\alpha A_\alpha\delta(\lambda-\lambda_\alpha).
\label{eq:measures}
\end{equation}

These two measures are generally distinct. The construction in
Eq.~\eqref{eq:construction} is chosen so that the prescribed $w_\alpha$ are
exactly realizable as eigenvector-overlap weights, up to floating-point error;
it is not a simulation of spontaneous spectral organization.

\subsection{Two complementary microscopic constructions}

\subsubsection{A: fixed logarithmic eigenvalue grid with variable collective weights}

In the A\_weighted construction, $N+1$ bin edges are logarithmically spaced
between $\lambda_{\rm IR}$ and $\lambda_{\max}$. The eigenvalues are placed at
the geometric centers of these bins,
\begin{equation}
  \lambda_\alpha^{(A)}=\sqrt{e_\alpha e_{\alpha+1}},\qquad
  w_\alpha^{(A)}=F_\beta(e_{\alpha+1})-F_\beta(e_\alpha),
\label{eq:A}
\end{equation}
where $e_\alpha$ denotes the bin edges. Consequently, the eigenvalues and the
matrix $L_0^{(A)}$ remain unchanged as $\beta$ varies, whereas the collective
coupling vector $\bm c^{(A)}$ changes with $\beta$. In the infrared regime,
$w_\alpha^{(A)}\simeq
p_\beta(\lambda_\alpha)\Delta\lambda_\alpha\propto\lambda_\alpha^{\beta+1}$
because $\Delta\lambda_\alpha\propto\lambda_\alpha$ on a logarithmic grid.
Thus, the individual weights need not have the same power-law exponent as the
density per unit $\lambda$.

\subsubsection{B: target-distributed eigenvalues with uniform collective weights}

In the B\_eigenvalue construction, the eigenvalues are placed at the midpoint
quantiles of the target distribution,
\begin{equation}
  \lambda_\alpha^{(B)}=F_\beta^{-1}\!\left(\frac{\alpha-\tfrac12}{N}\right),\qquad
  w_\alpha^{(B)}=\frac1N,\qquad \alpha=1,\ldots,N.
\label{eq:B}
\end{equation}
The matrix $L_0^{(B)}$ therefore varies with $\beta$, while the collective
coupling vector is fixed because $U$ and the weights are unchanged. For
$\beta=0$ and in the absence of the ultraviolet envelope, the quantile grid is
linearly spaced in $\lambda$; in general it is neither linearly nor
logarithmically spaced. This construction has $\rho_O^{(B)}=\rho_{\rm
raw}^{(B)}$ as discrete measures.

Both A and B approximate the same target collective spectral measure but use
different finite-$N$ discretizations. In particular, the number of eigenvalues
in a given infrared interval is not matched between the constructions;
agreement in integrated observables must therefore be distinguished from
pointwise agreement of binned spectral densities.

\subsection{Direct diagonalization and overlap validation}

Each $L_0$ is explicitly assembled and independently diagonalized using the
symmetric eigensolver \texttt{scipy.linalg.eigh}. Let $\lambda_\alpha^{\rm
diag}$ and $\bm v_\alpha^{\rm diag}$ denote the recovered eigenpairs, ordered
by eigenvalue. The recovered collective weights are calculated from the
eigenvectors rather than copied from their inputs:
\begin{equation}
  A_\alpha^{\rm diag}=\frac1N\left|\bm c^{\mathsf T}\bm v_\alpha^{\rm diag}\right|^2.
\label{eq:recovered}
\end{equation}
The maximum absolute eigenvalue error, absolute weight error, and relative
weight error are
\begin{align}
  \epsilon_\lambda&=\max_\alpha|\lambda_\alpha^{\rm diag}-\lambda_\alpha^{\rm input}|,\\
  \epsilon_A&=\max_\alpha|A_\alpha^{\rm diag}-w_\alpha|,\\
  \epsilon_A^{\rm rel}&=\max_\alpha\frac{|A_\alpha^{\rm diag}-w_\alpha|}{\max(w_\alpha,10^{-15})}.
\label{eq:errors}
\end{align}
The denominator floor in $\epsilon_A^{\rm rel}$ follows the implementation and
prevents division by extremely small ultraviolet-tail weights; this reported
quantity is therefore a floored relative-error diagnostic, not a conventional
relative error for weights below $10^{-15}$.
\begin{table}[htbp]
  \centering
\caption{Maximum eigenvalue and collective-weight recovery errors for the two
constructions. Values are obtained from the direct $N=2048$ diagonalization.}
\label{tab:accuracy}
  \small
\begin{tabular}{@{}rllrrr@{}}
  \toprule
  $\beta$ & Construction & & $\epsilon_\lambda$ & $\epsilon_A$ & $\epsilon_A^{\rm rel}$\\
  \midrule
  $-0.50$ & A\_weighted && $1.24\times10^{-14}$ & $5.92\times10^{-14}$ & $3.35\times10^{-9}$\\
  & B\_eigenvalue && $1.33\times10^{-15}$ & $7.28\times10^{-15}$ & $1.49\times10^{-11}$\\
  $-0.25$ & A\_weighted && $1.24\times10^{-14}$ & $6.28\times10^{-15}$ & $5.15\times10^{-9}$\\
  & B\_eigenvalue && $1.33\times10^{-15}$ & $1.02\times10^{-15}$ & $2.09\times10^{-12}$\\
  $-0.10$ & A\_weighted && $1.24\times10^{-14}$ & $1.54\times10^{-15}$ & $4.16\times10^{-9}$\\
  & B\_eigenvalue && $1.33\times10^{-15}$ & $4.71\times10^{-16}$ & $9.65\times10^{-13}$\\
  $0.00$  & A\_weighted && $1.24\times10^{-14}$ & $1.16\times10^{-15}$ & $5.13\times10^{-9}$\\
  & B\_eigenvalue && $1.44\times10^{-15}$ & $1.58\times10^{-15}$ & $3.23\times10^{-12}$\\
  $0.25$  & A\_weighted && $1.24\times10^{-14}$ & $1.10\times10^{-15}$ & $6.50\times10^{-9}$\\
  & B\_eigenvalue && $1.55\times10^{-15}$ & $1.86\times10^{-15}$ & $3.80\times10^{-12}$\\
  $0.50$  & A\_weighted && $1.24\times10^{-14}$ & $1.31\times10^{-15}$ & $7.08\times10^{-9}$\\
  & B\_eigenvalue && $1.55\times10^{-15}$ & $7.43\times10^{-16}$ & $1.52\times10^{-12}$\\
  \bottomrule
\end{tabular}
\end{table}

\begin{figure}[htbp]
\centering
\paperfigure[scale=0.63, trim=0cm 9.8cm 0cm 0cm]{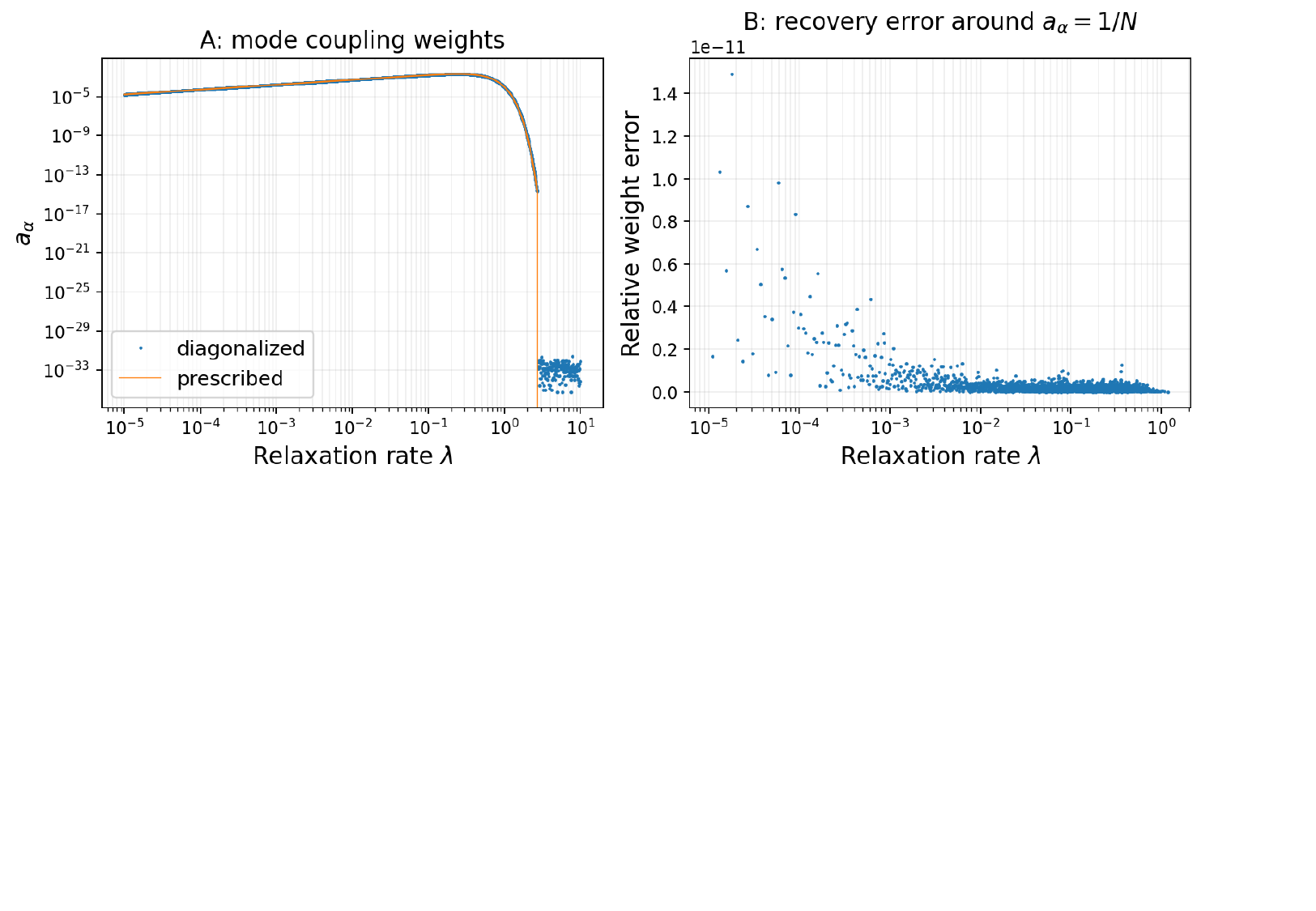}
\caption{\textbf{Numerical verification of collective-mode overlap weights at
$\beta=-0.5$.} Left: prescribed weights $w_\alpha$ (line) and weights
$A_\alpha^{\rm diag}=|\bm c^{\mathsf T}\bm v_\alpha^{\rm diag}|^2/N$ recovered
by direct diagonalization (points) for A\_weighted. Right: relative deviations
of the recovered weights from $w_\alpha=1/N$ for B\_eigenvalue. The agreement
tests the consistency of the constructed matrix, collective coupling vector,
and eigensolver; it does not establish that the target distribution arises
dynamically.}
\label{fig:overlap}
\end{figure}

Figure~\ref{fig:overlap} illustrates the eigenvector-overlap check for
$\beta=-0.5$. In A, the prescribed mode weights are reproduced by the
independently diagonalized eigenvectors over the resolved range. The steep
decline of the weights at large $\lambda$ reflects the imposed ultraviolet
envelope; extremely small values in that tail are susceptible to
floating-point roundoff. In B, the recovered weights remain close to $1/N$,
with relative deviations near the numerical precision of the eigensolver.

\subsection{Static response and finite-size considerations}

For each diagonalized matrix, the zero-frequency collective response and
the corresponding quadratic instability threshold are evaluated as

\begin{equation}
  \chi_0(0)=\sum_\alpha\frac{A_\alpha^{\rm diag}}{\lambda_\alpha^{\rm diag}},\qquad
  J_c=\frac1{\chi_0(0)}.
\label{supp:eq:critical}
\end{equation}

These are the quantities compared in the main analysis. The similarity of the A
and B results reflects their construction from the same normalized target
collective measure, but the agreement is not exact at finite $N$. In
particular, logarithmic spacing in A resolves the infrared region with many
low-weight modes, whereas quantile spacing in B can leave relatively few
eigenvalues in the same interval. For example, at $\beta=0$ and
$\lambda<10^{-3}$, A contains 683 modes and B contains 5 modes. Their static
responses are $\chi_0^{(A)}(0)=23.766879$ and $\chi_0^{(B)}(0)=20.782395$,
respectively. Thus, the small diagonalization errors in
Table~\ref{tab:accuracy} should not be confused with the discretization error
of a continuum spectral integral, which can be enhanced by the factor
$1/\lambda$ in Eq.~\eqref{supp:eq:critical}.

The present checks validate the algebraic construction and numerical recovery
of the prescribed finite-dimensional operators and overlap weights. They do
not constitute a systematic $N\to\infty$ convergence study, nor do they
demonstrate spontaneous formation of the target infrared distribution. Within
these stated limits, they provide an independent matrix-level consistency
check for the collective spectral representation used in the main analysis.

\subsection{Eigenvector Overlaps and Infrared Contributions
to Collective Polarization}

\label{sec:supp_eigenvector}

The preceding sections established the numerical construction
of the microscopic relaxation matrices and verified the
accuracy of their eigenvalues and collective-mode weights.
We now examine how the eigenvector overlaps determine the
spectral distribution of collective response and how
different infrared exponents redistribute the static
response among relaxation modes.

We use the collective-mode overlap weights and the
collective-weighted spectral measure defined in
Eqs.~\eqref{eq:polarization} and \eqref{eq:measures}.

\subsubsection{Mode overlaps and eigenvector participation}

\begin{figure}[htbp]
\centering
\paperfigure[scale=0.61, trim=0cm 1.3cm 0cm 0cm]{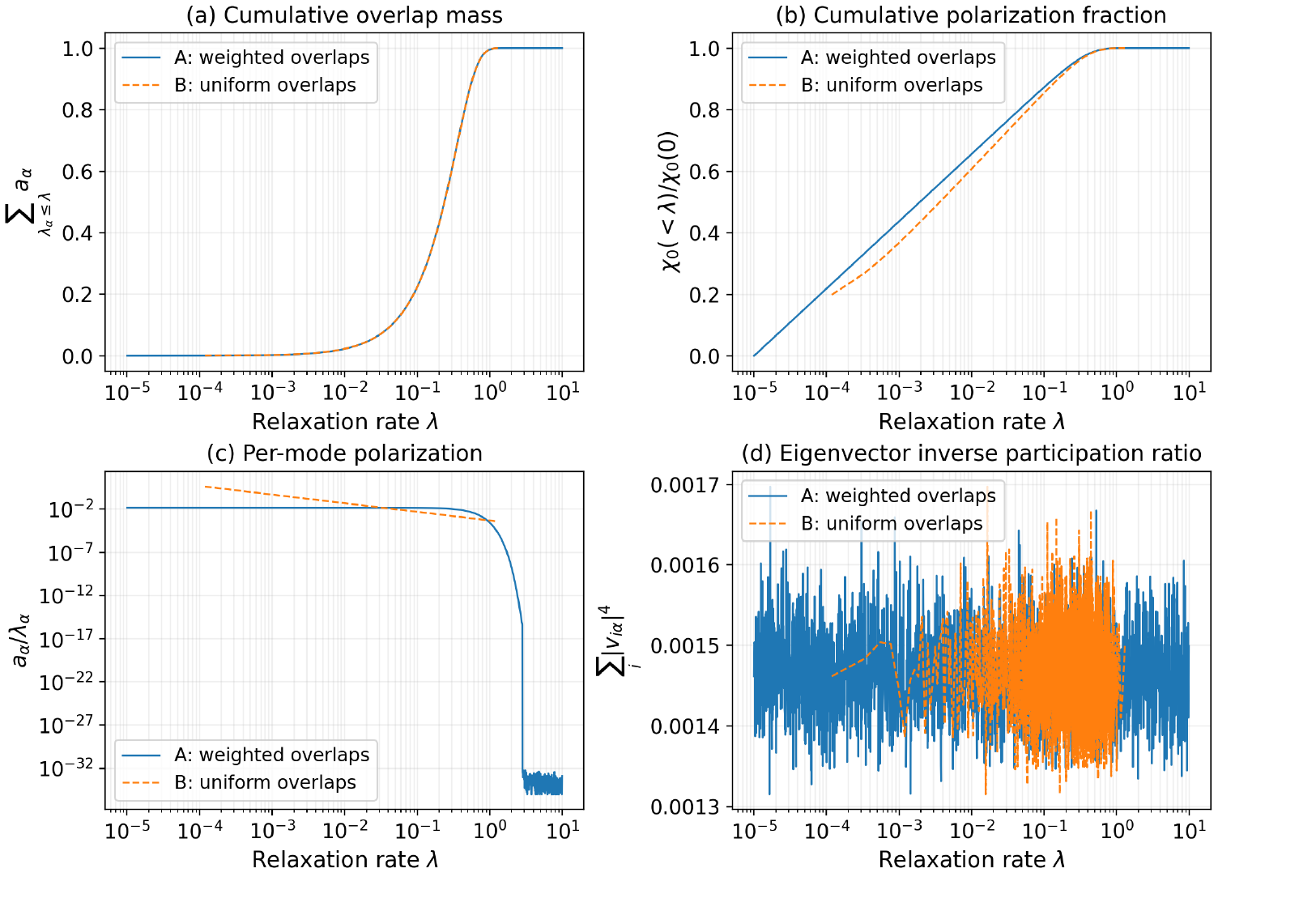}
\caption{
Eigenvector and collective-channel diagnostics for
$N=2048$ and $\beta=0$.
(a) Cumulative collective overlap mass
$\sum_{\lambda_\alpha\leq\lambda}A_\alpha$.
(b) Normalized cumulative static response
$F_\chi(\lambda)=\chi_0(<\lambda)/\chi_0(0)$.
(c) Individual mode contributions to the static response,
$A_\alpha/\lambda_\alpha$.
(d) Eigenvector inverse participation ratios,
$\mathrm{IPR}_\alpha=\sum_i |v_{i\alpha}|^4$.
The A-weighted and B-eigenvalue constructions approximate
the same collective spectral measure using different
discretizations.
Their participation statistics are consistent with
extended random eigenvectors.
}
\label{fig:supp_eigenvector}
\end{figure}

Figure~\ref{fig:supp_eigenvector} presents the eigenvector
and collective-channel diagnostics for $N=2048$ and
$\beta=0$. The A-weighted construction employs
logarithmically spaced relaxation rates with nonuniform
collective-mode weights, whereas the B-eigenvalue
construction distributes the relaxation rates according
to the prescribed target spectrum and assigns uniform
weights $A_\alpha=1/N$.

Although the individual mode weights differ substantially
between the two constructions, their cumulative overlap
distributions closely agree. This agreement is expected
because both constructions approximate the same normalized
collective spectral measure.

Using the static response in Eq.~\eqref{supp:eq:critical},
we identify the contribution from relaxation modes below
a given rate $\lambda$ by defining
\begin{equation}
F_\chi(\lambda)
=
\frac{\chi_0(<\lambda)}{\chi_0(0)}
=
\frac{
\displaystyle
\sum_{\lambda_\alpha<\lambda}
A_\alpha/\lambda_\alpha
}{
\displaystyle
\sum_\alpha A_\alpha/\lambda_\alpha
}.
\label{eq:supp_cumulative_pi}
\end{equation}
The middle panels of Fig.~\ref{fig:supp_eigenvector}
compare this cumulative response with the
mode-resolved contributions $A_\alpha/\lambda_\alpha$.
The small differences between the two constructions
reflect their distinct finite-size discretizations,
which become particularly relevant when the spectral
measure is weighted by $1/\lambda_\alpha$.

We additionally characterize the spatial extent of the
eigenvectors using the inverse participation ratio
\begin{equation}
\mathrm{IPR}_\alpha
=
\sum_{i=1}^{N}|v_{i\alpha}|^4
\label{eq:supp_ipr}
\end{equation}
and the normalized participation ratio
\begin{equation}
p_\alpha
=
\frac{1}{N\,\mathrm{IPR}_\alpha}.
\label{eq:supp_pr}
\end{equation}
For extended real random eigenvectors, the expected
inverse participation ratio is approximately
$3/(N+2)$, corresponding to a normalized participation
ratio near $1/3$ for large $N$.

The numerical results are consistent with these
expectations, indicating spatially extended random
eigenvectors rather than strongly localized modes.
Because the same orthogonal matrix is used in both
constructions, the eigenvector participation statistics
are not independent realizations of spatial organization.
The comparison instead isolates the effects of
eigenvalue placement and collective-channel overlaps.

\subsubsection{Infrared redistribution of static response}

We next examine how the infrared exponent $\beta$
controls the spectral distribution of collective
response. The prescribed collective-weighted
spectrum has the form
\begin{equation}
\rho_O(\lambda)
\propto
\lambda^\beta
\exp\left[
-\left(\frac{\lambda}{\lambda_{\mathrm{UV}}}\right)^2
\right],
\label{eq:supp_beta_spectrum}
\end{equation}
within the finite interval
$\lambda_{\mathrm{IR}}\leq\lambda\leq\lambda_{\max}$.

In the infrared regime, the static response receives
contributions proportional to
\begin{equation}
\chi_0(0)
\sim
\int_{\lambda_{\mathrm{IR}}}
\lambda^{\beta-1}\,d\lambda.
\label{eq:supp_ir_polarization}
\end{equation}
Consequently, the response exhibits logarithmic
infrared sensitivity for $\beta=0$ and stronger
power-law sensitivity for $\beta<0$, whereas positive
$\beta$ suppresses the relative contribution of the
slowest relaxation modes.

\begin{figure}[htbp]
\centering
\paperfigure[scale=0.56, trim=0.2cm 8.7cm 0cm 0cm]{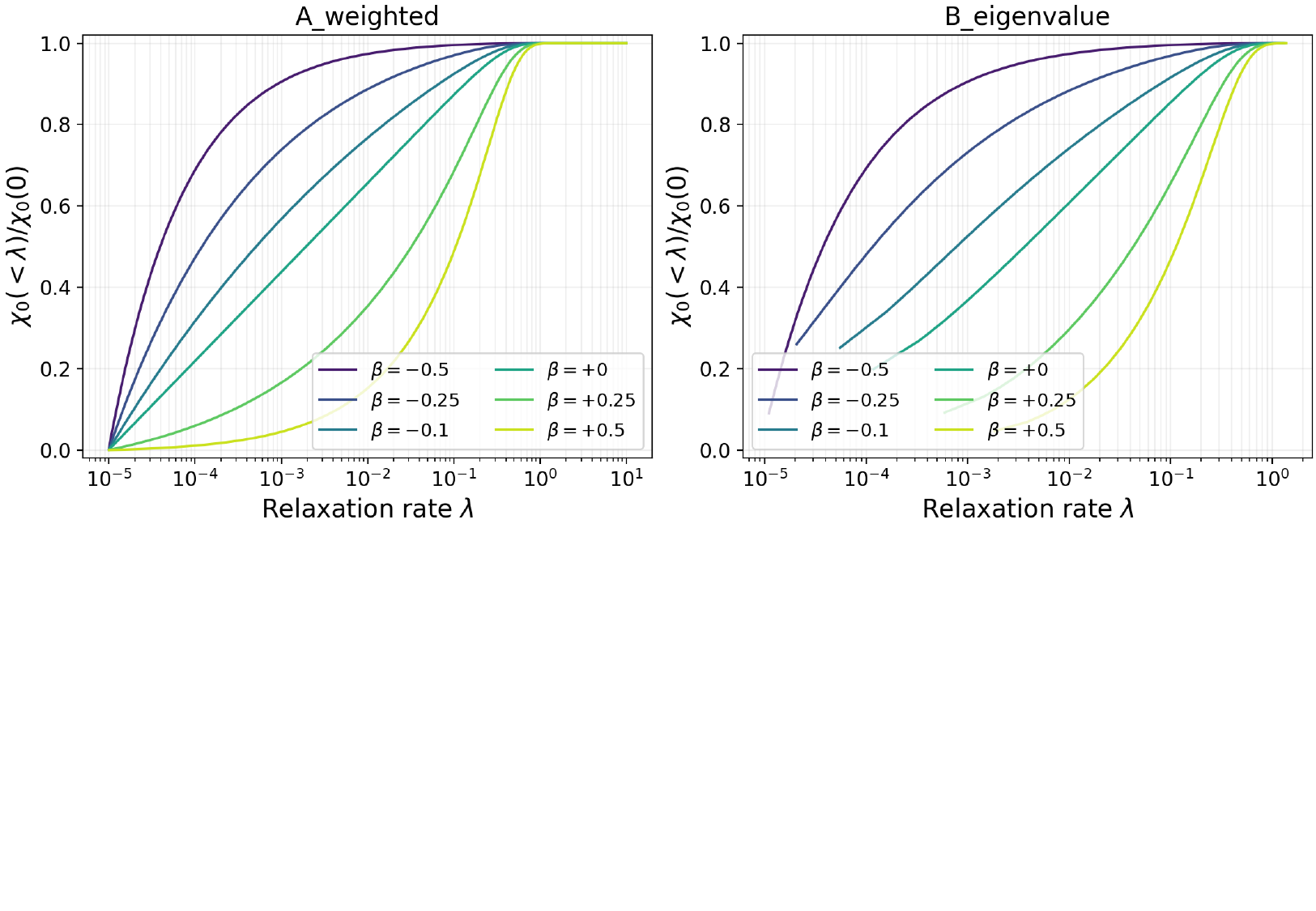}
\caption{
Normalized cumulative static response
$F_\chi(\lambda)=\chi_0(<\lambda)/\chi_0(0)$
for six infrared exponents $\beta$ at $N=2048$.
Left: A-weighted construction with logarithmically
spaced relaxation rates and nonuniform collective
overlap weights.
Right: B-eigenvalue construction with
spectrum-dependent relaxation rates and uniform
collective overlap weights.
As $\beta$ decreases, an increasing fraction of the
static collective response originates from
slow infrared modes.
Differences between the two constructions reflect
finite-size discretization effects.
}
\label{fig:supp_cumulative_beta}
\end{figure}

Figure~\ref{fig:supp_cumulative_beta} compares
$F_\chi(\lambda)$ for six infrared exponents,
$\beta=-0.5,-0.25,-0.1,0,0.25,$ and $0.5$.
In both microscopic constructions, decreasing $\beta$
systematically shifts the cumulative response
toward smaller relaxation rates. For negative $\beta$,
a substantial fraction of the total static response
is accumulated within the infrared sector.

The A-weighted and B-eigenvalue results exhibit the
same overall ordering with respect to $\beta$.
Quantitative differences are most apparent when
the infrared sector is sparsely sampled in the
B-eigenvalue construction. These differences arise
from finite-size spectral resolution rather than
from distinct prescribed continuum limits.

The results establish that the infrared distribution
of collective-mode spectral weight directly controls
the static collective response and therefore
the critical coupling in Eq.~\eqref{supp:eq:critical}.
This numerical comparison provides a controlled
microscopic illustration of the connection between
infrared relaxation spectra and collective instability.
The infrared distributions are prescribed in the
present construction; their spontaneous formation
under microscopic learning or interaction dynamics
is not assumed.

\subsection{Temporal Response Kernels and Finite-Size
Infrared Resolution}

The retarded response of the symmetric
overdamped relaxation operator
$L_0=U\operatorname{diag}(\lambda_\alpha)U^{\mathsf T}$
has the spectral representation
\begin{equation}
\chi_0^R(\omega)
=
\sum_{\alpha=1}^{N}
\frac{A_\alpha}{\lambda_\alpha-i\omega},
\qquad
A_\alpha
=
\frac{1}{N}
\left|
\mathbf c^{\mathsf T}\mathbf v_\alpha
\right|^2.
\end{equation}
The corresponding causal reference response kernel is
\begin{equation}
K_O(t)
=
\chi_0^R(t)
=
\Theta(t)
\sum_{\alpha=1}^{N}
A_\alpha e^{-\lambda_\alpha t}
=
\Theta(t)
\int_0^\infty d\lambda\,
\rho_O(\lambda)e^{-\lambda t},
\label{eq:supp_temporal_kernel}
\end{equation}
where
$\rho_O(\lambda)
=\sum_\alpha A_\alpha\delta(\lambda-\lambda_\alpha)$
is the collective-weighted relaxation spectrum.

For an infrared density
$\rho_O(\lambda)\simeq A_\beta\lambda^\beta$
with $\beta>-1$, the intermediate-time kernel
obeys
\begin{equation}
K_O(t)
\simeq
A_\beta
\int_{\lambda_{\rm IR}}^\Lambda
d\lambda\,\lambda^\beta e^{-\lambda t}
\simeq
A_\beta\Gamma(1+\beta)t^{-(1+\beta)},
\label{eq:supp_kernel_scaling}
\end{equation}
for
$\Lambda^{-1}\ll t\ll\lambda_{\rm IR}^{-1}$.
The marginal case $\beta=0$ therefore gives
$K_O(t)\propto t^{-1}$.

\begin{figure*}[t]
\centering
\paperfigure[scale=0.61, trim=-0.3cm 1.5cm 0cm 0cm]{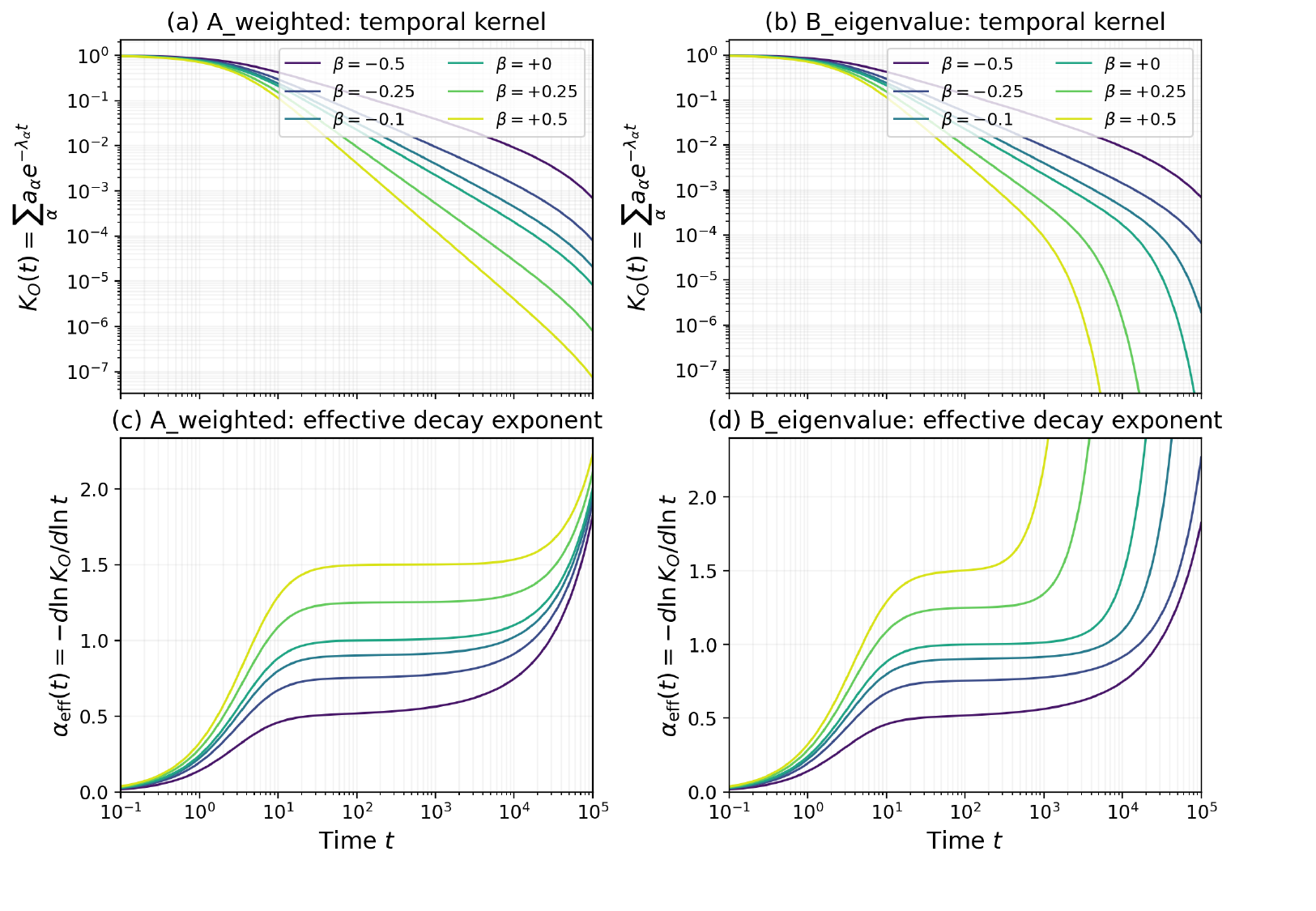}
\caption{Temporal response kernels from
collective-weighted relaxation spectra.
(a,b) Temporal kernels
$K_O(t)=\sum_\alpha A_\alpha e^{-\lambda_\alpha t}$
for the A-weighted and B-eigenvalue constructions,
respectively, with $N=2048$ and
$\beta=-0.5,-0.25,-0.1,0,0.25,0.5$.
(c,d) Corresponding effective decay exponents
$\alpha_{\rm eff}(t)=-d\ln K_O/d\ln t$.
Both constructions exhibit intermediate-time
power-law relaxation with
$\alpha_{\rm eff}\simeq1+\beta$.
For positive $\beta$, construction B shows
earlier finite-size crossovers due to
the sparse sampling of infrared modes
in the eigenvalue-based discretization.
}
\label{fig:supp_temporal_memory}
\end{figure*}

We evaluate the discrete kernels using the
A-weighted and B-eigenvalue constructions
of Sec.~\ref{app:numerics}, with
$N=2048$, $\lambda_{\rm IR}=10^{-5}$,
$\lambda_{\max}=10$, and
$\lambda_{\rm UV}=0.5$.
The infrared exponents are
$\beta=-0.5,-0.25,-0.1,0,0.25,0.5$.
Figures~\ref{fig:supp_temporal_memory}(a,b)
show comparable intermediate-time power-law
behavior in both constructions, despite their
different eigenvalue and overlap distributions.

The effective decay exponent is evaluated
directly from the discrete spectrum:
\begin{equation}
\alpha_{\rm eff}(t)
\equiv
-\frac{d\ln K_O(t)}{d\ln t}
=
t
\frac{
\sum_\alpha A_\alpha\lambda_\alpha
e^{-\lambda_\alpha t}
}{
\sum_\alpha A_\alpha e^{-\lambda_\alpha t}
}.
\label{eq:supp_alpha_eff}
\end{equation}
As shown in
Figs.~\ref{fig:supp_temporal_memory}(c,d),
both constructions develop intermediate-time
plateaus consistent with
$\alpha_{\rm eff}\simeq1+\beta$.
The agreement is particularly close for
$\beta\leq0$.

For $\beta>0$, construction B exhibits an
earlier finite-size crossover because its
quantile-based eigenvalue sampling becomes
sparse in the deep infrared.
Construction A instead retains logarithmically
spaced infrared eigenvalues with reduced
collective overlap weights.
This difference affects the duration of the
power-law regime rather than its continuum
scaling exponent.

For any finite spectrum with strictly
positive relaxation rates, the long-time
response is dominated by the slowest mode
with nonzero collective weight,
\begin{equation}
K_O(t)
\sim
a_{\alpha_*}e^{-\lambda_{\alpha_*}t},
\qquad
\lambda_{\alpha_*}
=
\min_{\alpha:A_\alpha>0}\lambda_\alpha,
\end{equation}
giving
$\alpha_{\rm eff}(t)\sim\lambda_{\alpha_*}t$.

The temporal kernel is determined by the
collective-weighted relaxation spectrum
rather than by the unweighted eigenvalue
density alone.

\section{Spectral organization and memory dynamics during learning}
\label{app:learning}
\label{sec:supp_learning_critical}

We examine the transient enhancement of the infrared spectral response
observed during Transformer learning. We first introduce a phenomenological
model separating the progressive organization of the background spectrum
from a transient critical-like enhancement. We then apply the spectral
reconstruction developed in the main text to distinguish the reference
response from the irreducible memory self-energy within a symmetric
uniform-weight relaxation surrogate.

\subsection{Learning-induced spectral reorganization}

A relaxation spectrum is already present before learning. Let the propagation
of the internal neural state between adjacent layers and its local Jacobian
be given by
\begin{align}
  \bm{x}_{l+1}&=F_l(\bm{x}_l;\bm{\theta}_s),\\
  \mathcal{J}_l(s)&=\frac{\partial F_l(\bm{x}_l;\bm{\theta}_s)}
  {\partial\bm{x}_l},
\end{align}
where $\bm{\theta}_s$ denotes the network parameters at training step $s$.
Denoting the eigenvalues of $\mathcal{J}_l(s)$ by $\zeta_\alpha(s)$, we define
the effective relaxation rates and the time-scale density of states (TDOS) as
\begin{align}
  \lambda_\alpha(s)&=-\frac{1}{\Delta l}\ln|\zeta_\alpha(s)|,\\
  \rho(\lambda;s)&=\frac{1}{N}\sum_{\alpha=1}^{N}
  \delta(\lambda-\lambda_\alpha(s)),
  \label{eq:supp_learning_tdos}
\end{align}
where $\Delta l=1$ for adjacent-layer propagation and $N$ is the Jacobian
dimension. Learning reorganizes an already existing spectrum through the
evolution of the network parameters, rather than creating a relaxation
spectrum from an initially absent dynamical structure.

For the stable sector, we define the regularized inverse-rate estimator
using the full-Jacobian normalization,
\begin{equation}
  \chi_{{\rm est},\epsilon}(0;s)
  =A_c\int_{0^+}^{\infty}d\lambda\,
  \frac{\rho(\lambda;s)}{\lambda+\epsilon}
  \label{eq:supp_learning_self_energy}
\end{equation}
or, equivalently,
\begin{equation}
  \chi_{{\rm est},\epsilon}(0;s)=\frac{A_c}{N}\sum_{\lambda_\alpha(s)>0}
  \frac{1}{\lambda_\alpha(s)+\epsilon},
  \label{eq:supp_numerical_learning_self_energy}
\end{equation}
where $A_c$ is a fixed collective-weight normalization and $\epsilon>0$
is an infrared regulator. Let $N_s$ denote the number of stable modes,
$f_{\rm st}(s)=N_s/N$ their fraction, and define the conditional spectral
average over these modes by
\begin{equation}
  \langle f(\lambda)\rangle_s\equiv\frac{1}{N_s}
  \sum_{\lambda_\alpha(s)>0}f(\lambda_\alpha(s)).
  \label{eq:supp_memory_spectral_average}
\end{equation}
For $N_s>0$, the stable-sector response estimator and its relation to the
full-Jacobian estimator are
\begin{align}
  P_\epsilon(s)&=\left\langle\frac{1}{\lambda+\epsilon}\right\rangle_s,
  \label{eq:supp_memory_regularized_response}\\
  \chi_{{\rm est},\epsilon}(0;s)&=A_cf_{\rm st}(s)P_\epsilon(s).
  \label{eq:supp_learning_normalization_relation}
\end{align}
The full-Jacobian estimator therefore combines changes in the stable-mode
population with changes in the distribution of rates within that population.
The conditional normalization isolates the latter and will be used in the
memory reconstruction below. Neither estimator incorporates the actual
collective-channel overlap weights or oscillatory components of the
Transformer Jacobian.

\subsection{Phenomenological model of transient spectral enhancement}

We phenomenologically decompose the full TDOS into a background spectrum
and a transient spectral correction,
\begin{equation}
  \rho(\lambda;s)=\rho_{\rm bg}(\lambda;s)
  +\Delta\rho_{\rm crit}(\lambda;s).
  \label{eq:supp_spectral_decomposition}
\end{equation}
The correction need not be positive at every rate, and the total spectrum
must remain normalized. By linearity of the inverse-rate integral,
\begin{equation}
  \chi_{{\rm est},\epsilon}(0;s)=\chi_{\rm bg}(s)
  +\Delta\chi_{\rm crit}(s),
  \label{eq:supp_self_energy_decomposition}
\end{equation}
where the two contributions are obtained by substituting the corresponding
spectral components into Eq.~\eqref{eq:supp_learning_self_energy}. This
separation is phenomenological and is not uniquely determined by the
measured total spectrum.

As a minimal approximation, we describe the background spectral evolution
by a first-order relaxation process in training-step space,
\begin{equation}
  \partial_s\rho_{\rm bg}(\lambda;s)
  =-\frac{\rho_{\rm bg}(\lambda;s)-\rho_f(\lambda)}{\tau},
  \label{eq:supp_background_dynamics}
\end{equation}
where $\rho_f$ is the effective late-training background spectrum and $\tau$
is a characteristic learning scale. With $\rho_i(\lambda)=\rho_{\rm bg}(\lambda;0)$,
the solution is
\begin{equation}
  \rho_{\rm bg}(\lambda;s)=\rho_f(\lambda)
  +[\rho_i(\lambda)-\rho_f(\lambda)]e^{-s/\tau}.
  \label{eq:supp_background_spectrum}
\end{equation}
Applying the same inverse-rate integral to this solution gives
\begin{equation}
  \chi_{\rm bg}(s)=\chi_f+(\chi_i-\chi_f)e^{-s/\tau},
  \label{eq:supp_background_self_energy}
\end{equation}
where $\chi_{i,f}=A_c\int_{0^+}^{\infty}d\lambda\,
\rho_{i,f}(\lambda)/(\lambda+\epsilon)$. For $\chi_f>\chi_i$, the background
response increases and subsequently saturates. The learning scale $\tau$
characterizes spectral evolution under training and must be distinguished
from the intrinsic inverse rate $\lambda_\alpha^{-1}$ at a fixed checkpoint.

To motivate the transient contribution, consider a hypothetical effective
collective model satisfying the assumptions of the main text. Its static
quadratic kernel and collective instability threshold are
\begin{align}
  \Gamma_2(0;s)&=\frac{1}{J_{\rm eff}(s)}-\chi_0^R(0;s),
  \label{eq:supp_learning_quadratic_kernel}\\
  J_c(s)&=\frac{1}{\chi_0^R(0;s)}.
\end{align}
Neither the effective coupling $J_{\rm eff}$ nor the actual reference
response $\chi_0^R$ is independently measured in the Transformer analysis.
On the disordered side, the dimensionless distance from the collective
instability and the Gaussian collective susceptibility are
\begin{align}
  \delta_J(s)&=1-J_{\rm eff}(s)\chi_0^R(0;s)
  =1-\frac{J_{\rm eff}(s)}{J_c(s)},
  \label{eq:supp_learning_critical_distance}\\
  \chi_m^R(0;s)&=\frac{\chi_0^R(0;s)}{\delta_J(s)}.
\end{align}
Provided that the numerator remains finite and regular, the Gaussian
response scales as $\delta_J^{-1}$. More generally, we introduce the
phenomenological susceptibility exponent $\gamma$ through
\begin{equation}
  \chi_m^R(0;s)\sim\delta_J(s)^{-\gamma},
  \label{eq:supp_learning_critical_scaling}
\end{equation}
with $\gamma_{\rm MF}=1$ at the Gaussian mean-field level. Suppose that a
smooth learning trajectory approaches the instability and subsequently
moves away without crossing it. Near a nondegenerate minimum at $s_c$,
where $\delta_J'(s_c)=0$ and $\delta_J''(s_c)>0$, the leading expansion is
\begin{equation}
  \delta_J(s)=\delta_{\min}+a(s-s_c)^2+O((s-s_c)^3),
  \label{eq:supp_learning_taylor}
\end{equation}
so that, to quadratic order,
\begin{equation}
  \delta_J(s)\simeq\delta_{\min}+a(s-s_c)^2,
  \label{eq:supp_learning_critical_trajectory}
\end{equation}
where $\delta_{\min}=\delta_J(s_c)>0$ and $a=\delta_J''(s_c)/2$. Substitution
into the critical-response form yields
\begin{equation}
  \chi_m^R(0;s)\simeq\frac{C}{[\delta_{\min}+a(s-s_c)^2]^\gamma}
  \label{eq:supp_learning_local_response}
\end{equation}
or, equivalently,
\begin{equation}
  \chi_m^R(0;s)\simeq\frac{B_\chi}{[1+((s-s_c)/w)^2]^\gamma},
  \label{eq:supp_learning_generalized_response}
\end{equation}
where $C$ is a locally regular amplitude, $B_\chi=C\delta_{\min}^{-\gamma}$,
and the characteristic width is
\begin{equation}
  w=\sqrt{\delta_{\min}/a}.
  \label{eq:supp_learning_width}
\end{equation}
The quadratic power follows from the local expansion of the trajectory,
whereas $\gamma$ characterizes the assumed susceptibility scaling. For
$\gamma=1$, the resulting profile is Lorentzian in the training coordinate.

Motivated by this profile, we approximate the transient spectral correction
by
\begin{equation}
  \Delta\chi_{\rm crit}(s)\simeq
  \frac{B_\chi}{[1+((s-s_c)/w)^2]^\gamma}.
  \label{eq:supp_transient_self_energy}
\end{equation}
This is an ansatz for the measured spectral estimator, not a derivation
of its critical scaling from Transformer dynamics. Combining it with the
background response gives
\begin{equation}
  \chi_{{\rm est},\epsilon}(0;s)\simeq
  \chi_f+(\chi_i-\chi_f)e^{-s/\tau}
  +\frac{B_\chi}{[1+((s-s_c)/w)^2]^\gamma}.
  \label{eq:supp_total_learning_self_energy}
\end{equation}
For numerical comparison, we normalize by the maximum measured estimator
$\chi_{\rm est,max}$ over the reference training trajectory. The fitting
function is then
\begin{equation}
  \begin{aligned}
    \overline{\chi}_{\rm est}(s)&=\overline{\chi}_f
    +(\overline{\chi}_i-\overline{\chi}_f)e^{-s/\tau}\\
    &\quad+\frac{B}{[1+((s-s_c)/w)^2]^\gamma},
  \end{aligned}
  \label{eq:supp_final_learning_fit}
\end{equation}
where $\overline{\chi}_{i,f}=\chi_{i,f}/\chi_{\rm est,max}$ and
$B=B_\chi/\chi_{\rm est,max}$. Because the transient term generally remains
finite at $s=0$, imposing the initial measured value gives the constraint
\begin{equation}
  \overline{\chi}_i=\overline{\chi}_{\rm est}(0)
  -\frac{B}{[1+(s_c/w)^2]^\gamma}.
  \label{eq:supp_initial_background_constraint}
\end{equation}
Thus $\overline{\chi}_i$ denotes the initial background amplitude rather than
the total initial response. Similarly, the fitted center $s_c$ of the
transient component need not coincide with the maximum of the total
estimator because the background itself evolves during learning.

\subsection{Numerical illustration using Transformer training checkpoints}

We analyze adjacent-layer propagation from layer 5 to layer 6 in the
Pythia-410M-deduped Transformer. This layer pair is selected because its
transient enhancement extends over a relatively broad training-step
interval, allowing better resolution of the peak shape with the available
public checkpoints. The full-Jacobian estimator in
Eq.~\eqref{eq:supp_learning_self_energy} is evaluated with $A_c=1$ and
$\epsilon=10^{-3}$, retaining all positive rates. The original numerical
sum omitted the prefactor $1/N$; for fixed $N$ and $A_c$, this constant
factor cancels exactly upon normalization by the maximum.

We compare Model I, with $\gamma=1$, and Model II, with $\gamma$ free,
using Eq.~\eqref{eq:supp_final_learning_fit}. In both models the quadratic
power is fixed, the initial total estimator is constrained by
Eq.~\eqref{eq:supp_initial_background_constraint}, and the late-training
background amplitude $\overline{\chi}_f$ is fixed to the same data-derived
estimate. The fit uses nine measurements over $0\leq s\leq30000$.
The resulting parameters are summarized in Table~\ref{tab:supp_learning_critical_fit}.
\begin{table}[htbp]
\centering
\caption{
Phenomenological fitting parameters for the
normalized infrared spectral estimator of
Pythia-410M-deduped, layer $5\to6$.
Model I uses the mean-field exponent $\gamma=1$,
whereas Model II allows $\gamma$ to vary.
The uncertainty in $\gamma$ is the conditional
standard error obtained from the local fit.
}
\label{tab:supp_learning_critical_fit}
\begin{tabular}{@{}lrr@{}}
\toprule
Parameter & Model I & Model II \\
\midrule
$\overline{\chi}_i$       & 0.2063 & 0.3046 \\
$\overline{\chi}_f$       & 0.6386 & 0.6386 \\
$\tau$      & 1532.8 & 1016.0 \\
$B$         & 0.4994 & 0.4331 \\
$s_c$       & 1430.3 & 1578.3 \\
$w$         & 1884.8 & 1123.7 \\
$\gamma$    & 1 (fixed) & $0.63\pm0.51$ \\
RMSE        & 0.02792 & 0.02679 \\
$R^2$       & 0.95895 & 0.96219 \\
\bottomrule
\end{tabular}
\end{table}

Both models describe the transient enhancement with comparable accuracy.
The free exponent, $\gamma_{\rm fit}=0.63\pm0.51$, is compatible with the
mean-field value within its substantial conditional uncertainty, and the
additional parameter produces only a small reduction in RMSE. The finite
width is consistent with a rounded critical-like profile, but the fit does
not establish proximity to an actual collective instability.

Subtracting the fitted background and introducing $x=(s-s_c)/w$ gives
\begin{equation}
  \frac{\overline{\chi}_{\rm fit}(s)-\overline{\chi}_{\rm bg}(s)}{B}
  =\mathcal{F}_\gamma(x),\qquad
  \mathcal{F}_\gamma(x)=(1+x^2)^{-\gamma},
  \label{eq:supp_learning_scaling_form}
\end{equation}
where $\overline{\chi}_{\rm fit}$ denotes the fitted curve. Its formal tail
is $\mathcal{F}_\gamma(x)\sim|x|^{-2\gamma}$, provided that the local
quadratic approximation remains applicable. This representation may guide
future comparisons across layers and model sizes, but a collapse using
the same fitted data is not an independent test of universality. Denser
checkpoints, independent estimates of the collective instability distance,
and direct collective-channel response measurements would be required to
determine a susceptibility exponent. The present fit is therefore a
phenomenological illustration rather than a determination of a universality
class of Transformer learning.

\subsection{Reconstruction of the irreducible memory self-energy}
\label{supp:transformer_memory_reconstruction}

The exact inverse-response relation developed in the main text provides a
route to reconstructing the irreducible memory self-energy within a
symmetric uniform-weight surrogate. Unlike the full-Jacobian estimator
used in the preceding fit, the reconstruction uses the conditional
stable-sector measure
\begin{equation}
  \rho_{O,\mathrm{est}}(\lambda;s)=\frac{A_c}{N_s}
  \sum_{\lambda_\alpha(s)>0}\delta(\lambda-\lambda_\alpha(s)),
\end{equation}
whose total weight is $A_c$. The corresponding reference response and
direct relaxation scale are
\begin{align}
  \chi_{0,\mathrm{est}}(z;s)&=A_c
  \left\langle\frac{1}{z+\lambda}\right\rangle_s,
  \label{eq:supp_memory_reference_response}\\
  r_{0,\mathrm{est}}(s)&=\langle\lambda\rangle_s.
  \label{eq:supp_memory_direct_rate}
\end{align}
Using $\Sigma(z)=z+r_0-A_c/\chi_0(z)$, we obtain
\begin{equation}
  \Sigma_{\mathrm{est}}(z;s)=z+\langle\lambda\rangle_s
  -\left\langle\frac{1}{z+\lambda}\right\rangle_s^{-1}.
  \label{eq:supp_memory_reconstruction}
\end{equation}
In particular, the static self-energy is the difference between the
arithmetic and harmonic means of the retained rates,
\begin{equation}
  \Sigma_{\mathrm{est}}(0;s)=\langle\lambda\rangle_s
  -\langle\lambda^{-1}\rangle_s^{-1}.
  \label{eq:supp_memory_static_reconstruction}
\end{equation}
Defining $H_s=\langle\lambda^{-1}\rangle_s^{-1}$, this expression can
also be written as
\begin{equation}
  \Sigma_{\mathrm{est}}(0;s)=
  \left\langle\frac{(\lambda-H_s)^2}{\lambda}\right\rangle_s\geq0.
  \label{eq:supp_memory_positive_representation}
\end{equation}
This representation makes explicit that a spectrum consisting of a single
relaxation rate generates no irreducible memory feedback within the
symmetric surrogate. The reconstruction also determines the frequency
response by analytic continuation $z=0^+-i\omega$ in
Eq.~\eqref{eq:supp_memory_reconstruction}.

To reduce sensitivity to individual near-zero modes, we evaluate the
reference response and self-energy consistently at $z=\epsilon>0$.
Using $P_\epsilon=\chi_{0,\mathrm{est}}(\epsilon;s)/A_c$ as defined in
Eq.~\eqref{eq:supp_memory_regularized_response}, we obtain
\begin{align}
  \Sigma_{\mathrm{est}}(\epsilon;s)&=r_{0,\mathrm{est}}(s)
  +\epsilon-P_\epsilon^{-1}(s),
  \label{eq:supp_memory_regularized_sigma}\\
  \Delta_\epsilon(s)&=r_{0,\mathrm{est}}(s)+\epsilon
  -\Sigma_{\mathrm{est}}(\epsilon;s)=P_\epsilon^{-1}(s).
  \label{eq:supp_memory_inverse_response}
\end{align}
The relative strength of memory dressing and the complementary
inverse-response ratio are
\begin{align}
  D_\epsilon(s)&=\frac{\Sigma_{\mathrm{est}}(\epsilon;s)}
  {r_{0,\mathrm{est}}(s)+\epsilon}
  =1-\frac{1}{[r_{0,\mathrm{est}}(s)+\epsilon]P_\epsilon(s)},
  \label{eq:supp_memory_dressing_ratio}\\
  \delta_\epsilon(s)&=1-D_\epsilon(s)
  =\frac{\Delta_\epsilon(s)}{r_{0,\mathrm{est}}(s)+\epsilon}.
  \label{eq:supp_memory_relative_inverse_response}
\end{align}
For positive rates, $0\leq D_\epsilon<1$. At $\epsilon=0$, these ratios
are invariant under a uniform rescaling of all rates; at finite $\epsilon$,
the same invariance holds if the regulator is rescaled together with the
rates. They characterize relative spectral organization rather than the
absolute relaxation scale.

The reference response and irreducible memory self-energy need not exhibit
the same learning dynamics. At fixed $\epsilon$, differentiating
Eq.~\eqref{eq:supp_memory_regularized_sigma} with respect to a smooth
interpolation of the training coordinate gives
\begin{equation}
  \frac{d\Sigma_{\mathrm{est}}(\epsilon;s)}{ds}
  =\frac{dr_{0,\mathrm{est}}}{ds}
  +\frac{1}{P_\epsilon^2}\frac{dP_\epsilon}{ds}.
  \label{eq:supp_memory_learning_derivative}
\end{equation}
At a smooth interior maximum $s_p$ of $P_\epsilon$, the inverse-response
scale $\Delta_\epsilon=P_\epsilon^{-1}$ attains a minimum and
\begin{equation}
  \left.\frac{d\Sigma_{\mathrm{est}}(\epsilon;s)}{ds}\right|_{s_p}
  =\left.\frac{dr_{0,\mathrm{est}}}{ds}\right|_{s_p}.
  \label{eq:supp_memory_competing_rates}
\end{equation}
The transient response maximum therefore does not generally coincide with
a maximum of the irreducible memory self-energy. It reflects the
competition between direct relaxation and memory feedback. Here $s_p$
refers to the maximum of the conditional response, not the fitted center
$s_c$ of the transient full-Jacobian estimator.

\begin{figure*}[t]
  \centering
  \paperfigure[scale=0.78, trim=3cm 1.3cm 0cm 0cm]{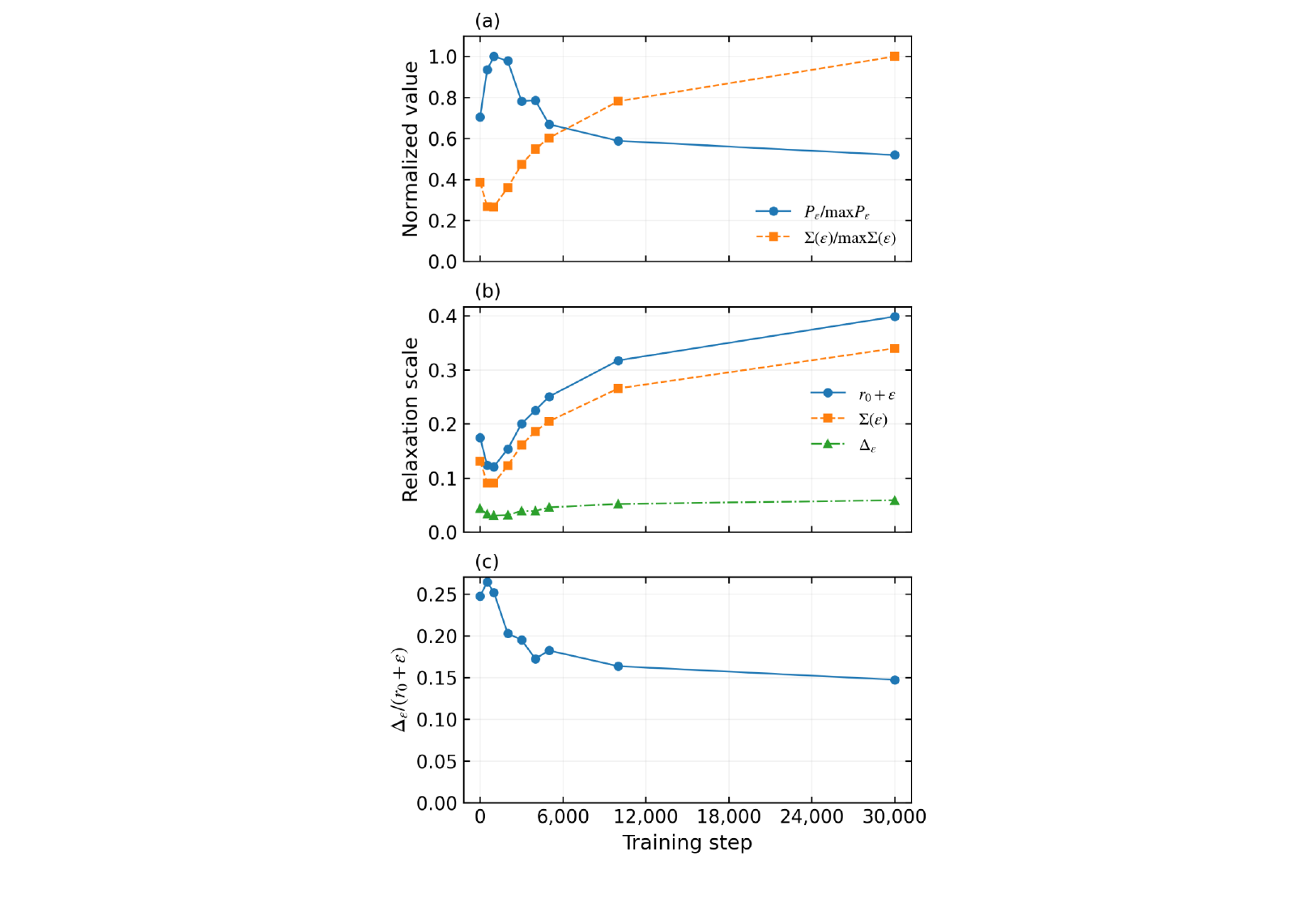}
  \caption{Reconstruction of the irreducible memory self-energy from
  Transformer Jacobian relaxation spectra during learning for Pythia-410M,
  layer transition $5\rightarrow6$, with eight input tokens, using the
  symmetric uniform-weight surrogate and $\epsilon=10^{-3}$.
  (a) Conditional reference-response estimator $P_\epsilon$ and reconstructed
  self-energy $\Sigma_{\mathrm{est}}(\epsilon)$, each normalized by its
  maximum over the measured interval. (b) Direct relaxation scale
  $r_{0,\mathrm{est}}+\epsilon$, self-energy $\Sigma_{\mathrm{est}}(\epsilon)$,
  and inverse-response scale $\Delta_\epsilon$. (c) Relative inverse-response
  scale $\delta_\epsilon=\Delta_\epsilon/(r_{0,\mathrm{est}}+\epsilon)$.
  The transient response maximum reflects the competition between direct
  relaxation and memory feedback rather than a maximum of the self-energy.}
  \label{fig:supp_memory_learning}
\end{figure*}

Figure~\ref{fig:supp_memory_learning} illustrates this distinction for
Pythia-410M, layer transition $5\rightarrow6$, with an input sequence length
of eight tokens and $\epsilon=10^{-3}$. Among the measured checkpoints,
the conditional response reaches its maximum near $s=10^3$, whereas the
reconstructed self-energy initially decreases and subsequently increases
throughout the remaining measured interval. The full-Jacobian estimator
peaks near $s=2\times10^3$. The difference is consistent with the additional
stable-mode population factor in
Eq.~\eqref{eq:supp_learning_normalization_relation}.

The regularized inverse-response scale increases from approximately $0.0305$
at $s=10^3$ to $0.0587$ at $s=3\times10^4$, while the relative memory-dressing
strength increases from $0.748$ to $0.853$. Thus, the absolute inverse-response
scale increases during later learning even as the relative suppression of
the direct relaxation scale becomes stronger. This behavior reflects the
simultaneous evolution of direct relaxation and memory feedback, rather
than a simple weakening of memory after the response maximum.

\subsection{Infrared marginality, collective instability, and spectral sensitivity}

Within the microscopic collective-interaction model of the main text,
the reference inverse-response scale $\Delta_0=r_0-\Sigma(0)$ and the
interaction-dressed inverse-response scale $\Delta_J=\Delta_0-JA_c$ play
distinct roles. The collective instability threshold introduced above
can equivalently be written as
\begin{equation}
  J_c=\frac{r_0-\Sigma(0)}{A_c}=\frac{1}{\chi_0(0)}.
  \label{eq:supp_memory_collective_threshold}
\end{equation}
A reduction of the static reference inverse-response scale lowers the
coupling required for a collective instability. However, the measured
regularized scale $\Delta_\epsilon$ and its relative ratio $\delta_\epsilon$
are not the static interaction-dressed inverse susceptibility or the
collective instability distance $\delta_J$.

As established by the flat-spectrum analysis in Sec.~I, a response-weighted
spectrum $\rho_O(\lambda)=A$ extending to zero rate yields, in the gapless
continuum limit and for $\epsilon\ll\Lambda$,
\begin{equation}
  \chi_0(\epsilon)\sim A\ln(\Lambda/\epsilon),\qquad
  \Delta_\epsilon\sim\frac{A_c}{A\ln(\Lambda/\epsilon)}.
  \label{eq:supp_memory_logarithmic_softening}
\end{equation}
The infrared singularity appears through the suppression of the reference
inverse-response scale, while the irreducible self-energy remains finite
and approaches $r_0$. Infrared marginality concerns the organization of
the internal spectrum, whereas collective criticality concerns the
vanishing of the interaction-dressed inverse susceptibility. The finite,
regularized Transformer estimates do not establish the gapless continuum
limit or independently determine the collective coupling $J$.

The inverse-rate moments can be strongly influenced by a small number of
near-zero modes. To quantify this sensitivity, define the normalized
contribution of each stable mode by
$q_{\alpha,\epsilon}=(\lambda_\alpha+\epsilon)^{-1}/
\sum_{\gamma\in\mathrm{stable}}(\lambda_\gamma+\epsilon)^{-1}$.
The effective participation number is
\begin{equation}
  N_{\mathrm{eff},\epsilon}=\frac{1}{\sum_\alpha q_{\alpha,\epsilon}^2}
  =N_s\frac{\langle(\lambda+\epsilon)^{-1}\rangle_s^2}
  {\langle(\lambda+\epsilon)^{-2}\rangle_s}.
  \label{eq:supp_memory_participation}
\end{equation}
In the equilibrium Gaussian surrogate with uniformly shifted rates
$\lambda_\alpha+\epsilon$ and $J=0$, the integrated correlation time is
\begin{equation}
  \tau_{\mathrm{int},\epsilon}^{(0)}=
  \frac{\langle(\lambda+\epsilon)^{-2}\rangle_s}
  {\langle(\lambda+\epsilon)^{-1}\rangle_s},\qquad
  N_{\mathrm{eff},\epsilon}=
  \frac{N_sP_\epsilon}{\tau_{\mathrm{int},\epsilon}^{(0)}}.
\end{equation}
At $\epsilon=0$, this reduces to the reference correlation time derived
in the main text. At finite $\epsilon$, it describes the shifted surrogate,
not the physical correlation time of the unshifted system.

For the unregularized spectrum at $s=3\times10^4$, the effective participation
number is approximately $2.6$, despite the presence of approximately
$6\times10^3$ stable modes. This indicates strong concentration of the
inverse-rate contributions, with the second inverse-rate moment even more
sensitive to the smallest rates. The regularized estimates nevertheless
retain the qualitative distinction between the transient reference-response
maximum and the subsequent growth of the reconstructed self-energy.
Further analysis of regulator dependence and numerical precision is needed
to distinguish continuum infrared scaling from individual near-zero-mode
contributions.

Finally, the reconstruction is exact for the specified symmetric surrogate,
not for the actual Transformer Jacobian. The latter is generally nonnormal,
requires biorthogonal response weights, and describes propagation along a
computational coordinate rather than physical-time relaxation. The present
results therefore constitute spectral estimates of reference response and
memory dressing, rather than direct measurements of physical collective
susceptibility, memory self-energy, temporal persistence, or a collective
ordering transition.

\end{document}